\documentclass[aip, amsmath, amssymb, reprint, longbibliography]{revtex4-2}
\usepackage{graphicx}
\usepackage{dcolumn}
\usepackage{bm}
\usepackage{color}
\usepackage{braket}
\usepackage{amsmath,amssymb,amsfonts,bm}

\def\ben{\begin{equation}}
\def\een{\begin{equation}}
\def\mr{\mathbf{r}}

\begin{document}

\title{Density functional descriptions of interfacial electronic structure}
\author{Zhen-Fei Liu}
\email{zfliu@wayne.edu}
\affiliation{Department of Chemistry, Wayne State University, Detroit, Michigan 48202 USA}

\date{\today}

\begin{abstract}
Heterogeneous interfaces are central to many energy-related applications in the nanoscale. From the first-principles electronic structure perspective, one of the outstanding problems is accurately and efficiently calculating how the frontier quasiparticle levels of one component are aligned in energy with those of another at the interface, i.e., the so-called interfacial band alignment or level alignment. The alignment or the energy offset of these frontier levels is phenomenologically associated with the charge-transfer barrier across the interface and therefore dictates the interfacial dynamics. Although many-body perturbation theory provides a formally rigorous framework for computing the interfacial quasiparticle electronic structure, it is often associated with a high computational cost and is limited by its perturbative nature. It is therefore of great interest to develop practical alternatives, preferably based on density functional theory (DFT), which is known for its balance between efficiency and accuracy. However, conventional developments of density functionals largely focus on total energies and thermodynamic properties, and the design of functionals aiming for interfacial electronic structure is only emerging recently. This Review is dedicated to a self-contained narrative of the interfacial electronic structure problem and the efforts of the DFT community in tackling it. Since interfaces are closely related to surfaces, we first discuss the key physics behind the surface and interface electronic structure, namely the image potential and the gap renormalization. This is followed by a review of early examinations of the surface exchange-correlation hole and the exchange-correlation potential, which are central quantities in DFT. Lastly, we survey two modern endeavors in functional development that focus on the interfacial electronic structure, namely the dielectric-dependent hybrids and local hybrids.
\end{abstract}

\maketitle

\tableofcontents

\section{Introduction}

Interfaces are intrinsically heterogeneous, where two different materials meet and interact. Interfaces are key in energy conversion applications in the nanoscale, including photovoltaics \cite{GHS14}, photocatalysis \cite{WLD19}, field-effect transistors \cite{LHD19}, energy storage devices such as batteries \cite{GP13}, and many others. To understand the mechanism behind each type of energy conversion and charge dynamics, one needs an accurate description of the underlying electronic structure at the interface. To this end, first-principles calculations are indispensable in providing microscopic structure-property relationships that could validate, interpret, and guide experimental measurements.

\begin{figure*}[htp]
\centering
\includegraphics[width=5.5in]{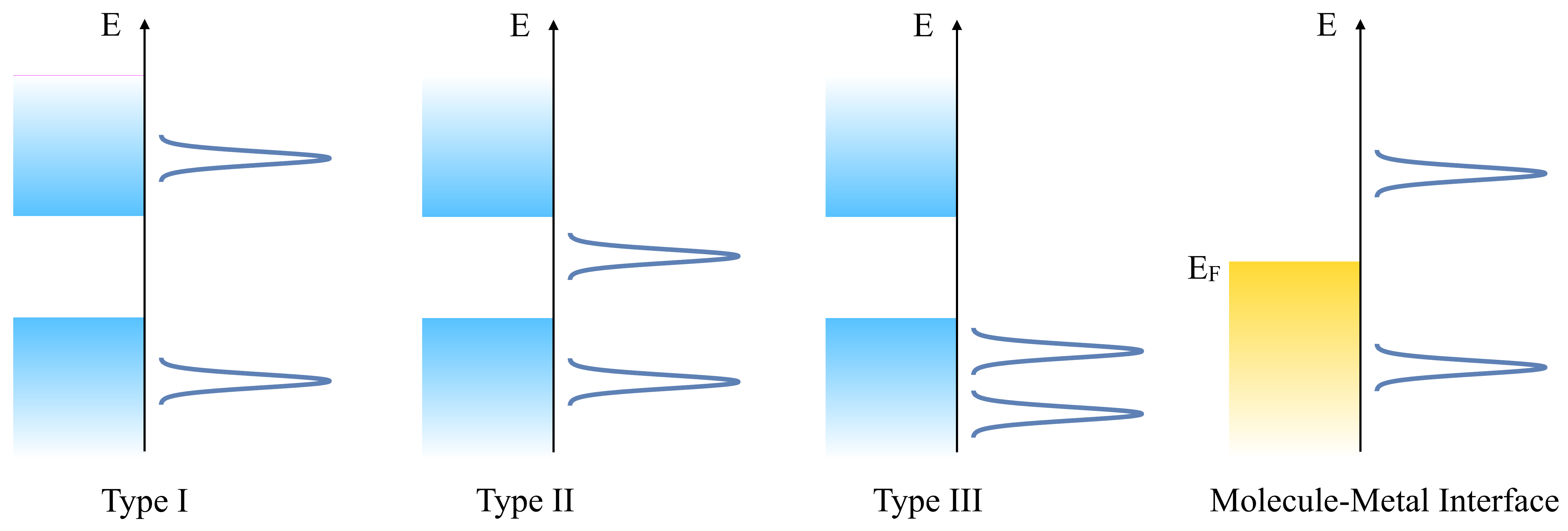}
\caption{The three types of molecule-semiconductor interfaces (the first three panels) and the molecule-metal interface (the last panel). Blue blocks denote the valence and conduction bands of semiconductor substrates. The yellow block denotes the occupied bands of a metal substrate, whose Fermi level is denoted by $E_{\rm F}$. Broadened lines denote the HOMO and LUMO energy levels of an adsorbed molecule on a semiconductor or a metal substrate. If one replaces the HOMO and LUMO with the VBM and the CBM of another semiconductor material, one would obtain semiconductor-semiconductor or metal-semiconductor interfaces.}
\label{fig:types}
\end{figure*}

From the electronic structure perspective, the properties of frontier orbitals/bands determine reactivity. For a heterogeneous interface, how the frontier levels of each component align in energy determines charge and energy flow across the interface. In the simplest picture of the interfacial electronic structure, for systems with a finite transport gap, i.e., semiconductors and molecules, we limit our scope to two frontier energy levels, one occupied and one unoccupied. While for metals, we use a single energy level, namely the Fermi level $E_{\rm F}$, to describe them. Based on the energy offsets of these frontier orbitals/bands, the possible energy level alignments at molecule-substrate interfaces are schematically represented in Fig. \ref{fig:types}. Note that if one replaces the molecular energy levels with the frontier bands of another semiconductor material, one would obtain the energy level alignments of semiconductor-semiconductor or metal-semiconductor interfaces.

In Fig. \ref{fig:types}, the colored blocks represent bands of the semiconductor or metal substrate, and the broadened lines represent the highest occupied molecular orbitals (HOMOs) and the lowest unoccupied molecular orbitals (LUMOs) of the molecular adsorbate. The first three panels are analogous to the three types of heterojunctions of semiconductors \cite{I10}: (i) straddled bands in Type I, where the LUMO (HOMO) is above (below) the conduction band minimum (CBM) [valence band maximum (VBM)] of the semiconductor substrate. An alternative situation, where both the HOMO and LUMO are within the VBM-CBM gap of the semiconductor substrate, is not shown; (ii) staggered bands in Type II, where the LUMO is within the VBM-CBM gap and the HOMO is below the VBM of the semiconductor substrate. An alternative situation, where the LUMO is above the CBM and the HOMO is within the VBM-CBM gap of the semiconductor substrate, is not shown; and (iii) broken bands in Type III, where both HOMO and LUMO are below the VBM of the semiconductor substrate. An alternative situation, where both HOMO and LUMO are above the CBM of the semiconductor substrate, is not shown. The last panel shows a molecule-metal interface, where the LUMO (HOMO) is above (below) the $E_{\rm F}$ of the metal substrate. Because electrons tend to move downward and holes tend to move upward in energy during a charge transfer, different level or band alignment patterns lead to different charge and energy transfer dynamics across the interface.

Strictly speaking, all the frontier levels of interest, namely the VBM and CBM of the semiconductor substrates, the HOMO and LUMO of the molecular adsorbates, and $E_{\rm F}$ of the metal substrates, are charged quasiparticle energy levels, i.e., the energy cost of removing one electron from or adding one electron to the system. In density functional theory (DFT), the Hohenberg-Kohn theorem \cite{HK64} does (formally) guarantee that all properties of a system are determined by the ground-state density, $n(\mathbf{r})$. However, under the Kohn-Sham (KS) formulation \cite{KS65}, only the form of $E[n(\mathbf{r})]$, the ground-state total energy as a functional of the density, is explicitly developed. Importantly, although the eigenvalues of the KS Hamiltonian are often interpreted as quasiparticle energies, there are no rigorous justifications for doing so, with the exception of the highest occupied level thanks to Koopmans' theorem. Even for the latter, in the context of the interface, it refers to the highest occupied energy level of \emph{the composite interface system}, rather than the VBM or the HOMO of each \emph{individual} component within the interface. In other words, depending on the specific type in Fig. \ref{fig:types}, the ``highest occupied energy level'' may be the VBM of the semiconductor substrate, the $E_{\rm F}$ of the metal substrate, the HOMO of the adsorbate, or even a linear combination of these in more complex situations. The accuracy of the other frontier energy levels of interest in Fig. \ref{fig:types}, therefore, is not formally justified in DFT. Lastly, we comment that the level alignment problem at interfaces is intimately relevant to the ``gap problem'' that has been outstanding in the DFT community for decades \cite{PPLB82}. The level alignment values predicted by local and semi-local density functionals are often too small compared to the true values, where the discrepancy is typically on the order of 1 eV.

It has been known that the charged quasiparticle excitations can be formally described in terms of Green's functions in the framework of many-body perturbation theory (MBPT) \cite{ORR02}. The simplest and most popular approximation by far has been the so-called $GW$ approximation \cite{H65}, where $G$ stands for the Green's function and $W$ stands for the screened Coulomb interaction. With significant advancements in methodology and computational packages developed in the era of high-performance computing \cite{yambo,BerkeleyGW,WEST}, first-principles $GW$ calculations are becoming routine, even for large-scale interfaces. However, from the perspective of the author of this Review, the development of DFT-based approaches for accurate interfacial energy level alignments is still meaningful: compared to their MBPT counterparts, density functionals are easier to be implemented in a self-consistent manner; they are easier to converge due to fewer converging parameters; and they are typically lower in scaling, enabling calculations of larger systems with similar computational cost. Moreover, the journey of developing new functionals and approximations generates new insight into the underlying physical principles specific to heterogeneous interfaces and pushes the boundaries of electronic structure methods.

This Review is intended to reflect on previous works done in the DFT community in understanding and tackling the interfacial electronic structure problem. We note that there have been a number of excellent reviews over the years dedicated to the formalism and developments of DFT \cite{CMY12,B12,B14,J15,YLT16,DFTexchange}, as well as many comprehensive reviews of the rich physics and chemistry associated with different types of interfaces \cite{HZS18,XNK20,CZL20,GBN21}. Neither of the above is the goal here, so we have to inevitably leave out some works that share similar interests in a broader sense. In this Review, we specifically focus on the methodological advancements in the framework of DFT to tackle the electronic structure problem at heterogeneous interfaces. Given the intimate relationship between an interface and a substrate surface (i.e., an interface between an extended material and the vacuum), earlier attempts in the field involve studies of metal surfaces. We hope our narrative of this well-defined field of research will guide the future development of functionals for accurate and efficient calculations of heterogeneous systems.

The outline of this Review is as follows. In Sec. \ref{sec:physics}, we give a brief overview of the physics behind the interfacial electronic structure. Two aspects are discussed: the image potential in Sec. \ref{sec:image} and the gap renormalization in Sec. \ref{sec:renorm}. In Sec. \ref{sec:early}, we review early attempts in understanding the problem using DFT quantities. We discuss the exchange-correlation (XC) hole at a metal surface in Sec. \ref{sec:xchole} and the XC potential in Sec. \ref{sec:kspot}. In Sec. \ref{sec:func}, we survey the modern developments of functionals toward accurate calculations of interfacial electronic structure, which falls into two categories: dielectric-dependent hybrid functionals in Sec. \ref{sec:dielec} and local-hybrid functionals in Sec. \ref{sec:lhybrid}. We make concluding remarks in Sec. \ref{sec:outlook}.

\section{The Physics Behind Surface and Interface Electronic Structure}
\label{sec:physics}
We mentioned above that the quasiparticle energy levels of interest in Fig. \ref{fig:types} are charged excitation energies. From the perspective of classical electrostatics, charge addition/removal in the adsorbate (i.e., the species near the substrate surface) will induce a change in the charge distribution within the substrate and near the surface, which in turn affects the Coulomb interaction within the adsorbate. This is known as the dielectric screening due to the substrate, and the Coulomb interaction within the adsorbate is said to be screened by the substrate, such that the electron-electron repulsion in the adsorbate is weaker than the bare Coulomb interaction. In the quantum mechanical description, the screened Coulomb interaction $W$ contains key information about the substrate dielectric screening, a long-range correlation effect. The simplest interface is a surface, where a periodic substrate meets the vacuum. The simplest substrate is a metal, whose classical dielectric constant is infinity. Thus metal surfaces were naturally the first type of heterogeneous systems studied in history. In this section, we review early studies of surfaces and interfaces, which centered on the understanding of the spatial dependence and orbital dependence of the screened Coulomb interaction $W$.

\subsection{Spatial dependence of the screened Coulomb interaction: The image potential}
\label{sec:image}

A few years after the DFT formalism was formally proposed, Lang and Kohn applied this theory to metal surfaces, resulting in three seminal papers \cite{LK70,LK71,LK73}, referred to as LK70, LK71, and LK73 below. Among other results that are less relevant to this Review, LK70 \cite{LK70} emphasized the importance of XC effects and self-consistency in solving the KS equations, which lead to Friedel oscillations in the density that is missing in the Thomas-Fermi approach. LK71 \cite{LK71} focused on the work function and presented the screening charge density induced by a weak external electric field along the outward surface normal. This screening charge density is subsequently used in LK73 \cite{LK73} to define an ``effective position of the metal surface'', $z_0$, which is now known as the image plane. To be specific, $z_0$ is defined as the center of mass of the screening charge density, $n(z)$:
\begin{equation}
z_0 = \frac{\int_{-\infty}^{\infty}zn(z)dz}{\int_{-\infty}^\infty n(z)dz},        
\label{eq:lk73_1}
\end{equation}
where $z$ is the direction perpendicular to the surface (note that the original paper used $x$. Here we use $z$ to align with recent conventions). 

Using linear-response theory, LK73 \cite{LK73} showed that the change in the electrostatic potential well outside of $n(z)$ is $-\mathcal{E}(z-z_0)$, where $\mathcal{E}$ is the magnitude of the external electric field. Therefore, $z_0$ is recognized as the effective position of the metal surface. $z_0$ is found to be closer to $z_b$ for larger $r_s$, where $z_b$ is the edge of the uniform positive-charge background (jellium) used to model the metal ions and $r_s$ is the Wigner-Seitz radius, i.e., $(4\pi/3)[r_s(n)]^3=1/n$. Besides the case of an external electric field, LK73 also studied the interaction between a small point charge $q$ with position $z_1$ and its induced surface charge. The paper showed that this interaction assumes the form of an image potential
\begin{equation}
U = -\frac{q^2}{4(z_1-z_0)}+\mathcal{O}\left(\frac{q^2}{(z_1-z_0)^3}\right).
\label{eq:lk73_2}
\end{equation}
Remarkably, the classical image plane $z_0$ in this equation is identical to that defined in Eq. \eqref{eq:lk73_1}, which clarifies mysteries raised by contemporary work \cite{AH72}. Note that LK73 argued that the positions of the positive ions are nearly unaffected by the external field, because the top-layer ions are located $1/2d$ ($d$ is the inter-layer spacing) below $z_b$, such that the external field has been largely screened out at the position of the ions.

While the LK series focused on metal surfaces, Inkson studied metal-semiconductor interfaces, which resulted in four papers in the early 1970s, referred to as I71a \cite{I71a}, I71b \cite{I71b}, I72 \cite{I72}, and I73 \cite{I73} below. This series of work largely followed the idea of Newns \cite{N69}, who used a similar approach of linear response to treat metal surfaces as Lang and Kohn did. The difference between Newns and LK lies in that the former used a linearized Thomas-Fermi approximation and the latter used DFT. 

I71a used static model dielectric functions \cite{I71a} of the bulk metal and the bulk semiconductor to derive the image potential in the vicinity of the metal-semiconductor interface. This approach neglects the effect of the charge redistribution near the interface on the dielectric functions of each component. The model dielectric functions are characterized by Thomas-Fermi screening lengths, and the behaviors of the resulting image potential depend on the \emph{relative} Thomas-Fermi screening lengths of the two components. Inkson found that the image potential has a long-range (well inside the semiconductor) part that recovers the Newns result \cite{N69} of metal-vacuum divided by the classical dielectric constant of the semiconductor. However, at short range (close to the interface), the form of the image potential is more complicated and can be either positive or negative. 

I71b constructed the screened Coulomb interaction $W$ at the interface \cite{I71b}, which goes into the expression for the first-order self-energy (i.e., the $GW$ approximation \cite{H65}):
\begin{equation}
\Sigma(\mr,\mr';\omega) = \frac{i}{(2\pi)^4}\int G(\mr,\mr';\omega-\omega')W(\mr,\mr';\omega')e^{-i\delta \omega'}\,d\omega'.
\label{eq:gw}
\end{equation}
The $\Sigma$ is the self-energy that is part of the quasiparticle wavefunction equation:
\begin{equation}
\begin{split}
\left(-\frac{1}{2}\nabla^2+v_{\rm ext}+v_{\rm H}\right)\phi_i(\mr)+\int \Sigma(\mr,\mr';E_i) & \phi_i(\mr')\,d\mr' \\
& =E_i\phi_i(\mr).
\end{split}
\label{eq:qp}
\end{equation}
Here, $v_{\rm ext}$ is the potential due to the ions and is called the external potential (for the electrons). $v_{\rm H}$ is the (classical) Hartree potential that depends on the electron density: $v_{\rm H}(\mr)=\int n(\mr')/|\mr-\mr'|\,d\mr'.$ Both sides of Eq. \eqref{eq:qp} depend on the quasiparticle energy $E_i$.

We note that the first three terms on the left-hand side of Eq. \eqref{eq:qp} is local and the last term on the left-hand side of Eq. \eqref{eq:qp} is non-local in $\mathbf{r}$. One can, nevertheless, define an \emph{orbital-dependent} effective potential phenomenologically based on the non-local term:
\begin{equation}
v_{i}^{\rm eff}(\mr)\phi_i(\mr)=\int \Sigma(\mr,\mr';E_i)\phi_i(\mr')\,d\mr'.
\label{eq:effpot}
\end{equation}

Based on these relationships and the model $W$ at the interface, I71b concluded that the asymptotic behavior of $v^{\rm eff}$ should recover the classical image potential $-e^2/(4z)$. Remarkably, because the image potential results from the non-locality of $W$ even at a very large distance from the interface, once the non-locality is built into $W$, the image potential asymptote follows naturally. 

The $W$ at the metal-semiconductor interface has rich pole structures that are responsible for surface plasmons, which are analyzed in I72 \cite{I72}. In particular, two types of plasmons are identified: a metal-metal type that has higher energy and a metal-dielectric type that has very low long-wavelength energy, lying well within the gap of the semiconductor. The latter was believed to be important in tunneling experiments. 

Similar calculations of the jellium model for metal surfaces include Refs. \citenum{GG85,GG86,SR87,WL87a,WL87b,SSG88}. Beyond the jellium model where the atomistic details of the metal nuclei are missing, first-principles calculations using pseudopotentials started emerging in the 1980s and early 1990s, including Refs. \citenum{HHL80,KBHL81,FH89,AI89,I89,F91,KWL92,LN92,LN93}, many of which studied the response of metal surfaces to external electric fields and charges. Here, we provide a brief account of the results from Ref. \citenum{LN93}, representative of contemporary works of that time. 

\begin{figure}[htp]
\centering
\includegraphics[width=3in]{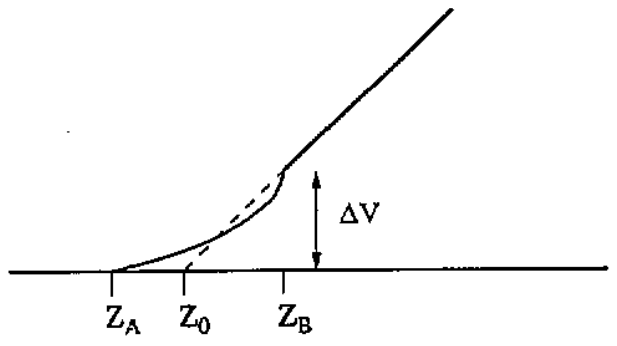}
\caption{The electrostatic potential in the direction perpendicular to a semi-infinite metal with an applied electric field. The screening charge density is non-zero only in the region between $z_A$ and $z_B$. Deep in the bulk of the metal ($z<z_A$), the potential is taken to be zero. Far out in the vacuum region ($z>z_B$), the potential has the form $\mathcal{E}(z-z_0)$, where $z_0$ is the position of the ``electrical surface''. $\Delta V$ is the potential step across the region where the screening charge density is non-zero. Reproduced with
permission from Ref. \citenum{LN93}: S. C. Lam and R. J. Needs, \emph{J. Phys.: Condens. Matter} \textbf{5}, 2101 (1993). Copyright 1993 IOP Publishing Ltd.}
\label{fig:ln93}
\end{figure}

Using calculations based on the local density approximation (LDA), Lam and Needs \cite{LN93} studied the dependence of the ``electrical surface'', $z_0$, as a function of the magnitude of the external field, $\mathcal{E}$, for two surfaces: Al(111) and Al(110). The authors first re-derived (based on different but more heuristic arguments) the LK73 result that $z_0$, the center of mass of the screening charge density as defined in Eq. \eqref{eq:lk73_1}, is also the point where the linear potential of the external field appears to originate. A schematic illustration of $z_0$ is shown in Fig. \ref{fig:ln93}.

Ref. \citenum{LN93} fitted $z_0$ as a function of the external electric field, $\mathcal{E}$, into a linear form for Al(110) and a quadratic form for Al(111). Importantly, LK73 defined $z_0$ with respect to the ``geometrical surface'' $z_b$, with the latter being the edge of the jellium and defined as $1/2d$ ($d$ is the inter-layer spacing) outside the top layer of atomic nuclei. Here, Lam and Needs argued that $z_0$ should instead be defined directly with respect to the outmost atomic layer, such that one could consider the situation of fully relaxed atomic layers. Compared to jellium, the atomic nuclei pull $z_0$ inwards, and the screening charge density is stiffer, in the sense that $dz_0/d\mathcal{E}$ is smaller than that calculated using the jellium model.

\subsection{Orbital dependence of the screened Coulomb interaction: The gap renormalization}
\label{sec:renorm}

In addition to the spatial dependence of $W$ as discussed in I71b, I73 discussed the state dependence of the $W$ and the self-energy \cite{I73}, by studying metal-semiconductor interfaces. To do that, Green's functions are needed and the author used the bulk Green's functions up to the surface and neglected any effects from the charge redistribution upon the formation of the interface. A spherically symmetric two-band model \cite{P62} for the wavefunctions was employed. The goal was to analyze how the effective potential defined in Eq. \eqref{eq:effpot} depends on each band, in particular, whether the band is occupied or unoccupied (note that there are only two bands in the model). The motivation behind Eq. \eqref{eq:effpot} is to describe the effect of the non-local self-energy $\Sigma(\mr,\mr')$ using a local potential $v^{\rm eff}_i(\mr)$.

The conclusion is that the correlation energy obeys the classical image-potential value divided by the classical dielectric constant and uniformly applies to every band, both occupied and unoccupied: $v_{\rm c}=-e^2/(4z\epsilon_0)$, while the screened exchange energy is twice the image potential and of opposite sign and applies only to the occupied band: $v_{\rm sx}=+e^2/(2z\epsilon_0)$. The net result is that the occupied band energy is shifted up compared to the unoccupied band, leading to a decrease in the fundamental gap upon formation of the interface, known as the gap renormalization.

\begin{figure}[htp]
\centering
\includegraphics[width=3in]{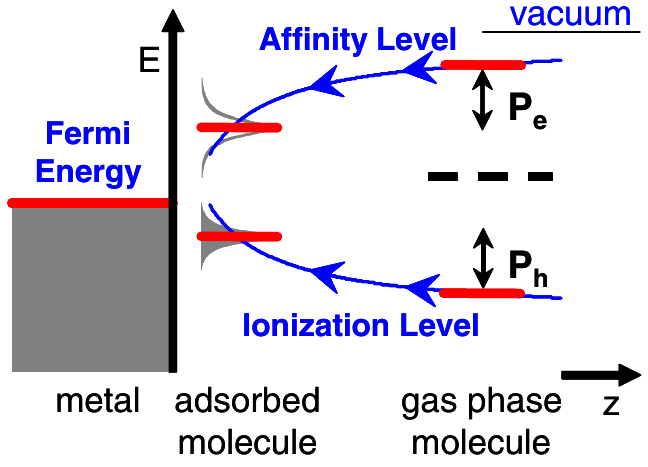}
\caption{Schematic energy level diagram at a weakly coupled molecule-metal interface. Blue lines show the molecular orbital energy and gap renormalizations. $P_h$ ($P_e$) is the surface polarization energy for occupied (unoccupied) molecular orbitals, which is approximated by the image-charge model. Reproduced with
permission from Ref. \citenum{NHL06}: J. B. Neaton, M. S. Hybertsen, and S. G. Louie, \emph{Phys. Rev. Lett.} \textbf{97}, 216405 (2006). Copyright 2006 American Physical Society.}
\label{fig:nhl06}
\end{figure}

Although Inkson's work in 1973 laid the foundation for understanding the orbital energy and gap renormalization at an interface using model $G$ and model $W$, it was not until the early 2000s that such effects are explicitly elucidated based on the first-principles $GW$ approach. With advancements in first-principles methodologies and high-performance computing, Neaton, Hybertsen, and Louie \cite{NHL06} studied the renormalization of molecular levels at a weakly coupled molecule-metal interface, benzene adsorbed on graphite (0001) surface. This work verified the basic conclusions of I73 using first-principles $GW$, i.e., the energy level shifts and gap renormalizations due to the metal surface polarization, as schematically illustrated in Fig. \ref{fig:nhl06}.

More importantly, Ref. \citenum{NHL06} proposed a simple ``image-charge model'' that corrects the quantitatively inaccurate KS energy level alignments from local and semi-local functionals. The basic idea is to attribute the molecular energy level renormalization to the combined effect of the changes in the screened exchange and Coulomb-hole correlation upon adsorption, which altogether is also referred to as surface polarization in Ref. \citenum{NHL06}. In certain limits, this surface polarization could be well approximated by a classical image-charge interaction between a point charge (or a collection of point charges) placed on the molecule and the induced image charge(s) in the metal slab. Then, an approximation is made such that all occupied (unoccupied) molecular levels are shifted upward (downward) by the amount of surface polarization, compared to gas-phase quasiparticle orbital energies (note that the latter are \emph{not} gas-phase KS orbital energies), shown as $P_h$ ($P_e$) in Fig. \ref{fig:nhl06}. This idea led to the development of the so-called ``DFT+$\Sigma$'' approach, which has been particularly successful in quantitative predictions of transport properties in molecular junctions \cite{QVCL07,LWYA14}. In the original version, the image-plane position is defined according to Ref. \citenum{LN93}, based on the response of a metal surface to an external electric field\cite{LK73}. Alternatively, the image-plane position can be defined by matching the long-range asymptote of the XC potential (see Sec. \ref{sec:kspot} for details) to an image potential \cite{LLG09,ELNK15}.

Apart from first-principles studies, Thygesen and Rubio explained the same physics using site-model Hamiltonians in Ref. \citenum{TR09}. This work focused on qualitative trends under different binding and interaction strengths, which are easily tunable by changing parameters in the model Hamiltonian. The Hamiltonian consists of three parts, $\hat{H}=\hat{H}_{\rm metal}+\hat{H}_{\rm mol}+\hat{V}$, describing the metal surface as a semi-infinite tight-binding chain, the molecule as an interacting two-level system (``H'' for HOMO and ``L'' for LUMO), and the interactions between the metal and the molecule:
\begin{equation}
\hat{H}_{\rm metal}=\sum_{i=-\infty}^{0}\sum_{\sigma=\uparrow,\downarrow}t\left(c_{i\sigma}^\dagger c_{i-1,\sigma}+c^\dagger_{i-1,\sigma}c_{i\sigma}\right),
\label{eq:tr09_1}
\end{equation}
\begin{equation}
\hat{H}_{\rm mol}=\xi_H\hat{n}_H+\left(\xi_H+\Delta_0\right)\hat{n}_L+\hat{U}_{\rm mol},
\label{eq:tr09_2}
\end{equation}
\begin{equation}
\hat{V}=\sum_{\nu=H,L}\sum_{\sigma=\uparrow,\downarrow}t_{\rm hyb}\left(c_{0\sigma}^\dagger c_{\nu\sigma}+c^\dagger_{\nu\sigma}c_{0\sigma}\right)+U_{\rm ext}\delta\hat{n}_{0}\delta\hat{n}_{\rm mol}.
\label{eq:tr09_3}
\end{equation}
The meanings of most symbols in these equations should be self-explanatory. In Eq. \eqref{eq:tr09_2}, $\Delta_0$ is the HOMO-LUMO gap of the molecule, and $\hat{U}_{\rm mol}=U_0\hat{n}_{H\uparrow}\hat{n}_{H\downarrow}+U_0\hat{n}_{L\uparrow}\hat{n}_{L\downarrow}+U_{HL}\hat{n}_{H}\hat{n}_{L}$. Notably, the interaction between the metal and the molecule $\hat{V}$ consists of two terms. The first term is a one-body interaction, in the form of a standard hopping $t_{\rm hyb}$ between the two molecular sites and the end site (``0'') of the tight-binding chain that models the metal. The second term is a many-body interaction, in the form of an \emph{inter-site} $U_{\rm ext}$ (note that this is conceptually different from the on-site Hubbard $U$), where $\delta \hat{n}_0=(\hat{n}_0-1)$ and $\delta \hat{n}_{\rm mol}=(\hat{n}_{\rm mol}-2)$ is the deviation of charge from the ground state on the end-site of the chain and the molecule, respectively. It is this inter-site $U_{\rm ext}$ term that distinguishes this Hamiltonian from the Anderson model \cite{A61} (which has an \emph{on-site} $U$) and other models alike. It is also this inter-site $U_{\rm ext}$ term that gives rise to the surface polarization or screening that is responsible for molecular energy level renormalizations.

Based on this model, Ref. \citenum{TR09} studied molecular level renormalization as a function of $U_{\rm ext}$, $\Delta_0$, and $t$ using different approaches, including total energy difference, Hartree-Fock, $GW$, and ``exact'' KS DFT. The latter is defined as a constructed KS potential via reverse engineering to reproduce the $GW$ occupation. It is interesting - and encouraging in a certain sense - to see that although not quantitative, the ``exact'' KS DFT is able to qualitatively capture the trend in energy level renormalization. This model is further generalized to describe a semiconducting substrate \cite{GRRT09} and to adopt other forms of the molecular Hamiltonian \cite{ST12}.

\section{Early Examinations of Surfaces from DFT Perspectives}
\label{sec:early}

Before first-principles calculations of nanoscale heterogeneous interfaces (or even metal and semiconductor surfaces) were possible, early works in the DFT community focused on jellium models to understand the origin of the image charge and the image potential at metal surfaces, as well as their relationships with the XC hole and the XC potential.

\subsection{Exchange-correlation hole at a metal surface}
\label{sec:xchole}
For completeness, we first briefly provide necessary definitions and explanations of the concept of the XC hole, to facilitate the review of studies of this quantity at a metal surface.

The pair density is defined as the probability density of finding an electron of spin $\sigma$ within volume element $d\mathbf{r}$ and a second electron of spin $\sigma'$ within volume element $d\mathbf{r}'$:
\begin{equation}
P(x,x')=N(N-1)\int\left|\Psi(x,x',x_3,\cdots,x_N)\right|^2\,dx_3\cdots dx_N,
\label{eq:pairdensity}
\end{equation}
where $x\equiv(\mathbf{r},\sigma)$ is the spin-coordinate and $N$ is the total electron number. $P(x,x')$ is also known as the diagonal second-order density matrix and is a symmetric function of $x$ and $x'$. One can also define the conditional probability
\begin{equation}
\Omega(x,x')=\frac{P(x,x')}{n(x)},
\label{eq:condprob}
\end{equation}
where $n(x)$ is the electron density. $\Omega(x,x')$ is then the probability density of finding any electron of spin $\sigma'$ within volume element $d\mathbf{r}'$, if there is already an electron of spin $\sigma$ within volume element $d\mathbf{r}$. The XC hole is defined as
\begin{equation}
h_{\rm XC}(x,x')=\Omega(x,x')-n(x')=\frac{P(x,x')}{n(x)}-n(x').
\label{eq:xchole}
\end{equation}
In other words, the XC hole is the change in the conditional probability compared to the uncorrelated system. Alternatively, one can define the pair-correlation function
\begin{equation}
g(x,x')=\frac{P(x,x')}{n(x)n(x')}
\end{equation}
that contains the same information as the hole. It is easy to see that $h_{\rm XC}(x,x')/n(x')=g(x,x')-1$.

The exchange hole arises from the KS orbitals and can be expressed as:
\begin{equation}
h_{\rm X}(x,x')=-\frac{\left|\Gamma_{\rm s}(x,x')\right|^2}{n(x)}.
\label{eq:xhole}
\end{equation}
Here, $\Gamma_{\rm s}(x,x')$ is the KS density matrix and is diagonal in spin:
\begin{equation}
\Gamma_{\rm s}^{\sigma=\sigma'}(x,x')\equiv\Gamma_{\rm s}^{\sigma}(\mathbf{r},\mathbf{r}')=\sum_{i=1}^{N_\sigma}\phi_{i\sigma}^*(\mathbf{r})\phi_{i\sigma}(\mathbf{r}').
\label{eq:ksdm}
\end{equation}
So $h_{\rm X}(x,x')$ is diagonal in spin and is negative everywhere. The on-top exchange hole, i.e., 
 setting $x'\to x$, is the negative of the electron density: $h_{\rm X}(x,x'\to x)=-n(x)$. The difference between $h_{\rm XC}(x,x')$ and $h_{\rm X}(x,x')$ is the correlation hole $h_{\rm C}(x,x')$. 
 
 Both the $h_{\rm XC}(x,x')$ and $h_{\rm X}(x,x')$ normalize to $-1$:
\begin{equation}
\int h_{\rm XC}(x,x')\,dx' = -1; \hspace{0.2in} \int h_{\rm X}(x,x')\,dx' = -1,
\label{eq:hxcnorm}
\end{equation}
such that $h_{\rm C}(x,x')$ normalizes to 0:
\begin{equation}
\int h_{\rm C}(x,x')\,dx' = 0.
\label{eq:hcnorm}
\end{equation}

Lastly, the XC energy is the Coulomb interaction between the electron density and the XC hole:
\begin{equation}
E_{\rm XC}[n]=\frac{1}{2}\iint\frac{n(\mathbf{r}_1)h_{\rm XC}(\mathbf{r}_1,\mathbf{r}_2)}{\left|\mathbf{r}_1-\mathbf{r}_2\right|}\,d\mathbf{r}_1d\mathbf{r}_2.
\label{eq:excinhole}
\end{equation}
Therefore, any approximation to the XC hole is formally associated with an approximation to $E_{\rm XC}[n]$.

The study of the exchange-only hole (also known as the exchange charge density in the old literature) of jellium can be dated back to Bardeen \cite{B36} and Juretschke \cite{J53}, before the formal development of DFT. Based on the infinite-barrier model of the metal surface, the exchange hole was originally believed to be localized near the metal surface when the reference electron moves away from the surface. Together with the information about the XC hole obtained from the random-phase approximation (RPA) \cite{IM78} using single-particle orbitals from the infinite-barrier model, the image potential was believed to be a result of both the exchange and the correlation holes. 

This understanding was challenged by the works of Sahni and co-authors \cite{SB84,SB85, HS88}. The authors first showed that the previously used infinite-barrier model for the jellium leads to a surface-localized exchange hole. Rather, the authors considered a linear-potential model for the metal surface (under a certain limit, it is reduced to the infinite-barrier model) in Refs. \citenum{SB84,SB85} and further extended the study to the step-potential model \cite{HS88}. The authors concluded that using these models, as the reference electron moves away from the surface, the exchange hole is not only left behind, but also becomes wider rather than narrower (as predicted by the infinite-barrier model) along the direction of the surface normal. Furthermore, the exchange hole spreads in directions parallel to the surface. As a result, the exchange hole ``takes the shape of equally spaced disks'' \cite{HS88}. 

\begin{figure}[htp]
\centering
\includegraphics[width=3in]{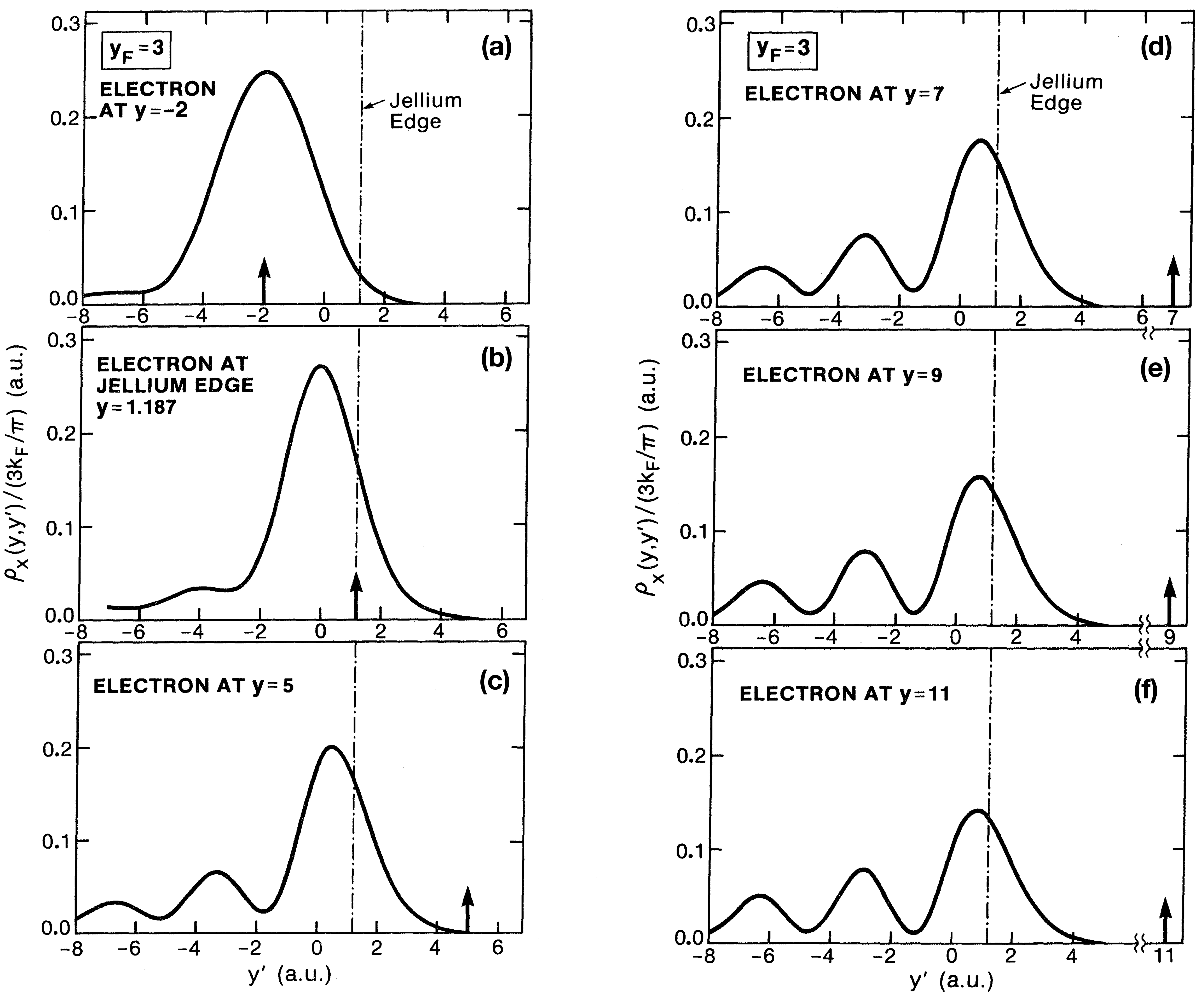}
\caption{The plane-averaged exchange hole $h_{\rm X}(y,y')$ at a surface. $\vec{y}$ is the direction of the surface normal. The up arrow indicates the position of the reference electron and the dashed line indicates the jellium edge. From (a) to (f), the reference electron moves from the interior [(a)] to the surface [(b)] and away from the surface [(c) to (f)]. One can see that the height of the main peak diminishes and the weights of smaller peaks increase as the reference electron moves away from the surface, indicating that the exchange hole becomes more delocalized and eventually spread over the entire crystal. Reproduced and adapted with
permission from Ref. \citenum{SB84}: V. Sahni and K.-P. Bohnen, \emph{Phys. Rev. B} \textbf{29}, 1045 (1984). Copyright 1984 American Physical Society.}
\label{fig:sb84}
\end{figure}

The details can be seen in Fig. \ref{fig:sb84}, where the plane-averaged exchange hole $h_{\rm X}(y,y')$ is plotted as a function of $y'$, for different choices of the positions of the reference electron $y$. In the asymptotic limit, the exchange hole will be delocalized throughout the crystal, not only over its entire length (perpendicular to the surface), but also over its entire width (parallel to the surface). Crucially, since the XC hole is localized at the surface, the authors conjectured that the correlation hole, which normalizes to 0 according to Eq. \eqref{eq:hcnorm}, must add constructively to the exchange hole for a deeper XC hole near the surface and cancel out the exchange hole deep in the bulk.

\begin{figure}[htp]
\centering
\includegraphics[width=3in]{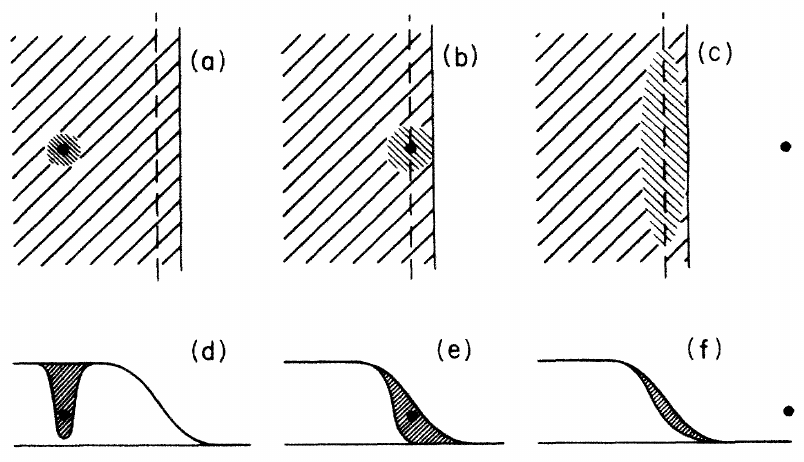}
\caption{(a)-(c) Schematic representation of the XC hole shape for different electron positions (solid circle). The solid line marks the end of the electron density and the dashed line marks the image plane, where electron-hole separation begins. Before separation, the hole follows the electron, and after the separation, it remains on the image plane and spreads laterally. (d)-(f) Schematic electron density profile at the surface, and hole density profiles, showing the same effect. Reproduced with
permission from Ref. \citenum{SSG86}: P. A. Serena, J. M. Soler, and N. Garc\'{i}a, \emph{Phys. Rev. B} \textbf{34}, 6767 (1986). Copyright 1986 American Physical Society.}
\label{fig:ssg86}
\end{figure}

The evolution of the XC hole $h_{\rm XC}(\mathbf{r},\mathbf{r}')$ as the reference electron $\mathbf{r}$ moves through a metal surface is schematically summarized in Ref. \citenum{SSG86} and the result is reproduced in Fig. \ref{fig:ssg86}. Here, the solid circle represents the electron, and one considers how the shape of the XC hole (shaded area) evolves when the electron moves from inside the metal to the image plane and then to the outside of the metal. Inside the metal [(a)(d)], the XC hole is localized near the electron. The electron and the hole start to separate when the reference electron is at the image-plane position. When the reference electron is outside the metal surface, the hole stays near the surface within the metal but spreads laterally. Note that this figure contains the full XC hole, as compared to the exchange-only hole in Fig. \ref{fig:sb84}.

In addition to various jellium models for the metal surface where the single-particle orbitals can be obtained analytically, more sophisticated approaches have been used to analyze the pair correlation function $g(\mr,\mr')$ and the XC hole. Ref. \citenum{AC96} performed diffusion Monte Carlo calculations of $g(\mr,\mr')$ for the jellium surface of different $r_s$. Diffusion and variational quantum Monte Carlo approaches have proved to be useful in computing the XC holes in molecules \cite{HCL06} and bulk solids \cite{HCWR98,RTC02}. Ref. \citenum{GACT00} used the weighted density approximation (WDA) \cite{GJL79} to study the jellium surface, focusing on both $g(\mr,\mr')$ and the XC hole, where the results of Ref. \citenum{SSG86} shown in Fig. \ref{fig:ssg86} for the XC hole are verified (i.e., the hole is left behind when the reference electron is moved outside the surface). The WDA has also been used in Ref. \citenum{JC07}. Notably, Ref. \citenum{JC07} not only verified the previous results on jellium, but also applied the approach to the Cu(100) surface where the atomistic details of the surface were considered. The XC holes are analyzed for different vertical planes through the atop, hollow, and bridge sites, respectively. Although the results are qualitatively similar to previous results, this work did not show a complete separation between the reference electron and the XC hole for the Cu(100) surface (although it did so for the jellium surface), perhaps due to the fact that the reference electron was placed not too far from the Cu(100) surface.

\begin{figure}[htp]
\centering
\includegraphics[width=3in]{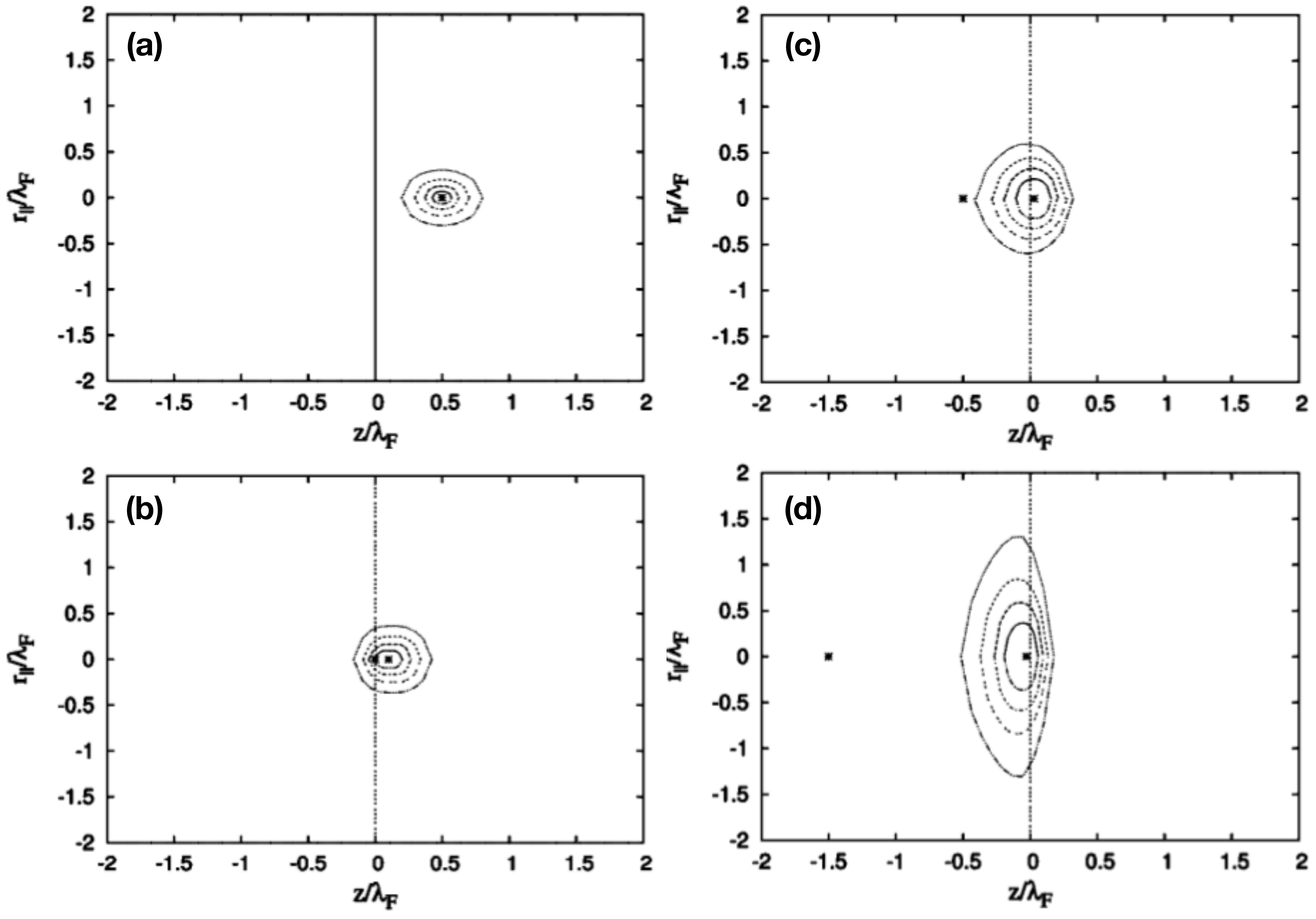}
\caption{Contour plots of the XC hole $h_{\rm XC}(r_\parallel,z,z')$. From (a) to (d), the reference electron $z'$ moves from inside the jellium to far outside the surface in the vacuum. The bulk parameter is $r_s=2.07$ and the jellium surface is at $z=0$. $r_\parallel=\pm|\mr_\parallel-\mr_\parallel'|$ and $\lambda_F=2\pi/k_F$ is the bulk Fermi wavelength. Reproduced and adapted with
permission from Ref. \citenum{CP09}: L. A. Constantin and J. M. Pitarke, \emph{J. Chem. Theory Comput.} \textbf{5}, 895 (2009). Copyright 2009 American Chemical Society.}
\label{fig:cp09}
\end{figure}

Lastly, the XC hole could be computed from the density-response function, via \cite{NP01}
\begin{equation}
\begin{split}
g(\mathbf{r},\mathbf{r}')=1+\frac{1}{n(\mr)n(\mr')}\left[-\frac{1}{\pi} \right. & \int_0^\infty dE \, \chi(\mr,\mr';iE) \\ 
& \left. \frac{}{} -n(\mr)\delta(\mr-\mr')\right], \\
\end{split}
\label{eq:holeinchi}
\end{equation}
where $\chi$ is the coupling-constant-averaged retarded density-response function. One can then use, e.g., DFT-based RPA to compute the $\chi$ and then the XC hole afterward. This is the approach taken by Refs. \citenum{NP01,CP09}. The advantage of this approach is that one can build in self-consistency \cite{PE98,PE01} and use different approximations for $\chi$ \cite{CP09}, compared to the pre-1990 calculations using single-particle orbitals. The results of Refs. \citenum{NP01,CP09} for the XC hole once again confirmed the qualitative results in the 1980s shown in Fig. \ref{fig:ssg86}. The results of Ref. \citenum{CP09} are reproduced in Fig. \ref{fig:cp09}, where contour plots of the XC hole are generated (very similar results were presented in Ref. \citenum{NP01}). Additionally, Ref. \citenum{CP09} analyzed the on-top correlation hole and concluded that it is accurately described by local and semi-local density-functional approximations. However, Ref. \citenum{CP09} showed that the exchange-only hole is also localized near the surface, which is contradictory to the results of Refs. \citenum{SB84,SB85}. It is not entirely clear whether the vanishing density assumption (used in Ref. \citenum{CP09} and intentionally avoided in Refs. \citenum{SB84,SB85}) and/or the self-consistency in the calculations led to this contradiction.

\subsection{Exchange-correlation potential at a metal surface}
\label{sec:kspot}

In the KS formulation of DFT, the XC potential $v_{\rm xc}(\mr)$ is a local, multiplicative potential that is part of the KS Hamiltonian and applies to every KS orbital. The KS equation reads:
\begin{equation}
\left[-\frac{1}{2}\nabla^2+v_{\rm ext}(\mr)+v_{\rm H}(\mr)+v_{\rm xc}(\mr)\right]\phi_i(\mr)=E_i\phi_i(\mr).
\label{eq:ks}
\end{equation}
Note that there is a fundamental difference between the $v_{\rm xc}(\mr)$ in Eq. \eqref{eq:ks} and the effective orbital-dependent potential defined in Eq. \eqref{eq:effpot}. The $v_{\rm xc}(\mr)$ is defined as the functional derivative of the XC energy: $v_{\rm xc}(\mr)=\delta E_{\rm xc}[n(\mr)]/\delta n(\mr)$.

Different approaches have been taken to analyze the asymptotic behavior of the $v_{\rm xc}(z)$ near a metal surface, where $z$ is the direction perpendicular to the surface. We review three different perspectives below.

The first perspective is through the analysis of the so-called Sham-Schl\"{u}ter equation \cite{SS83}:
\begin{equation}
\begin{split}
\int & d\mr' v_{\rm xc}(\mr')\int \frac{d\omega}{2\pi}G_0(\mr,\mr';\omega)G(\mr',\mr;\omega)= \\
& \int d\mr_1\int d\mr_2 \int \frac{d\omega}{2\pi}G_0(\mr,\mr_1;\omega)\Sigma_{\rm xc}(\mr_1,\mr_2;\omega)G(\mr_2,\mr;\omega). \\
\end{split}
\label{eq:ss}
\end{equation}
In this equation, $G_0$ is the KS Green's function, $G$ is the fully dressed one-particle Green's function, and $\Sigma_{\rm xc}$ is the self-energy $\Sigma$ less the Hartree term. 

Based on Eq. \eqref{eq:ss} and the infinite-barrier model of the metal surface, Sham \cite{S85}, citing the work of Rudnick \cite{R70}, pointed out that
\begin{equation}
\int dz' \Sigma_{\rm xc}(\mathbf{K}=0;z,z';\mu)\phi_k(z')=-\frac{e^2}{4z}\phi_k(z).
\label{eq:s85}
\end{equation}
Here, $\mu$ is the chemical potential representing the highest occupied state that decays the slowest as $z\to\infty$. Note that comparing Eq. \eqref{eq:s85} and Eq. \eqref{eq:effpot} reveals that the effective potential decays as $-1/(4z)$, consistent with LK73\cite{LK73}. From a term-by-term examination of the $\Sigma_{\rm xc}$, Ref. \citenum{S85} concluded that the $v_{\rm x}$ yields a $1/z^2$ behavior as $z\to\infty$, so it is the $v_{\rm c}$ that is responsible for the overall $1/z$ behavior.

Under the same theoretical framework, Ref. \citenum{EHFH92} investigated the resulting XC potential from Eq. \eqref{eq:ss} for a jellium surface and the Al(100) surface. Ref. \citenum{EHFH92} compared the $v_{\rm xc}$ derived from Eq. \eqref{eq:ss} where $\Sigma_{\rm xc}$ is from a $GW$ calculation of a jellium slab and the LDA $v_{\rm xc}$. For the latter, Ref. \citenum{EHFH92} used the definition $v_{\rm xc}=\Sigma_{\rm xc}(k=k_F;E=E_F)$ with $\Sigma_{\rm xc}$ taken from the $GW$ self-energy for the bulk jellium (rather than jellium \emph{slab}). The authors of Ref. \citenum{EHFH92} noted that this definition of LDA $v_{\rm xc}$ numerically differs from the ``standard'' Ceperley-Alder formulation \cite{CA80} by a small amount. Notably, the $v_{\rm xc}$ derived from Eq. \eqref{eq:ss} becomes image-like outside the surface, while the LDA $v_{\rm xc}$ decays much faster as $z\to\infty$.

\begin{figure}[htp]
\centering
\includegraphics[width=3in]{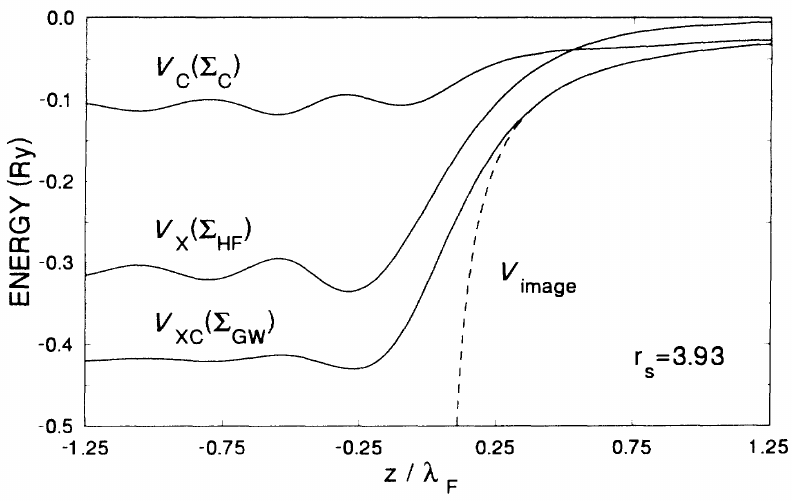}
\caption{Solution of Eq. \eqref{eq:ss} for the XC potential using three approximations to the self-energy: $\Sigma_{\rm HF}$ (Hartree-Fock), $\Sigma_{GW}$ (the $GW$ approximation), and $\Sigma_{\rm c}$ (which is $\Sigma_{GW}-\Sigma_{\rm HF}$). Reproduced with
permission from Ref. \citenum{EHFH92}: A. G. Eguiluz, M. Heinrichsmeier, A. Fleszar, and W. Hanke, \emph{Phys. Rev. Lett.} \textbf{68}, 1359 (1992). Copyright 1992 American Physical Society.}
\label{fig:ehfh92}
\end{figure}

Furthermore, Ref. \citenum{EHFH92} separated the exchange and correlation contributions, by computing $v_{\rm x}$ from $\Sigma_{\rm HF}$ and $v_{\rm c}$ from $\Sigma_{\rm c}=\Sigma_{GW}-\Sigma_{\rm HF}$. The result is reproduced in Fig. \ref{fig:ehfh92}. The authors showed that $v_{\rm x}(z)\to -a/z^2$ as $z\to\infty$, consistent with Sham \cite{S85}. Here, the coefficient $a$ gives rise to the exchange contribution to the image-plane position $z_0$ [recall the image potential is $-e^2/4(z-z_0)$]. The authors concluded that the position of the image plane includes a significant contribution from the exchange, because the full $v_{\rm xc}(z)$ merges with the image potential much closer to the surface than $v_{\rm c}(z)$ does, see Fig. \ref{fig:ehfh92}. Numerically, Ref. \citenum{EHFH92} showed that one can define the image-plane position by matching the asymptotic behavior of the $v_{\rm xc}(z)$ with an image potential, and the resulted $z_0$ is closer to the jellium surface than the value computed from the linear response of the surface to an external electric field \cite{LK73,LN93}. Note that this conclusion is different from that in Ref. \citenum{SSG86}.

Moreover, Ref. \citenum{EHFH92} constructed a point-by-point tabulation between the $v_{\rm xc}$ derived from Eq. \eqref{eq:ss} and the density $n$, denoted by $v_{\rm xc}(r_s;n^{1/3})$, where the $1/3$-order in $n$ is to ensure the correct density scaling. Effectively, one can carry out non-local calculations using this functional with the same ease as standard LDA calculations. The authors applied this numerically defined $v_{\rm xc}$ to the Al(100) surface, and successfully produced the Rydberg series of image states.

A related but fundamentally different piece of work is Ref. \citenum{WGRN98}, which performed $GW$ calculations of the Al(111) surface. Unlike Ref. \citenum{EHFH92} that studies the exact $v_{\rm xc}(\mr)$ from Eq. \eqref{eq:ss}, Ref. \citenum{WGRN98} analyzed the effective orbital-dependent potential defined in Eq. \eqref{eq:effpot}. The authors reached a similar conclusion as those of Ref. \citenum{EHFH92}: the $v_{i}^{\rm eff}$ crosses smoothly to the asymptotic image potential, and the $z_0$ defined in this way is closer to the surface than that defined from linear-response calculations using an external test charge\cite{LK73, LN93, SSG88}. This conclusion is consistent with Ref. \citenum{EH89}. Moreover, Ref. \citenum{WGRN98} studied the exchange part of the $v_{i}^{\rm eff}$ for unoccupied orbitals and concluded that it decays exponentially into the vacuum and it is the correlation that is responsible to describe the image potential felt by occupied states.

The second perspective in the literature to analyze the asymptotic behavior of the $v_{\rm xc}(z)$ near a surface was proposed by Almbladh and von Barth \cite{AB85}. The authors approached the problem based on the decay of one-electron orbitals, density matrix, and spectral function in the asymptotic region. The authors concluded that for the exact DFT to reproduce the density profile, $v_{\rm xc}(\mr) \sim \alpha(\mr,\mr)/2$ as $z\to\infty$, where $\alpha(\mr,\mr)$ is the diagonal element of the polarizability, and
\begin{equation}
\alpha(\mr,\mr)\sim\begin{cases}
-\frac{1}{2z} & \mbox{for a metal surface;} \\
-\frac{1}{2z}\frac{\epsilon_0-1}{\epsilon_0+1} & \mbox{for a semiconductor surface,}
\end{cases}
\label{eq:ab85}
\end{equation}
with $\epsilon_0$ being the macroscopic dielectric constant of the semiconductor. It is then straightforward to see that the $v_{\rm xc}(\mr)$ assumes the image potential form as $z\to\infty$ for a metal surface, and the form is revised for a semiconductor surface. Note that Ref. \citenum{AB85} did not separate exchange and correlation contributions.

The third perspective in the literature to analyze the asymptotic behavior of the $v_{\rm xc}(z)$ near a metal surface was proposed by Sahni and co-workers \cite{HS89a,HS89b,SS96}. Refs. \citenum{HS89a,HS89b} focused on numerical calculations of the jellium surface, with an analytical study presented in Ref. \citenum{SS96}. The key idea is to recognize that the potential of the electron is the work done in bringing it from infinity to its final position against the electric field of the XC hole. Using the finite linear effective potential model for the jellium surface, Ref. \citenum{HS89b} showed that the exact exchange potential decays as $-1/(4z)$, consistent with the classical picture of the image potential being a consequence of the Coulomb interaction between a test charge and its image \cite{LK73}. As a result, according to Sahni and co-workers, although the exchange hole does not constitute part of the \emph{image charge} at the semi-infinite jellium surface (recall that it is delocalized over the crystal \cite{HS88}), work done against this delocalized exchange hole  - which is the exact exchange potential - corresponds to the \emph{image potential} asymptotically outside a jellium surface \cite{HS89a}. Additionally, Ref. \citenum{HS89b} showed that the Slater potential \cite{S51} is also image-potential-like, but with a different coefficient. 

Notably, there is the controversy between Refs. \citenum{S85,EHFH92} and Refs. \citenum{HS89a,HS89b,SS96}  regarding the asymptotic behavior of the exact exchange potential as well as the physical origin of the image potential. Ref. \citenum{WGRN98} pointed out that the former perspective dealt with \emph{slab} geometries, where the exchange potential goes as $-1/z^2$, while the latter perspective dealt with \emph{semi-infinite} jellium, where the exchange is responsible for the image-potential. 

Lastly, we note in passing that the $v_{\rm x}$ discussed here in Sec. \ref{sec:kspot} is the exact exchange potential, rather than the \emph{screened} exchange potential $v_{\rm sx}$ mentioned in I73 \cite{I73} and similar works that we discussed in Sec. \ref{sec:renorm}. The former is defined within the DFT framework while the latter is defined in the MBPT framework and the difference is a correlation effect.

\section{Modern Developments of Density Functionals for Interfaces}
\label{sec:func}

We discuss two types of density functionals developed in recent years, namely the dielectric-dependent hybrid and the local hybrid functionals, which are designed to improve the description of the electronic structure at heterogeneous interfaces compared to conventional local and semilocal functionals. We will also see that some functionals possess features of both categories.

In a certain sense, the issue of interfacial energy level alignment and the famous ``gap problem'' in DFT \cite{PPLB82} share the same origin: both are properties of quasiparticle energy levels and problems arise when one intends to approximate these energy levels using eigenvalues from the KS equation. In light of the success of hybrid functionals - or more generally, the generalized Kohn-Sham (GKS) scheme \cite{SGVM96} - in tackling the ``gap problem'', most modern density functionals targeting accurate interfacial energy level alignments are hybrid. Needless to say, this Review is not intended to be a comprehensive narrative of the ``gap problem'' and its solutions, for which many excellent review articles and accounts exist \cite{YCM12,PYBY17,PR18}. Here, we merely list the core concepts of hybrid functionals for completeness to facilitate our discussion below.

Introduced by Becke in 1993 \cite{B93}, hybrid functionals mix a fraction of non-local Fock exchange into the XC energy. The simplest form is
\begin{equation}
E_{\rm xc}=\alpha E_{\rm x}^{\rm ex}+(1-\alpha)E_{\rm x}^{\rm sl}[n]+E_{\rm c}^{\rm sl}[n],
\label{eq:hyb}
\end{equation}
where $E_{\rm x}^{\rm ex}$ is the non-local Hartree-Fock exchange energy that explicitly depends on orbitals, $E_{\rm x}^{\rm sl}[n]$ and $E_{\rm c}^{\rm sl}[n]$ are the semi-local approximation to the exchange and correlation, respectively, which explicitly depend on the density. $\alpha$ is the mixing parameter that is between 0 and 1.

Apart from this simplest form, the idea of range separation has led to success in a variety of systems. In range-separated hybrid (RSH) functionals\cite{BLS10,KSRB12,KB14}, the Coulomb interaction is separated into long-range and short-range components \cite{HSE03,TCS04,YTH04}. A general form for the range separation used in most RSH functionals is
\begin{equation}
\begin{split}
\frac{1}{|\mr-\mr'|}= \,& \frac{\alpha+\beta \mbox{erf}(\gamma |\mr-\mr'|)}{|\mr-\mr'|}\\
& +\frac{1-[\alpha+\beta \mbox{erf}(\gamma |\mr-\mr'|)]}{|\mr-\mr'|}. \\
\end{split}
\label{eq:rsh}
\end{equation}
Here, $\alpha$, $\beta$, and $\gamma$ are parameters (all between 0 and 1) and $\mbox{erf}(\cdot)$ is the error function. In this form, the first term is treated using the non-local Fock exchange that involves orbitals, and the second term is treated using a semi-local exchange that only involves the density. As a result, the corresponding XC energy is
\begin{equation}
\begin{split}
E_{\rm xc} = \, & \alpha E_{\rm x,SR}^{\rm ex}+(1-\alpha)E_{\rm x,SR}^{\rm sl}[n]+(\alpha+\beta)E_{\rm x,LR}^{\rm ex} \\
& +(1-\alpha-\beta)E_{\rm x,LR}^{\rm sl}[n]+E_{\rm c}^{\rm sl}[n]. \\
\end{split}
\label{eq:rshexc}
\end{equation}
Here, SR (LR) denotes the short-range (long-range) component of the Coulomb interaction.

Hybrid functionals are most commonly implemented in the so-called GKS scheme \cite{SGVM96}, with the GKS equation
\begin{equation}
\left[-\frac{1}{2}\nabla^2+v_{\rm ext}(\mr)+v_{\rm H}(\mr)+\hat{O}\right]\phi_i(\mr)=E_i\phi_i(\mr),
\label{eq:gks}
\end{equation}
where the $\hat{O}$ is an operator that corresponds to the energy expression of the hybrid functional. In general, part of $\hat{O}$ can be a non-local XC operator, whose physical meaning is clear by comparing Eq. \eqref{eq:ks} and Eq. \eqref{eq:gks}. Generally, the $\hat{O}$ has form
\begin{equation}
\hat{O}=v_{\rm x}^{\rm non-local}(\mr,\mr')+v_{\rm x}^{\rm sl}[n](\mr)+v_{\rm c}^{\rm sl}[n](\mr),
\label{eq:ohyb}
\end{equation}
where $v_{\rm x}^{\rm sl}$ and $v_{\rm c}^{\rm sl}$ are the semilocal exchange and correlation potential from KS DFT, respectively. The $v_{\rm x}^{\rm non-local}$ has the form of the non-local Fock operator:
\begin{equation}
v_{\rm x}^{\rm non-local}\phi_i(\mr)=-\sum_j\phi_j(\mr)\int d\mr'\,\phi_j^*(\mr')\phi_i(\mr')v(|\mr-\mr'|).
\label{eq:vx}
\end{equation}
In the case of a global hybrid functional in Eq. \eqref{eq:hyb}, $v(|\mr-\mr'|)$ in Eq. \eqref{eq:vx} will have the form of $\alpha/|\mr-\mr'|$, i.e., bare Coulomb interaction scaled by the mixing parameter $\alpha$. $v_{\rm x}^{\rm sl}[n](\mr)$ will have the form of a standard semi-local functional multiplied by $1-\alpha$. $v_{\rm c}^{\rm sl}[n](\mr)$ is from a standard semi-local functional. As a result, for a global hybrid functional in Eq. \eqref{eq:hyb}, the $\hat{O}$ has form
\begin{equation}
\hat{O}=\alpha v_{\rm x}^{\rm ex}(\mr,\mr')+(1-\alpha)v_{\rm x}^{\rm sl}[n](\mr)+v_{\rm c}^{\rm sl}[n](\mr).
\label{eq:ohybfr}
\end{equation}

In the case of a RSH in Eq. \eqref{eq:rshexc}, $v(|\mr-\mr'|)$ in Eq. \eqref{eq:vx} will have the form of $[\alpha+\beta \mbox{erf}(\gamma |\mr-\mr'|)]/|\mr-\mr'|$, i.e., the first term in Eq. \eqref{eq:rsh}. $v_{\rm x}^{\rm sl}[n](\mr)$ will be a modified version of the semi-local functional used where the exchange hole is scaled \cite{HSE03} by the second term in Eq. \eqref{eq:rsh}. $v_{\rm c}^{\rm sl}[n](\mr)$ is from a standard semi-local functional. As a result, for a RSH, the $\hat{O}$ has form
\begin{equation}
\begin{split}
\hat{O}= & \alpha v_{\rm x,SR}^{\rm ex}(\mr,\mr';\gamma)+(\alpha+\beta) v_{\rm x,LR}^{\rm ex}(\mr,\mr';\gamma) \\
& + (1-\alpha) v_{\rm x,SR}^{\rm sl}[n](\mr;\gamma)+(1-\alpha-\beta) v_{\rm x,LR}^{\rm sl}[n](\mr;\gamma) \\
& +v_{\rm c}[n](\mr), \\
\label{eq:ohybrsh}
\end{split}
\end{equation}
where all exchange quantities parametrically depend on $\gamma$, the range-separation parameter.

\subsection{Dielectric-dependent hybrid functionals}
\label{sec:dielec}

In dielectric-dependent hybrid functionals, the parameters of a hybrid functional (especially the mixing parameter for the Fock exchange, among others) are expressed in terms of the dielectric properties of the material. There have been two main strategies in developing dielectric-dependent functionals: (i) start with a model dielectric function and then construct the corresponding XC potential, and (ii) start with a fixed functional form and then express the parameters in terms of the dielectric properties. We discuss these two different strategies below.

The idea of incorporating the dielectric function into a density functional was first proposed by Shimazaki and Asai \cite{SA08}, where the authors followed the first strategy mentioned above. The authors proposed a model for the dielectric function:
\begin{equation}
\epsilon(\mathbf{k})=1+\left[(\epsilon_{\rm s}-1)^{-1}+\alpha\left(\frac{k^2}{k^2_{\rm TF}}\right)\right]^{-1}.
\label{eq:sa08}
\end{equation}
In this equation, $\epsilon_{\rm s}$ is the electronic part of the static dielectric constant, $\alpha=1.563$ following Bechstedt \cite{BSCR92,CDLB93}, $k$ is the momentum, and $k_{\rm TF}$ is the Thomas-Fermi wave number. This is a simplified version of the Bechstedt model and behaves like the Thomas-Fermi model for large $k$. The Fourier transform of Eq. \eqref{eq:sa08} yields a screened Coulomb interaction
\begin{equation}
\begin{split}
v^{\rm scr}(\mr)& =\frac{1}{(2\pi)^3}\int \frac{4\pi}{k^2\epsilon(\mathbf{k})}\exp{(i\mathbf{k}\cdot\mr)}d\mathbf{k} \\
& = \left(1-\frac{1}{\epsilon_{\rm s}}\right)\frac{\exp{(-\tilde{k}_{\rm TF}r)}}{r}+\frac{1}{\epsilon_{\rm s}}\frac{1}{r} \\
& \approx \left(1-\frac{1}{\epsilon_{\rm s}}\right)\frac{\mbox{erfc}(2\tilde{k}_{\rm TF}r/3)}{r}+\frac{1}{\epsilon_{\rm s}}\frac{1}{r}.
\end{split}
\label{eq:sa09}
\end{equation}
Here, $\tilde{k}_{\rm TF}=(k^2_{\rm TF}/\alpha)[1/(\epsilon_{\rm s}-1)+1]$. The resulting $v^{\rm scr}(\mr)$ contains a Yukawa-type potential in the second line of Eq. \eqref{eq:sa09}, which is further replaced by the $\mbox{erfc}(\cdot)$ function for computational simplicity in the third line of Eq. \eqref{eq:sa09}.

Substituting Eq. \eqref{eq:sa09} as the $v(|\mr-\mr'|)$ in Eq. \eqref{eq:vx}, one can see that the result is a dielectric-dependent non-local Fock exchange potential. It is then combined with a local correlation potential \cite{SA08}, either the Vosko-Wilk-Nusair or the Lee-Yang-Parr flavor, for a complete non-local XC potential [the $\hat{O}$ in Eq. \eqref{eq:gks}]. This approach was applied to compute the band structure of the diamond, yielding a band gap that is very close to the experimental value.

In a subsequent work \cite{SA09}, the authors further recognized that when the screening length $\tilde{k}_{\rm TF}$ is large, the nonlocal contribution in the first term of Eq. \eqref{eq:sa09} becomes small. Therefore, it can be approximated using a local potential, e.g., the Slater exchange \cite{S51}. As a result, the XC potential becomes
\begin{equation}
v_{\rm xc}(\epsilon_{\rm s})=\frac{1}{\epsilon_{\rm s}}v^{\rm ex}_{\rm x}+\left(1-\frac{1}{\epsilon_{\rm s}}\right)v^{\rm Slater}+v_{\rm c}.
\label{eq:sa09b}
\end{equation}
If one compares this equation with Eqs. \eqref{eq:ohyb} and \eqref{eq:hyb}, one recognizes the similarity: the non-local Fock exchange mixes with a semi-local exchange, with the mixing fraction being $1/\epsilon_{\rm s}$, hence a dielectric-dependent hybrid functional is constructed. A self-consistent procedure was also developed in Ref. \citenum{SA09} to determine $\epsilon_{\rm s}$, which is approximated using the fundamental gap from the GKS calculation. This approach was applied to a large set of materials \cite{SA10} to test its performance.

Besides starting from a model dielectric function and then deriving the corresponding XC potential form, as we mentioned above, an alternative strategy is to start from a fixed XC functional form, say, the RSH functional defined in Eqs. \eqref{eq:rshexc} and \eqref{eq:ohybrsh}, and then express the parameters $(\alpha,\beta,\gamma)$ in terms of the dielectric properties.

To this end, one naturally wonders if there is an optimal way of determining $(\alpha,\beta,\gamma)$. For finite systems, this is possible thanks to Koopmans' theorem: for an exact functional, the energy of the HOMO is negative of the ionization potential. This idea led to the development of optimally tuned range-separated hybrid functional (OT-RSH) \cite{RSGA12}. Here, $\alpha$ is chosen to be 0.25 as in the PBE0 functional (PBE=Perdew-Burke-Ernzerhof), $\alpha+\beta=1$ is used as a constraint to enforce the correct asymptotic potential \cite{RH08}, as full Fock exchange in the long range has proved to be essential for gap predictions of finite systems such as molecules \cite{KSRB12}. $\gamma$ is tuned to minimize the difference between the HOMO energy computed from the RSH functional and the ionization potential calculated from the energy difference between the cation and the neutral species, i.e., enforcing Koopmans' theorem.

However, this idea faces two challenges for extended or heterogeneous systems: (i) one cannot tune $\gamma$ based on Koopmans' theorem because the ionization potential can no longer be calculated from energy differences, and (ii) the long-range Coulomb interaction is screened by the dielectric environment of the extended material. In Ref. \citenum{RSJB13}, the OT-RSH was extended to molecular crystals, with the above two challenges addressed as follows: (i) the $\gamma$ is fixed to be that tuned for the constituting molecule in its gas phase, and (ii) with $\alpha$ fixed to be 0.25 as in PBE0, $\alpha+\beta$ is chosen to be $\epsilon^{-1}$ to reflect the long-range dielectric screening, where $\epsilon$ is the macroscopic dielectric constant of the molecular crystal. The result is then a dielectric-dependent hybrid functional.

This idea was further generalized to treat heterogeneous interfaces formed between molecules and metal surfaces in Ref. \citenum{LERK17}. To address the same two challenges as above, the $\gamma$ is fixed to be that tuned for the molecular adsorbate, and $\alpha+\beta$ is chosen to incorporate the dielectric screening effect from the metal substrate. For the latter, one tunes $\beta$ such that $E_{\rm HOMO}(\alpha=0.25,\mbox{tuned }\beta)-E_{\rm HOMO}(\alpha+\beta=1)$ is equivalent to the surface polarization energy that is approximated by the classical image-charge model \cite{NHL06,ELNK15}. Here, $E_{\rm HOMO}(\alpha+\beta=1)$ is simply the HOMO energy of the isolated molecule predicted by the original formulation of OT-RSH \cite{RSGA12}. $E_{\rm HOMO}(\alpha=0.25,\mbox{tuned }\beta)$ is also calculated for the gas-phase molecule only (rather than for the molecular adsorbate within the interface system), implicitly assuming that a hybrid functional with a fixed set of parameters does not lead to energy and gap renormalization after a molecule is adsorbed on a metal substrate\cite{BTNK11}. 

The essential idea is Ref. \citenum{LERK17} is to focus on the electronic structure of the molecular adsorbate, and assumes that the optical set of $(\alpha,\beta,\gamma)$ suitable for the molecular adsorbate does not deteriorate the properties of the metal substrate, which is typically well described by a semi-local functional. An additional challenge exists if one tends to apply the same approach to interfaces formed between a molecule and a semiconductor substrate. Here, one needs to find a common set of $(\alpha,\beta,\gamma)$ that works for both components of the interface. This is a non-trivial task and there is no \emph{a priori} guarantee that this is possible for an arbitrary combination of molecular adsorbate and substrate. In Ref. \citenum{ZLMD21}, this was accomplished via a multi-objective optimal tuning for a series of (metallo)phthalocyanine molecules adsorbed on two-dimensional (2D) MoS$_2$. One first tunes the $\alpha+\beta$ against surface polarization energy (the same procedure as Ref. \citenum{LERK17}), which is estimated by the classical image-charge interaction near a dielectric slab with a finite size along the surface normal \cite{NCXQ19,CB18}. Then, one tunes $(\alpha,\gamma)$ by minimizing the sum of the error in  the band gap of the freestanding MoS$_2$ and the surface polarization energy for the molecular adsorbate.

The relationship between the Fock exchange mixing parameter and the dielectric constant of the material can be formalized by comparing the hybrid functional and the $GW$ approximation. This is similar to the practice of comparing Eq. \eqref{eq:sa09b} and Eqs. \eqref{eq:ohyb} and \eqref{eq:hyb} as done in Ref. \citenum{SA09}.

Ref. \citenum{MVOR11} first drew a heuristic connection between hybrid functionals and $GW$, where the authors noted that if the screening in the screened-exchange (SEX) term in $GW$ is replaced by an effective static dielectric constant $\epsilon_\infty=1/\alpha$ ($\alpha$ is the mixing fraction of the Fock exchange in a hybrid functional) and then the Coulomb-hole (COH) term in $GW$ is modeled by the static and local parts of the hybrid functional, then the quasiparticle equation has the same form as the GKS equation for a hybrid functional. This idea was further explored in Ref. \citenum{KBT13}, where self-consistent and non-self-consistent hybrid calculations were carried out.

In a subsequent paper following Refs. \citenum{SA08,SA09} discussed above, Shimazaki and Nakajima \cite{SN14}  compared the non-local XC potential to the COHSEX approximation \cite{H65} of $GW$, and devised a local correlation potential that also depends on the dielectric function.

These ideas were further extensively formalized and discussed in Ref. \citenum{SGG14} by the Galli group. The authors noted that if one approximates $W(\mr,\mr')$ as
\begin{equation}
\begin{split}
W(\mr,\mr') & = \int \,d\mr'' \epsilon^{-1}(\mr,\mr'')v(\mr'',\mr') \\
& \approx \epsilon_\infty^{-1} v(\mr,\mr'), \\
\end{split}
\label{eq:sgg14a}
\end{equation}
where $v(\mr,\mr')=1/|\mr-\mr'|$ is the Coulomb interaction, then the SEX and COH terms of $GW$ can be approximated as
\begin{equation}
\begin{split}
\Sigma_{\rm SEX}(\mr,\mr') & =-\sum_{i}^{{\rm occ.}}\phi_i(\mr)\phi_i^*(\mr')W(\mr,\mr') \\
& \approx -\epsilon_\infty^{-1} \sum_{i}^{{\rm occ.}}\phi_i(\mr)\phi_i^*(\mr')v(\mr,\mr').
\end{split}
\label{eq:sgg14b}
\end{equation}

\begin{equation}
\begin{split}
\Sigma_{\rm COH}(\mr,\mr') & = -\frac{1}{2}\delta(\mr-\mr')[v(\mr,\mr')-W(\mr,\mr')] \\
& \approx -\left(1-\epsilon_\infty^{-1}\right)\frac{1}{2}\delta(\mr-\mr')v(\mr,\mr').
\end{split}
\label{eq:sgg14c}
\end{equation}

If one compares these two terms with the non-local operator $\hat{O}$ of a global hybrid functional [Eq. \eqref{eq:ohyb}], one recognizes that $\alpha=\epsilon_\infty^{-1}$ for the non-local term. Then one further approximates the $\Sigma_{\rm COH}$ as the local part of the exchange and the local correlation. The approach was made self-consistent by evaluating $\epsilon_\infty$ using the coupled perturbed KS equations \cite{FROD08}, where one considers the dielectric response of the system subject to a macroscopic electric field.

In a subsequent paper \cite{SGG16}, the dielectric-dependent hybrid functional was extended to a range-separated version. Within the framework of RSH, Eq. \eqref{eq:sgg14a} becomes
\begin{equation}
W(\mr,\mr')\approx \frac{\epsilon_\infty^{-1}}{|\mr-\mr'|}+(\alpha-\epsilon_\infty^{-1})\frac{\mbox{erfc}(\gamma|\mr-\mr'|)}{|\mr-\mr'|}.
\end{equation}
Then one compares this equation to the non-local operator $\hat{O}$ of an RSH [Eq. \eqref{eq:ohybrsh}, note that the notation we use is different from that used in Ref. \citenum{SGG16}] to determine the parameters or express them in terms of the dielectric function. Ref. \citenum{SGG16} chooses the long-range fraction $\alpha+\beta=\epsilon_{\infty}^{-1}$, and the short-range fraction $\alpha=0.25$ as in PBE0. The authors provided three means to determine the range-separation parameter $\gamma$, including one that fits the long-range decay of the diagonal elements of the dielectric matrix $\epsilon^{-1}(\mathbf{G},\mathbf{G}'=\mathbf{G})$, which can be computed from first principles\cite{WGG08}.

Both Refs. \citenum{SGG14} and \citenum{SGG16} targeted applications in bulk materials. The method was then generalized to finite systems \cite{BVGG16} and heterogeneous interfaces \cite{ZGG19}. For the latter, the mixing fraction for the Fock exchange depends not only on the dielectric function (hence a dielectric-dependent hybrid), but also on the spatial variable $\mr$. Therefore, it can also be considered as a local hybrid functional, which we reserve for Sec. \ref{sec:lhybrid} below.

\begin{figure}[htp]
\centering
\includegraphics[width=3in]{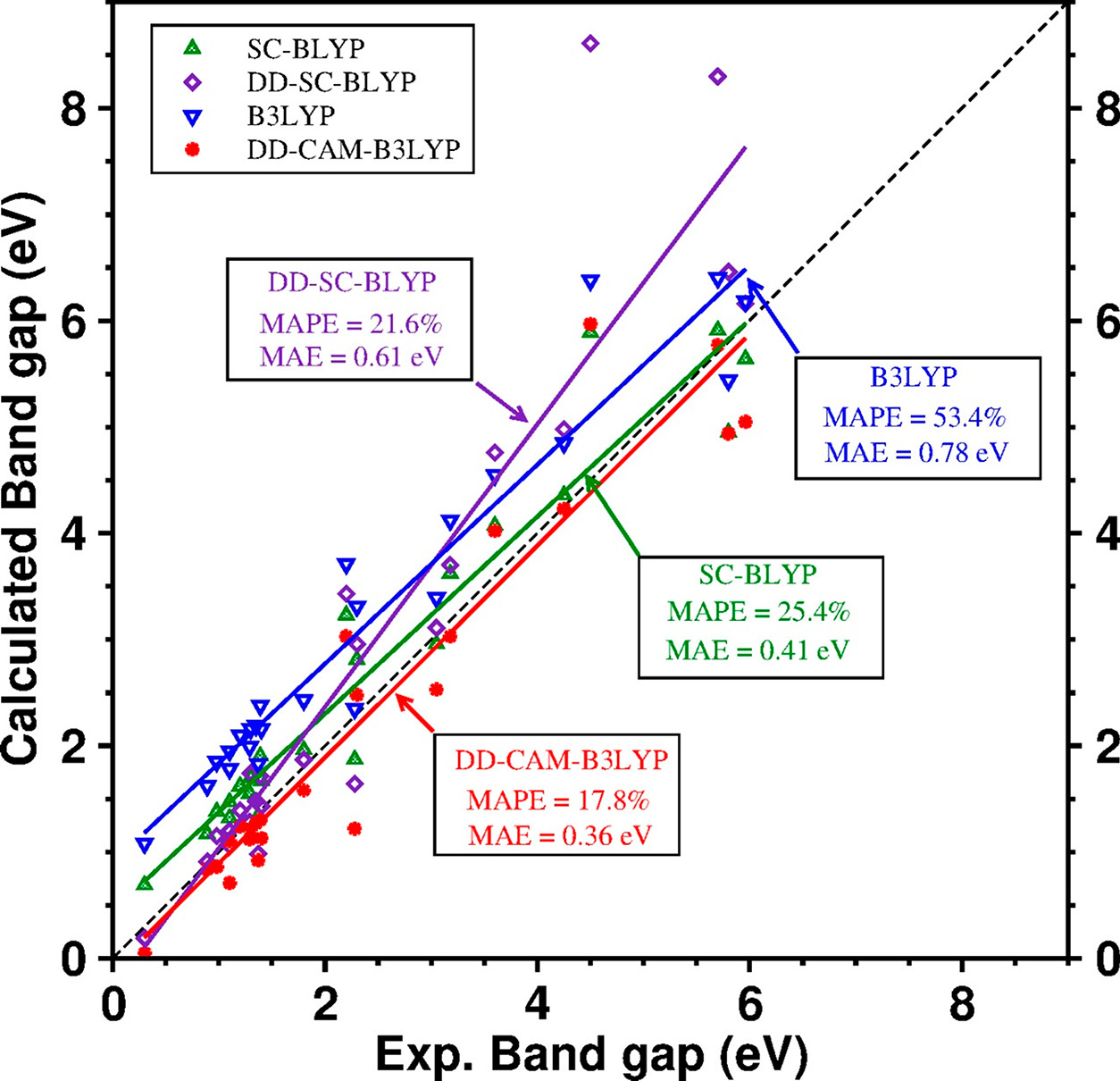}
\caption{Correlations between computed and measured band gaps for the series of 24 quasi-2D materials. The best performance is obtained for the dielectric-dependent hybrid functional (the red circles and line). Reproduced with
permission from Ref. \citenum{DDTP19}: T. Das, G. Di Liberto, S. Tosoni, and G. Pacchioni, \emph{J. Chem. Theory Comput.} \textbf{15}, 6294 (2019). Copyright 2019 American Chemical Society.}
\label{fig:ddtp19}
\end{figure}

Although most applications of dielectric-dependent hybrid functionals are for extended bulk systems \cite{GBCO15a,GBDO18,CMRP18,BWCP19,JPSC20,LFMK20,SGMP20}, there are notable applications in heterogeneous systems, including defects \cite{GBCO15b}, 2D or layered materials, \cite{DDTP19,HB20} and surfaces/interfaces \cite{HB18,HKTO17,HGO19,NHLJ22}. Ref. \citenum{DDTP19} analyzed 24 bulk metal oxides and 24 quasi-2D semiconductors, and concluded that layered materials benefit from the use of dielectric-dependent hybrid functionals more than 3D extended bulk systems. The findings are shown in Fig. \ref{fig:ddtp19}, where one can see that the dielectric-dependent hybrid functional (the red circles and line) leads to the best agreement with experiments. Ref. \citenum{HB18} self-consistently determined the range-separation parameter $\gamma$ from the surface polarizability tensor and found excellent agreement in band-edge energies of NaCl(100) surface compared to $GW$ results. Refs. \citenum{HKTO17,HGO19,NHLJ22} studied the band alignment of wide-gap semiconductors using dielectric-dependent functionals and found good agreement with MBPT or experimental measurements. Lastly, Refs. \citenum{KBT13,HKTO17,SGMP20} pointed out that non-self-consistent applications of the dielectric-dependent hybrid functionals lead to acceptable results compared to self-consistent calculations, which enables high-throughput screening of materials.

\subsection{Local hybrid functionals}
\label{sec:lhybrid}

The concept of local hybrid functional was first discussed in Ref. \citenum{CLB98} and was first realized in Ref. \citenum{JES03}. The essential idea is to generalize Eq. \eqref{eq:hyb} by introducing spatial dependent mixing parameters:
\begin{equation}
E_{\rm xc}=\int d^3\mr \, n(\mr)\left[a(\mr)e_{\rm x}^{\rm ex}(\mr)+\left(1-a(\mr)\right)e_{\rm x}^{\rm sl}(\mr)+e_{\rm c}^{\rm sl}(\mr)\right],
\label{eq:lhyb}
\end{equation}
where $e_{\rm x}^{\rm ex}$, $e_{\rm x}^{\rm sl}$, and $e_{\rm c}^{\rm sl}$ are exact-exchange energy density, semi-local exchange energy density, and correlation energy density, respectively. In other words, $E_{\rm x}^{\rm ex}=\int d^3\mr \, n(\mr)e_{\rm x}^{\rm ex}(\mr)$. Similar relationships hold for $E_{\rm x}^{\rm sl}$ and $E_{\rm c}^{\rm sl}$.

Eq. \eqref{eq:lhyb} provides additional flexibility than Eq. \eqref{eq:hyb}, due to the spatial dependence of $a(\mr)$. However, the efforts in designing a suitable $a(\mr)$ and making such functionals practically feasible are highly non-trivial, since the exchange integral in Eq. \eqref{eq:vx} becomes non-standard due to the spatial dependence of $a(\mr)$. Furthermore, the issue of gauge problem \cite{BCL98} arises: unlike exchange energies, the exchange-energy densities are not unambiguously defined, because one can add another function that integrates to zero to any energy density without affecting the integrated energy. Despite these fundamental and practical challenges, there has been considerable progress made in local hybrid functionals \cite{BRAK07,PSTS08,SKMK14,J14,HF22}, including range-separated ones \cite{JKS08,HS10} with position-dependent range separation function \cite{KSPS08,KB20}. Ref. \citenum{MAK18} provides an excellent review of local hybrid functionals.

Most local hybrid functionals focused on improving the thermodynamic properties. As far as interfacial quasiparticle properties are concerned, we discuss two representative works below, Ref. \citenum{BMB18} and Ref. \citenum{ZGG19}. The latter can also be considered as a dielectric-dependent functional as we briefly mentioned above.

From the discussion in Sec. \ref{sec:dielec}, it is apparent that the local mixing parameter $a(\mr)$ in Eq. \eqref{eq:lhyb} must be associated with the dielectric properties in some way, even if it is not explicitly expressed in terms of the dielectric function. Ref. \citenum{SN15a} is an early attempt to introduce a position-dependent dielectric function. Building upon Ref. \citenum{SN14} and Eq. \eqref{eq:sa09}, the dielectric constant $\epsilon_{\rm s}$ is partitioned into atomic contributions and is determined self-consistently. As a result, the $v_{\rm xc}$ is defined in terms of atomic orbital basis functions, which depends on the average of dielectric constants of atom $A$ and $B$, $\epsilon_{AB}=(\epsilon_A+\epsilon_B)/2$. A similar idea was presented in Ref. \citenum{R10} (although not in the DFT context), where density response functions (polarizabilities) are modeled for each atom, either with metallic response or non-metallic response. The dielectric function of the whole system is then related to the sum of all atomic density response functions.

Beyond the partition of dielectric function into atomic contributions, Ref. \citenum{MVOR11} proposed an estimator $\bar{g}$, which is an averaged value of the ``local estimator'' over the unit cell:
\begin{equation}
\bar{g}=\frac{1}{V_{\rm cell}}\int_{\rm cell}\,d\mr \sqrt{\frac{|\nabla n(\mr)|}{n(\mr)}}.
\label{eq:mvor11}
\end{equation}
The quantity $|\nabla n|/n$ is commonly used as a descriptor for the ``local gap'' in meta-GGAs (GGA=generalized gradient approximation) such as the one proposed by Tran and Blaha, known as TB09 \cite{TB09}, as well as in other local hybrids \cite{KSPS08}. Via the integration over the unit cell, $\bar{g}$ represents a global estimator of the band gap. Ref. \citenum{MVOR11} performed a fitting of the Fock exchange mixing parameter $\alpha$ in Eq. \eqref{eq:hyb} in terms of $\bar{g}$. Furthermore, the authors proposed a local form of $\bar{g}$, a convolution of $\bar{g}$ in Eq. \eqref{eq:mvor11} with a Gaussian of variance $\sigma$:
\begin{equation}
\bar{g}(\mr;\sigma)=\frac{1}{(2\pi \sigma^2)^{3/2}}\int d\mr \, \sqrt{\frac{|\nabla n(\mr')|}{n(\mr')}}\mbox{exp}\left(-\frac{|\mr-\mr'|^2}{2\sigma^2}\right).
\label{eq:mvor11b}
\end{equation}
The authors noted that the $\sigma$ should be large enough to allow for a proper estimation of the dielectric properties, but small enough to sample only the ``local'' environment. Ref. \citenum{MVOR11} did not proceed to develop a local hybrid functional based on Eq. \eqref{eq:mvor11b}, and only mentioned that it will be more meaningful to use Eq. \eqref{eq:mvor11b} than Eq. \eqref{eq:mvor11} for non-bulk systems.

Building on top of these preliminary ideas, Ref. \citenum{BMB18} proposed a local hybrid functional of the form
\begin{equation}
\begin{split}
E_{\rm xc} = & -\frac{1}{2}\sum_{ij}^{\rm occ.}\iint d\mr d\mr' \phi_i^*(\mr)\phi_j^*(\mr')\frac{\alpha(\mr,\mr')}{|\mr-\mr'|}\phi_j(\mr)\phi_i(\mr') \\
& + \int d\mr \, n(\mr)\left\{\left[1-\alpha(\mr,\mr')\right]e_{\rm x}^{\rm sl}(\mr)+e_{\rm c}^{\rm sl}(\mr)\right\}.
\end{split}
\label{eq:bmb18}
\end{equation}
Comparing this equation with the general form of local hybrid functionals in Eq. \eqref{eq:lhyb}, one finds that the mixing parameter is deliberately chosen to be $\alpha(\mr,\mr')$ rather than simply $a(\mr)$. We will see below that this choice greatly simplifies the implementation in plane-wave-based packages. In Ref. \citenum{BMB18}, $\alpha(\mr,\mr')$ was further made symmetric and separable
\begin{equation}
\alpha(\mr,\mr')=\sqrt{a(\mr)a(\mr')}.
\label{eq:bmb18b}
\end{equation}
The local mixing function $a(\mr;\sigma)$ is then expressed in terms of $\bar{g}(\mr;\sigma)$: $a(\mr;\sigma)=a_1+a_2[\bar{g}(\mr;\sigma)]^m$, where $a_1$ and $a_2$ are parameters inherited from Ref. \citenum{MVOR11}, and $m=1$ for a PBE0 form of the hybrid functional and $m=4$ for an HSE (HSE=Heyd-Scuseria-Ernzerhof) form of the hybrid functional. In the case of the latter, $|\mr-\mr'|$ in Eq. \eqref{eq:bmb18} needs to be replaced by the corresponding screened Coulomb interaction [c.f. the discussion of Eq. \eqref{eq:vx} and thereafter]. 

\begin{figure}[htp]
\centering
\includegraphics[width=3in]{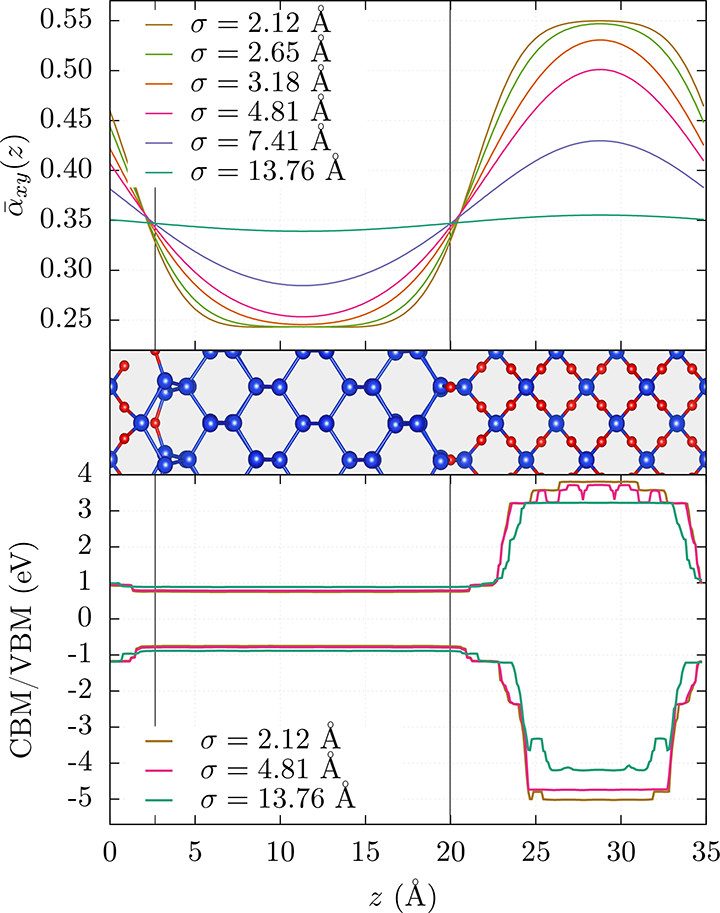}
\caption{Upper panel: planar averaged mixing parameter $\bar{\alpha}_{xy}(z;\sigma)$. Middle panel: Profile view of the Si/SiO$_2$ structure. Si atoms are blue and O atoms are red. Lower panel: band profiles of the CBM and VBM. The calculations are performed with the HSE-based local hybrid functional using different values of $\sigma$. Reproduced with
permission from Ref. \citenum{BMB18}: P. Borlido, M. A. L. Marques, and S. Botti, \emph{J. Chem. Theory Comput.} \textbf{14}, 939 (2018). Copyright 2018 American Chemical Society.}
\label{fig:bmb18}
\end{figure}

A few discussions of this functional are in order. First, Ref. \citenum{BMB18} neglected the gauge freedom \cite{BCL98} for the exchange energy densities. Second, the derivatives of $\alpha(\mr,\mr')$ with respect to $n(\mr)$ and $\nabla n(\mr)$ are neglected in the functional derivative $\delta E_{\rm xc}/\delta n(\mr)$. This is equivalent to the approximation that the mixing parameter $\alpha(\mr,\mr')$ is applied directly to the nonlocal Fock \emph{potential}. The consequence is that the XC potential is not a functional derivative of the XC energy. Both approximations are partially justified since the focus of Ref. \citenum{BMB18} is interfacial electronic structure rather than thermochemistry. Third, for a surface (i.e., an interface with the vacuum), the $\alpha \to 1$ limit is not correctly recovered.

Practically, neglecting the derivatives of $\alpha(\mr,\mr')$ with respect to $n(\mr)$ and $\nabla n(\mr)$ enables easy implementation and efficient calculations. Using a plane-wave basis, the exact exchange is typically computed \cite{SFH06} by a fast Fourier transform of the auxiliary codensities $\rho_{ij}(\mr)=\phi_i^*(\mr)\phi_j(\mr)$. The separable form of $\alpha(\mr,\mr')$ in Eq. \eqref{eq:bmb18b} simply alters the form of the codensities such that $\sqrt{a(\mr;\sigma)}$ is absorbed into $\rho_{ij}(\mr)$. The computational cost is therefore similar to that of a standard hybrid functional.

Ref. \citenum{BMB18} performed test calculations on a few heterogeneous interfaces formed between two semiconductors, Si/SiO$_2$, GaP/Si, AlP/GaP, and AlAs/GaAs. The results are typically on par with $G_0W_0$, with 0.1-0.2 eV difference in the VBM or CBM band offset. Fig. \ref{fig:bmb18} shows the planar average of the mixing parameter $\bar{\alpha}_{xy}(z;\sigma)=A^{-1}\int dxdy \, \alpha(\mr,\mr;\sigma)$, where $A$ is the unit area, for the Si/SiO$_2$ interface. One can see that the mixing parameter clearly depends on the position, i.e., different for the Si side and for the SiO$_2$ side, which is the unique strength of a local hybrid functional.

Following the same strategy of Ref. \citenum{BMB18}, Ref. \citenum{ZGG19} proposed another version of $\alpha(\mr,\mr')$ to be used in Eq. \eqref{eq:bmb18}:
\begin{equation}
\alpha(\mr,\mr')=\frac{1}{\sqrt{\epsilon(\mr)\epsilon(\mr')}},
\label{eq:zgg19}
\end{equation}
where $\epsilon(\mr)$ is the local dielectric function. Comparing this equation to Eq. \eqref{eq:bmb18b}, one realizes that Eq. \eqref{eq:zgg19} is motivated by the argument that the Fock exchange mixing parameter $\alpha$ in a global hybrid functional can be approximated by $\epsilon_\infty^{-1}$ [c.f. Eqs. \eqref{eq:sgg14b} and \eqref{eq:sgg14c} and discussions therein]. The resulting functional is then a dielectric-dependent local hybrid functional. Ref. \citenum{ZGG19} further proposed a self-consistent scheme in determining $\epsilon(\mr)$ based on the finite-field approach. This is achieved by defining a spatial dependent polarization $P(\mr)$, which can be computed from the shift of the centers of the Wannier functions of the unperturbed system when an external electric field is applied \cite{KV93}. The $\epsilon(\mr)$ is then computed from $P(\mr)$, completing the self-consistent loop.

\begin{figure}[htp]
\centering
\includegraphics[width=3in]{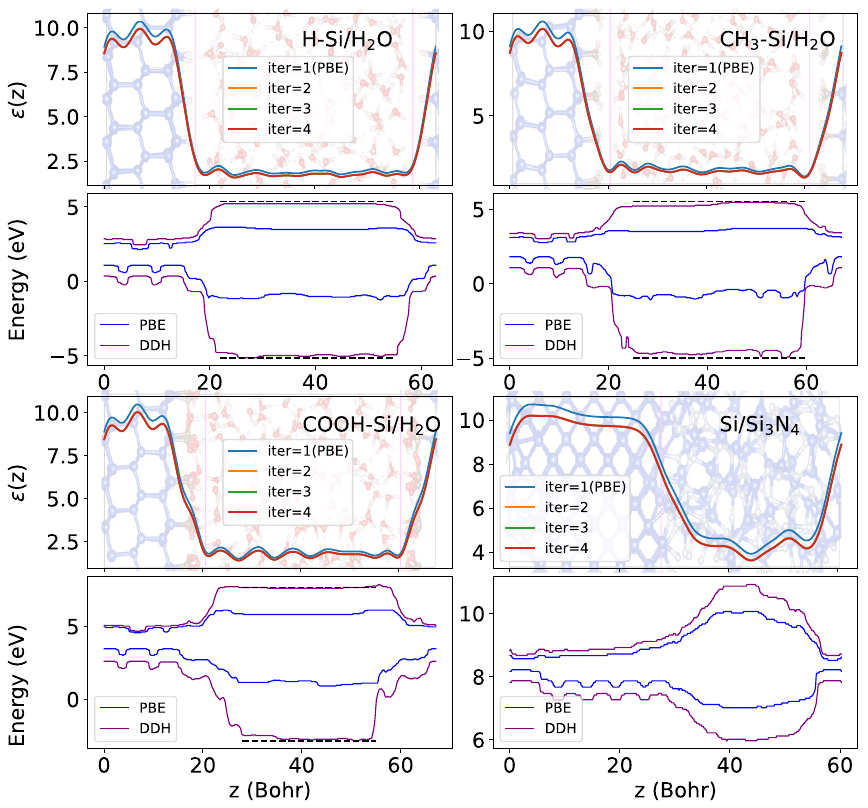}
\caption{$xy$-averaged dielectric function $\epsilon(z)$ and band offsets of four interfaces, based on Eqs. \eqref{eq:bmb18} and \eqref{eq:zgg19}. $\epsilon(z)$ results are shown as a function of the number of iterations. For the band offsets, comparisons between PBE results (blue lines) and those from the dielectric-dependent hybrid functional (purple lines) are provided. Reproduced with
permission from Ref. \citenum{ZGG19}: H. Zheng, M. Govoni, and G. Galli, \emph{Phys. Rev. Mater.} \textbf{3}, 073803 (2019). Copyright 2019 American Physical Society.}
\label{fig:zgg19}
\end{figure}

Fig. \ref{fig:zgg19} shows the performance of the dielectric-dependent local hybrid functional proposed in Ref. \citenum{ZGG19}. Similar to Fig. \ref{fig:bmb18} reproduced from Ref. \citenum{BMB18}, the $xy$-averaged dielectric function $\epsilon(z)$ depends on the spatial coordinate and the specific material. It also shows that the self-consistent determination of $\epsilon(\mr)$ converges in a few iterations.

Lastly, we discuss a local modified Becke-Johnson (BJ) XC potential designed for interfacial electronic structure \cite{RMB20}. Although this is not a local \emph{hybrid} functional, the idea of local mixing closely resembles that in a local hybrid functional, as we see below.

The modified BJ exchange potential was proposed by TB09 and has the following form\cite{TB09}
\begin{equation}
v_{\rm x}^{\rm mBJ} = cv_{\rm x}^{\rm BR}(\mr)+(3c-2)\frac{1}{\pi}\sqrt{\frac{5}{12}}\sqrt{\frac{2t(\mr)}{n(\mr)}},
\label{eq:RMB20}
\end{equation}
where $t(\mr)$ is the kinetic-energy density used in meta-GGAs, and $v_{\rm x}^{\rm BR}(\mr)$ is the Becke-Russel (BR) exchange potential \cite{BR89}
\begin{equation}
v_{\rm x}^{\rm BR}(\mr)=-\frac{1}{b(\mr)}\left[1-e^{-x(\mr)}-\frac{1}{2}x(\mr)e^{-x(\mr)}\right],
\label{eq:RMB20b}
\end{equation}
where $x(\mr)$ and $b(\mr)$ can be calculated from $n(\mr)$, $\nabla n(\mr)$, and $\nabla^2 n(\mr)$.

Eq. \eqref{eq:RMB20} is called the modified BJ exchange potential because the original BJ potential \cite{BJ06} uses the Slater potential \cite{S51} instead of $v_{\rm x}^{\rm BR}(\mr)$ in the first term and sets $c=1$. Note that $v_{\rm x}^{\rm BR}(\mr)$ is a model for the exact exchange potential, and the second term in Eq. \eqref{eq:RMB20} can be seen as a screening that corrects the error of the first term. Therefore, the parameter $c$ in Eq. \eqref{eq:RMB20} is reminiscent of the Fock exchange mixing parameter found in hybrid functionals, although Eq. \eqref{eq:RMB20} is, strictly speaking, a meta-GGA. It is due to this similarity that we decide to keep the discussion of Eq. \eqref{eq:RMB20} and Ref. \citenum{RMB20} in this part of the Review. TB09 further fits $c$ in terms of $\bar{g}$ defined in Eq. \eqref{eq:mvor11}: $c=\alpha + \beta \bar{g}$ with $\alpha=0.488$ and $\beta=0.5$ bohr.

In Ref. \citenum{RMB20}, the authors generalized Eq. \eqref{eq:RMB20} by making the parameter $c$ ``local'', i.e., replacing $c$ in Eq. \eqref{eq:RMB20} with $c(\mr)$. The $c(\mr)$ is related to $\bar{g}(\mr)$ defined in Eq. \eqref{eq:mvor11b} in the same way as $c$ is related to $\bar{g}$ in TB09. The authors further enforced $c(\mr) \to 1$ in the vacuum region (recall that this was not possible with the local hybrid functional proposed in Ref. \citenum{BMB18}), by slightly adjusting the definition of $g(\mr)$ such that $c(\mr) \to 1$ when the density is below a defined threshold.

The local modified BJ exchange potential shares the same limitation as TB09 and the local hybrid functionals proposed in Refs. \citenum{BMB18,ZGG19}, in that it is not a functional derivative of any density functional. This means that it violates a few exact conditions and could not be used to compute total-energy-related properties and thermochemistry. Nevertheless, these functionals seem to perform well as far as interfacial electronic structure is concerned. 

\begin{figure}[htp]
\centering
\includegraphics[width=3in]{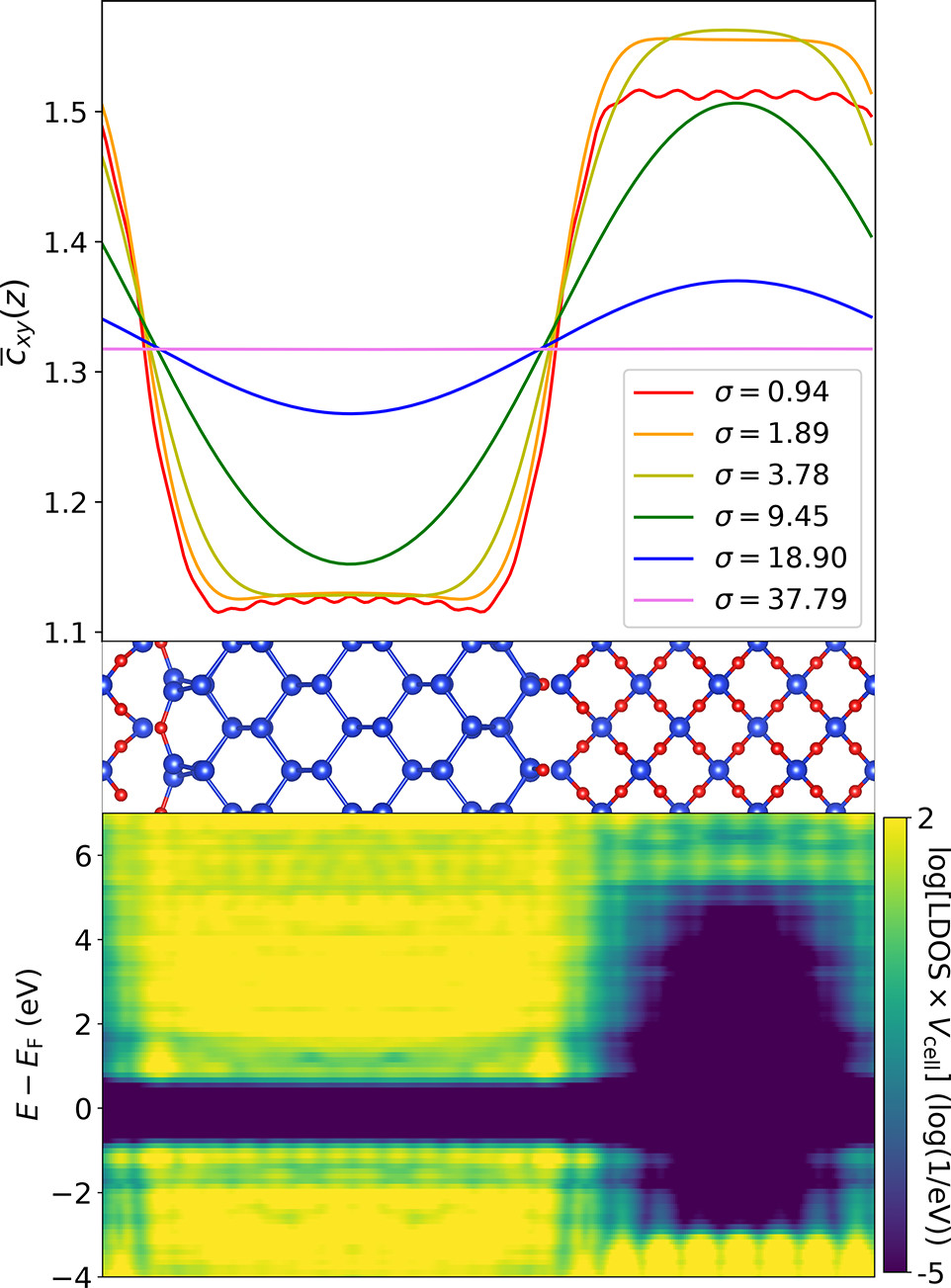}
\caption{Top: $xy$-averaged mixing parameter $\bar{c}_{xy}(z)$ for different values of $\sigma$ given in bohr. Middle: atomic structure of Si/SiO$_2$ interface (Si: blue; O: red). Bottom: logarithm of the local density of states averaged in the $xy$ plane (yellow: high value; violet: low value) calculated with $\sigma=3.78$ bohr, showing the band offset at the interface. Reproduced with
permission from Ref. \citenum{RMB20}: T. Rauch, M. A. L. Marques, and S. Botti, \emph{J. Chem. Theory Comput.} \textbf{16}, 2654 (2020). Copyright 2020 American Chemical Society.}
\label{fig:RMB20}
\end{figure}

Ref. \citenum{RMB20} used the same test systems as Ref. \citenum{BMB18}, and its performance is shown in Fig. \ref{fig:RMB20}. Similar to the behavior of $\bar{\alpha}_{xy}(z)$ in Fig. \ref{fig:bmb18}, here the $\bar{c}_{xy}(z)$ also shows a clear difference for the two sides of the interface, justifying the necessity of a \emph{local} mixing parameter.

\section{Concluding Remarks}
\label{sec:outlook}
We started this Review by explaining the physics behind surface and interface electronic structure using quantities defined in MBPT: the screened Coulomb interaction $W$. We discussed the spatial dependence of $W$, whose long-range limit admits the image-potential form. We also discussed the orbital dependence of $W$, which leads to the energy and gap renormalization at an interface. After that, we reviewed early efforts in the DFT community to understand the surface in terms of quantities defined in DFT: the XC hole and the XC potential. We discussed the evolution of the shapes of the exact exchange hole and the XC hole as the reference electron is taken away from the metal surface. We also discussed the relationship between the XC potential and the image potential, as well as different perspectives in the literature on understanding this relationship, making connections between quantities defined in DFT and those defined in MBPT. In the last part, we surveyed modern developments of density functionals for accurate interfacial electronic structure. We focused on two types of functionals: dielectric-dependent hybrids and local hybrids. In both types, the key strategy is to build in certain ingredients related to dielectric screening and its spatial variance, again making connections between MBPT concepts and the DFT language.

We stress again that although there is no formal justification for interpreting eigenvalues from static DFT calculations as quasiparticle energy levels, it is of great practical interest to develop density functionals as alternatives to MBPT to describe the quasiparticle electronic structure at heterogeneous interfaces. Compared to MBPT, it is easier to perform self-consistent density functional calculations with moderate computational cost. It is also easier to incorporate other physics than the dielectric screening into the functional development, such as van der Waals interactions and the strong correlation associated with transition metal elements. Moreover, it is technically simpler to reach convergence in density functional calculations, as they are governed by fewer parameters than in MBPT calculations.

Looking forward, novel functionals are being developed to improve the description of interfacial electronic structure within the DFT framework, without invoking MBPT but borrowing MBPT concepts into density functional developments, which blurs the boundary between these two theories in a healthy way. Most functionals we reviewed in Sec. \ref{sec:func} have been tested using semiconductor-semiconductor interfaces, and it is yet to be seen how these functionals, or their improved versions, work for molecule-semiconductor interfaces, which are more heterogeneous due to the vast difference in dielectric effects between a molecule and a semiconductor. Furthermore, it is yet to be seen how one can adopt the old wisdom of the surface XC hole and/or the XC potential discussed in Sec. \ref{sec:early} into functional developments. Can one use the expected behaviors of the surface XC hole and/or the XC potential as exact conditions? Are these conditions on the \emph{surface} XC hole and/or the XC potential enough to guarantee accuracy in describing the electronic structure at an \emph{interface}? Can one go beyond the idea of using the screened exchange to describe the long-range correlation across the interface, and directly develop non-local correlation functionals that capture the heterogeneity in the interfacial dielectric properties? We may not be able to reach satisfactory solutions to all these questions very soon, but the approaches taken to answer these questions will surely lead to the novel development of functionals for a better description of interfacial electronic structure within the framework of DFT.

\begin{acknowledgements}
Z.-F.L. acknowledges an NSF CAREER award, DMR-2044552.
\end{acknowledgements}

\section*{Author Declarations}
\subsection*{Conflict of Interest}
The author has no conflicts to disclose.

\section*{Data Availability}
Data sharing is not applicable to this article as no new data were created or analyzed in this study.
\bibliography{refs.bib}

\begin{thebibliography}{156}%
\makeatletter
\providecommand \@ifxundefined [1]{%
 \@ifx{#1\undefined}
}%
\providecommand \@ifnum [1]{%
 \ifnum #1\expandafter \@firstoftwo
 \else \expandafter \@secondoftwo
 \fi
}%
\providecommand \@ifx [1]{%
 \ifx #1\expandafter \@firstoftwo
 \else \expandafter \@secondoftwo
 \fi
}%
\providecommand \natexlab [1]{#1}%
\providecommand \enquote  [1]{``#1''}%
\providecommand \bibnamefont  [1]{#1}%
\providecommand \bibfnamefont [1]{#1}%
\providecommand \citenamefont [1]{#1}%
\providecommand \href@noop [0]{\@secondoftwo}%
\providecommand \href [0]{\begingroup \@sanitize@url \@href}%
\providecommand \@href[1]{\@@startlink{#1}\@@href}%
\providecommand \@@href[1]{\endgroup#1\@@endlink}%
\providecommand \@sanitize@url [0]{\catcode `\\12\catcode `\$12\catcode
  `\&12\catcode `\#12\catcode `\^12\catcode `\_12\catcode `\%12\relax}%
\providecommand \@@startlink[1]{}%
\providecommand \@@endlink[0]{}%
\providecommand \url  [0]{\begingroup\@sanitize@url \@url }%
\providecommand \@url [1]{\endgroup\@href {#1}{\urlprefix }}%
\providecommand \urlprefix  [0]{URL }%
\providecommand \Eprint [0]{\href }%
\providecommand \doibase [0]{https://doi.org/}%
\providecommand \selectlanguage [0]{\@gobble}%
\providecommand \bibinfo  [0]{\@secondoftwo}%
\providecommand \bibfield  [0]{\@secondoftwo}%
\providecommand \translation [1]{[#1]}%
\providecommand \BibitemOpen [0]{}%
\providecommand \bibitemStop [0]{}%
\providecommand \bibitemNoStop [0]{.\EOS\space}%
\providecommand \EOS [0]{\spacefactor3000\relax}%
\providecommand \BibitemShut  [1]{\csname bibitem#1\endcsname}%
\let\auto@bib@innerbib\@empty
\bibitem [{\citenamefont {Green}, \citenamefont {Ho-Baillie},\ and\
  \citenamefont {Snaith}(2014)}]{GHS14}%
  \BibitemOpen
  \bibfield  {author} {\bibinfo {author} {\bibfnamefont {M.~A.}\ \bibnamefont
  {Green}}, \bibinfo {author} {\bibfnamefont {A.}~\bibnamefont {Ho-Baillie}},\
  and\ \bibinfo {author} {\bibfnamefont {H.~J.}\ \bibnamefont {Snaith}},\
  }\bibfield  {title} {\enquote {\bibinfo {title} {The emergence of perovskite
  solar cells},}\ }\href@noop {} {\bibfield  {journal} {\bibinfo  {journal}
  {Nat. Photon.}\ }\textbf {\bibinfo {volume} {8}},\ \bibinfo {pages} {506}
  (\bibinfo {year} {2014})}\BibitemShut {NoStop}%
\bibitem [{\citenamefont {Wang}, \citenamefont {Li},\ and\ \citenamefont
  {Domen}(2019)}]{WLD19}%
  \BibitemOpen
  \bibfield  {author} {\bibinfo {author} {\bibfnamefont {Z.}~\bibnamefont
  {Wang}}, \bibinfo {author} {\bibfnamefont {C.}~\bibnamefont {Li}},\ and\
  \bibinfo {author} {\bibfnamefont {K.}~\bibnamefont {Domen}},\ }\bibfield
  {title} {\enquote {\bibinfo {title} {Recent developments in heterogeneous
  photocatalysts for solar-driven overall water splitting},}\ }\href@noop {}
  {\bibfield  {journal} {\bibinfo  {journal} {Chem. Soc. Rev.}\ }\textbf
  {\bibinfo {volume} {48}},\ \bibinfo {pages} {2109} (\bibinfo {year}
  {2019})}\BibitemShut {NoStop}%
\bibitem [{\citenamefont {Liu}, \citenamefont {Huang},\ and\ \citenamefont
  {Duan}(2019)}]{LHD19}%
  \BibitemOpen
  \bibfield  {author} {\bibinfo {author} {\bibfnamefont {Y.}~\bibnamefont
  {Liu}}, \bibinfo {author} {\bibfnamefont {Y.}~\bibnamefont {Huang}},\ and\
  \bibinfo {author} {\bibfnamefont {X.}~\bibnamefont {Duan}},\ }\bibfield
  {title} {\enquote {\bibinfo {title} {Van der waals integration before and
  beyond two-dimensional materials},}\ }\href@noop {} {\bibfield  {journal}
  {\bibinfo  {journal} {Nature}\ }\textbf {\bibinfo {volume} {567}},\ \bibinfo
  {pages} {323} (\bibinfo {year} {2019})}\BibitemShut {NoStop}%
\bibitem [{\citenamefont {Goodenough}\ and\ \citenamefont {Park}(2013)}]{GP13}%
  \BibitemOpen
  \bibfield  {author} {\bibinfo {author} {\bibfnamefont {J.~B.}\ \bibnamefont
  {Goodenough}}\ and\ \bibinfo {author} {\bibfnamefont {K.-S.}\ \bibnamefont
  {Park}},\ }\bibfield  {title} {\enquote {\bibinfo {title} {The {L}i-ion
  rechargeable battery: A perspective},}\ }\href@noop {} {\bibfield  {journal}
  {\bibinfo  {journal} {J. Am. Chem. Soc.}\ }\textbf {\bibinfo {volume}
  {135}},\ \bibinfo {pages} {1167} (\bibinfo {year} {2013})}\BibitemShut
  {NoStop}%
\bibitem [{\citenamefont {Ihn}(2010)}]{I10}%
  \BibitemOpen
  \bibfield  {author} {\bibinfo {author} {\bibfnamefont {T.}~\bibnamefont
  {Ihn}},\ }\href@noop {} {\emph {\bibinfo {title} {Semiconductor
  Nanostructures: Quantum states and electronic transport}}},\ \bibinfo
  {edition} {1st}\ ed.\ (\bibinfo  {publisher} {Oxford University Press},\
  \bibinfo {address} {Oxford},\ \bibinfo {year} {2010})\BibitemShut {NoStop}%
\bibitem [{\citenamefont {Hohenberg}\ and\ \citenamefont {Kohn}(1964)}]{HK64}%
  \BibitemOpen
  \bibfield  {author} {\bibinfo {author} {\bibfnamefont {P.}~\bibnamefont
  {Hohenberg}}\ and\ \bibinfo {author} {\bibfnamefont {W.}~\bibnamefont
  {Kohn}},\ }\bibfield  {title} {\enquote {\bibinfo {title} {Inhomogeneous
  electron gas},}\ }\href {https://doi.org/10.1103/PhysRev.136.B864} {\bibfield
   {journal} {\bibinfo  {journal} {Phys. Rev.}\ }\textbf {\bibinfo {volume}
  {136}},\ \bibinfo {pages} {B864} (\bibinfo {year} {1964})}\BibitemShut
  {NoStop}%
\bibitem [{\citenamefont {Kohn}\ and\ \citenamefont {Sham}(1965)}]{KS65}%
  \BibitemOpen
  \bibfield  {author} {\bibinfo {author} {\bibfnamefont {W.}~\bibnamefont
  {Kohn}}\ and\ \bibinfo {author} {\bibfnamefont {L.~J.}\ \bibnamefont
  {Sham}},\ }\bibfield  {title} {\enquote {\bibinfo {title} {Self-consistent
  equations including exchange and correlation effects},}\ }\href
  {https://doi.org/10.1103/PhysRev.140.A1133} {\bibfield  {journal} {\bibinfo
  {journal} {Phys. Rev.}\ }\textbf {\bibinfo {volume} {140}},\ \bibinfo {pages}
  {A1133} (\bibinfo {year} {1965})}\BibitemShut {NoStop}%
\bibitem [{\citenamefont {Perdew}\ \emph {et~al.}(1982)\citenamefont {Perdew},
  \citenamefont {Parr}, \citenamefont {Levy},\ and\ \citenamefont
  {Balduz}}]{PPLB82}%
  \BibitemOpen
  \bibfield  {author} {\bibinfo {author} {\bibfnamefont {J.~P.}\ \bibnamefont
  {Perdew}}, \bibinfo {author} {\bibfnamefont {R.~G.}\ \bibnamefont {Parr}},
  \bibinfo {author} {\bibfnamefont {M.}~\bibnamefont {Levy}},\ and\ \bibinfo
  {author} {\bibfnamefont {J.~L.}\ \bibnamefont {Balduz}},\ }\bibfield  {title}
  {\enquote {\bibinfo {title} {Density-functional theory for fractional
  particle number: {D}erivative discontinuities of the energy},}\ }\href@noop
  {} {\bibfield  {journal} {\bibinfo  {journal} {Phys. Rev. Lett.}\ }\textbf
  {\bibinfo {volume} {49}},\ \bibinfo {pages} {1691} (\bibinfo {year}
  {1982})}\BibitemShut {NoStop}%
\bibitem [{\citenamefont {Onida}, \citenamefont {Reining},\ and\ \citenamefont
  {Rubio}(2002)}]{ORR02}%
  \BibitemOpen
  \bibfield  {author} {\bibinfo {author} {\bibfnamefont {G.}~\bibnamefont
  {Onida}}, \bibinfo {author} {\bibfnamefont {L.}~\bibnamefont {Reining}},\
  and\ \bibinfo {author} {\bibfnamefont {A.}~\bibnamefont {Rubio}},\ }\bibfield
   {title} {\enquote {\bibinfo {title} {Electronic excitations:
  {D}ensity-functional versus many-body {G}reen's-function approaches},}\
  }\href {https://doi.org/10.1103/RevModPhys.74.601} {\bibfield  {journal}
  {\bibinfo  {journal} {Rev. Mod. Phys.}\ }\textbf {\bibinfo {volume} {74}},\
  \bibinfo {pages} {601} (\bibinfo {year} {2002})}\BibitemShut {NoStop}%
\bibitem [{\citenamefont {Hedin}(1965)}]{H65}%
  \BibitemOpen
  \bibfield  {author} {\bibinfo {author} {\bibfnamefont {L.}~\bibnamefont
  {Hedin}},\ }\bibfield  {title} {\enquote {\bibinfo {title} {New method for
  calculating the one-particle {G}reen's function with application to the
  electron-gas problem},}\ }\href@noop {} {\bibfield  {journal} {\bibinfo
  {journal} {Phys. Rev.}\ }\textbf {\bibinfo {volume} {139}},\ \bibinfo {pages}
  {A796} (\bibinfo {year} {1965})}\BibitemShut {NoStop}%
\bibitem [{\citenamefont {Marini}\ \emph {et~al.}(2009)\citenamefont {Marini},
  \citenamefont {Hogan}, \citenamefont {Gr\"{u}ning},\ and\ \citenamefont
  {Varsano}}]{yambo}%
  \BibitemOpen
  \bibfield  {author} {\bibinfo {author} {\bibfnamefont {A.}~\bibnamefont
  {Marini}}, \bibinfo {author} {\bibfnamefont {C.}~\bibnamefont {Hogan}},
  \bibinfo {author} {\bibfnamefont {M.}~\bibnamefont {Gr\"{u}ning}},\ and\
  \bibinfo {author} {\bibfnamefont {D.}~\bibnamefont {Varsano}},\ }\bibfield
  {title} {\enquote {\bibinfo {title} {yambo: An \emph{ab initio} tool for
  excited state calculations},}\ }\href@noop {} {\bibfield  {journal} {\bibinfo
   {journal} {Comput. Phys. Commun.}\ }\textbf {\bibinfo {volume} {180}},\
  \bibinfo {pages} {1392} (\bibinfo {year} {2009})}\BibitemShut {NoStop}%
\bibitem [{\citenamefont {Deslippe}\ \emph {et~al.}(2012)\citenamefont
  {Deslippe}, \citenamefont {Samsonidze}, \citenamefont {Strubbe},
  \citenamefont {Jain}, \citenamefont {Cohen},\ and\ \citenamefont
  {Louie}}]{BerkeleyGW}%
  \BibitemOpen
  \bibfield  {author} {\bibinfo {author} {\bibfnamefont {J.}~\bibnamefont
  {Deslippe}}, \bibinfo {author} {\bibfnamefont {G.}~\bibnamefont
  {Samsonidze}}, \bibinfo {author} {\bibfnamefont {D.~A.}\ \bibnamefont
  {Strubbe}}, \bibinfo {author} {\bibfnamefont {M.}~\bibnamefont {Jain}},
  \bibinfo {author} {\bibfnamefont {M.~L.}\ \bibnamefont {Cohen}},\ and\
  \bibinfo {author} {\bibfnamefont {S.~G.}\ \bibnamefont {Louie}},\ }\bibfield
  {title} {\enquote {\bibinfo {title} {Berkeley{GW}: A massively parallel
  computer package for the calculation of the quasiparticle and optical
  properties of materials and nanostructures},}\ }\href@noop {} {\bibfield
  {journal} {\bibinfo  {journal} {Comput. Phys. Commun.}\ }\textbf {\bibinfo
  {volume} {183}},\ \bibinfo {pages} {1269} (\bibinfo {year}
  {2012})}\BibitemShut {NoStop}%
\bibitem [{\citenamefont {Govoni}\ and\ \citenamefont {Galli}(2015)}]{WEST}%
  \BibitemOpen
  \bibfield  {author} {\bibinfo {author} {\bibfnamefont {M.}~\bibnamefont
  {Govoni}}\ and\ \bibinfo {author} {\bibfnamefont {G.}~\bibnamefont {Galli}},\
  }\bibfield  {title} {\enquote {\bibinfo {title} {Large scale {GW}
  calculations},}\ }\href@noop {} {\bibfield  {journal} {\bibinfo  {journal}
  {J. Chem. Theory Comput.}\ }\textbf {\bibinfo {volume} {11}},\ \bibinfo
  {pages} {2680} (\bibinfo {year} {2015})}\BibitemShut {NoStop}%
\bibitem [{\citenamefont {Cohen}, \citenamefont {Mori-Sanchez},\ and\
  \citenamefont {Yang}(2012)}]{CMY12}%
  \BibitemOpen
  \bibfield  {author} {\bibinfo {author} {\bibfnamefont {A.}~\bibnamefont
  {Cohen}}, \bibinfo {author} {\bibfnamefont {P.}~\bibnamefont
  {Mori-Sanchez}},\ and\ \bibinfo {author} {\bibfnamefont {W.}~\bibnamefont
  {Yang}},\ }\bibfield  {title} {\enquote {\bibinfo {title} {Challenges for
  density functional theory},}\ }\href@noop {} {\bibfield  {journal} {\bibinfo
  {journal} {Chem. Rev.}\ }\textbf {\bibinfo {volume} {112}},\ \bibinfo {pages}
  {289} (\bibinfo {year} {2012})}\BibitemShut {NoStop}%
\bibitem [{\citenamefont {Burke}(2012)}]{B12}%
  \BibitemOpen
  \bibfield  {author} {\bibinfo {author} {\bibfnamefont {K.}~\bibnamefont
  {Burke}},\ }\bibfield  {title} {\enquote {\bibinfo {title} {Perspective on
  density functional theory},}\ }\href@noop {} {\bibfield  {journal} {\bibinfo
  {journal} {J. Chem. Phys.}\ }\textbf {\bibinfo {volume} {136}},\ \bibinfo
  {pages} {150901} (\bibinfo {year} {2012})}\BibitemShut {NoStop}%
\bibitem [{\citenamefont {Becke}(2014)}]{B14}%
  \BibitemOpen
  \bibfield  {author} {\bibinfo {author} {\bibfnamefont {A.~D.}\ \bibnamefont
  {Becke}},\ }\bibfield  {title} {\enquote {\bibinfo {title} {Perspective:
  Fifty years of density-functional theory in chemical physics},}\ }\href@noop
  {} {\bibfield  {journal} {\bibinfo  {journal} {J. Chem. Phys.}\ }\textbf
  {\bibinfo {volume} {140}},\ \bibinfo {pages} {18A301} (\bibinfo {year}
  {2014})}\BibitemShut {NoStop}%
\bibitem [{\citenamefont {Jones}(2015)}]{J15}%
  \BibitemOpen
  \bibfield  {author} {\bibinfo {author} {\bibfnamefont {R.~O.}\ \bibnamefont
  {Jones}},\ }\bibfield  {title} {\enquote {\bibinfo {title} {Density
  functional theory: Its origins, rise to prominence, and future},}\
  }\href@noop {} {\bibfield  {journal} {\bibinfo  {journal} {Rev. Mod. Phys.}\
  }\textbf {\bibinfo {volume} {87}},\ \bibinfo {pages} {897} (\bibinfo {year}
  {2015})}\BibitemShut {NoStop}%
\bibitem [{\citenamefont {Yu}, \citenamefont {Li},\ and\ \citenamefont
  {Truhlar}(2016)}]{YLT16}%
  \BibitemOpen
  \bibfield  {author} {\bibinfo {author} {\bibfnamefont {H.~S.}\ \bibnamefont
  {Yu}}, \bibinfo {author} {\bibfnamefont {S.~L.}\ \bibnamefont {Li}},\ and\
  \bibinfo {author} {\bibfnamefont {D.~G.}\ \bibnamefont {Truhlar}},\
  }\bibfield  {title} {\enquote {\bibinfo {title} {Perspective: Kohn-{S}ham
  density functional theory descending a staircase},}\ }\href@noop {}
  {\bibfield  {journal} {\bibinfo  {journal} {J. Chem. Phys.}\ }\textbf
  {\bibinfo {volume} {145}},\ \bibinfo {pages} {130901} (\bibinfo {year}
  {2016})}\BibitemShut {NoStop}%
\bibitem [{\citenamefont {Teale}\ \emph {et~al.}(2022)\citenamefont {Teale},
  \citenamefont {Helgaker}, \citenamefont {Savin}, \citenamefont {Adamo},
  \citenamefont {Aradi}, \citenamefont {Arbuznikov}, \citenamefont {Ayers},
  \citenamefont {Baerends}, \citenamefont {Barone}, \citenamefont {Calaminici},
  \citenamefont {Cances}, \citenamefont {Carter}, \citenamefont {Chattaraj},
  \citenamefont {Chermette}, \citenamefont {Ciofini}, \citenamefont {Crawford},
  \citenamefont {De~Proft}, \citenamefont {Dobson}, \citenamefont {Draxl},
  \citenamefont {Frauenheim}, \citenamefont {Fromager}, \citenamefont
  {Fuentealba}, \citenamefont {Gagliardi}, \citenamefont {Galli}, \citenamefont
  {Gao}, \citenamefont {Geerlings}, \citenamefont {Gidopoulos}, \citenamefont
  {Gill}, \citenamefont {Gori-Giorgi}, \citenamefont {Gorling}, \citenamefont
  {Gould}, \citenamefont {Grimme}, \citenamefont {Gritsenko}, \citenamefont
  {Jensen}, \citenamefont {Johnson}, \citenamefont {Jones}, \citenamefont
  {Kaupp}, \citenamefont {Koster}, \citenamefont {Kronik}, \citenamefont
  {Krylov}, \citenamefont {Kvaal}, \citenamefont {Laestadius}, \citenamefont
  {Levy}, \citenamefont {Lewin}, \citenamefont {Liu}, \citenamefont {Loos},
  \citenamefont {Maitra}, \citenamefont {Neese}, \citenamefont {Perdew},
  \citenamefont {Pernal}, \citenamefont {Pernot}, \citenamefont {Piecuch},
  \citenamefont {Rebolini}, \citenamefont {Reining}, \citenamefont
  {Romaniello}, \citenamefont {Ruzsinszky}, \citenamefont {Salahub},
  \citenamefont {Scheffler}, \citenamefont {Schwerdtfeger}, \citenamefont
  {Staroverov}, \citenamefont {Sun}, \citenamefont {Tellgren}, \citenamefont
  {Tozer}, \citenamefont {Trickey}, \citenamefont {Ullrich}, \citenamefont
  {Vela}, \citenamefont {Vignale}, \citenamefont {Wesolowski}, \citenamefont
  {Xu},\ and\ \citenamefont {Yang}}]{DFTexchange}%
  \BibitemOpen
  \bibfield  {author} {\bibinfo {author} {\bibfnamefont {A.~M.}\ \bibnamefont
  {Teale}}, \bibinfo {author} {\bibfnamefont {T.}~\bibnamefont {Helgaker}},
  \bibinfo {author} {\bibfnamefont {A.}~\bibnamefont {Savin}}, \bibinfo
  {author} {\bibfnamefont {C.}~\bibnamefont {Adamo}}, \bibinfo {author}
  {\bibfnamefont {B.}~\bibnamefont {Aradi}}, \bibinfo {author} {\bibfnamefont
  {A.~V.}\ \bibnamefont {Arbuznikov}}, \bibinfo {author} {\bibfnamefont
  {P.~W.}\ \bibnamefont {Ayers}}, \bibinfo {author} {\bibfnamefont {E.~J.}\
  \bibnamefont {Baerends}}, \bibinfo {author} {\bibfnamefont {V.}~\bibnamefont
  {Barone}}, \bibinfo {author} {\bibfnamefont {P.}~\bibnamefont {Calaminici}},
  \bibinfo {author} {\bibfnamefont {E.}~\bibnamefont {Cances}}, \bibinfo
  {author} {\bibfnamefont {E.~A.}\ \bibnamefont {Carter}}, \bibinfo {author}
  {\bibfnamefont {P.~K.}\ \bibnamefont {Chattaraj}}, \bibinfo {author}
  {\bibfnamefont {H.}~\bibnamefont {Chermette}}, \bibinfo {author}
  {\bibfnamefont {I.}~\bibnamefont {Ciofini}}, \bibinfo {author} {\bibfnamefont
  {T.~D.}\ \bibnamefont {Crawford}}, \bibinfo {author} {\bibfnamefont
  {F.}~\bibnamefont {De~Proft}}, \bibinfo {author} {\bibfnamefont {J.~F.}\
  \bibnamefont {Dobson}}, \bibinfo {author} {\bibfnamefont {C.}~\bibnamefont
  {Draxl}}, \bibinfo {author} {\bibfnamefont {T.}~\bibnamefont {Frauenheim}},
  \bibinfo {author} {\bibfnamefont {E.}~\bibnamefont {Fromager}}, \bibinfo
  {author} {\bibfnamefont {P.}~\bibnamefont {Fuentealba}}, \bibinfo {author}
  {\bibfnamefont {L.}~\bibnamefont {Gagliardi}}, \bibinfo {author}
  {\bibfnamefont {G.}~\bibnamefont {Galli}}, \bibinfo {author} {\bibfnamefont
  {J.}~\bibnamefont {Gao}}, \bibinfo {author} {\bibfnamefont {P.}~\bibnamefont
  {Geerlings}}, \bibinfo {author} {\bibfnamefont {N.}~\bibnamefont
  {Gidopoulos}}, \bibinfo {author} {\bibfnamefont {P.~M.~W.}\ \bibnamefont
  {Gill}}, \bibinfo {author} {\bibfnamefont {P.}~\bibnamefont {Gori-Giorgi}},
  \bibinfo {author} {\bibfnamefont {A.}~\bibnamefont {Gorling}}, \bibinfo
  {author} {\bibfnamefont {T.}~\bibnamefont {Gould}}, \bibinfo {author}
  {\bibfnamefont {S.}~\bibnamefont {Grimme}}, \bibinfo {author} {\bibfnamefont
  {O.}~\bibnamefont {Gritsenko}}, \bibinfo {author} {\bibfnamefont {H.~J.~A.}\
  \bibnamefont {Jensen}}, \bibinfo {author} {\bibfnamefont {E.~R.}\
  \bibnamefont {Johnson}}, \bibinfo {author} {\bibfnamefont {R.~O.}\
  \bibnamefont {Jones}}, \bibinfo {author} {\bibfnamefont {M.}~\bibnamefont
  {Kaupp}}, \bibinfo {author} {\bibfnamefont {A.~M.}\ \bibnamefont {Koster}},
  \bibinfo {author} {\bibfnamefont {L.}~\bibnamefont {Kronik}}, \bibinfo
  {author} {\bibfnamefont {A.}~\bibnamefont {Krylov}, \bibfnamefont {I}},
  \bibinfo {author} {\bibfnamefont {S.}~\bibnamefont {Kvaal}}, \bibinfo
  {author} {\bibfnamefont {A.}~\bibnamefont {Laestadius}}, \bibinfo {author}
  {\bibfnamefont {M.}~\bibnamefont {Levy}}, \bibinfo {author} {\bibfnamefont
  {M.}~\bibnamefont {Lewin}}, \bibinfo {author} {\bibfnamefont
  {S.}~\bibnamefont {Liu}}, \bibinfo {author} {\bibfnamefont {P.-F.}\
  \bibnamefont {Loos}}, \bibinfo {author} {\bibfnamefont {N.~T.}\ \bibnamefont
  {Maitra}}, \bibinfo {author} {\bibfnamefont {F.}~\bibnamefont {Neese}},
  \bibinfo {author} {\bibfnamefont {J.~P.}\ \bibnamefont {Perdew}}, \bibinfo
  {author} {\bibfnamefont {K.}~\bibnamefont {Pernal}}, \bibinfo {author}
  {\bibfnamefont {P.}~\bibnamefont {Pernot}}, \bibinfo {author} {\bibfnamefont
  {P.}~\bibnamefont {Piecuch}}, \bibinfo {author} {\bibfnamefont
  {E.}~\bibnamefont {Rebolini}}, \bibinfo {author} {\bibfnamefont
  {L.}~\bibnamefont {Reining}}, \bibinfo {author} {\bibfnamefont
  {P.}~\bibnamefont {Romaniello}}, \bibinfo {author} {\bibfnamefont
  {A.}~\bibnamefont {Ruzsinszky}}, \bibinfo {author} {\bibfnamefont {D.~R.}\
  \bibnamefont {Salahub}}, \bibinfo {author} {\bibfnamefont {M.}~\bibnamefont
  {Scheffler}}, \bibinfo {author} {\bibfnamefont {P.}~\bibnamefont
  {Schwerdtfeger}}, \bibinfo {author} {\bibfnamefont {V.~N.}\ \bibnamefont
  {Staroverov}}, \bibinfo {author} {\bibfnamefont {J.}~\bibnamefont {Sun}},
  \bibinfo {author} {\bibfnamefont {E.}~\bibnamefont {Tellgren}}, \bibinfo
  {author} {\bibfnamefont {D.~J.}\ \bibnamefont {Tozer}}, \bibinfo {author}
  {\bibfnamefont {S.~B.}\ \bibnamefont {Trickey}}, \bibinfo {author}
  {\bibfnamefont {C.~A.}\ \bibnamefont {Ullrich}}, \bibinfo {author}
  {\bibfnamefont {A.}~\bibnamefont {Vela}}, \bibinfo {author} {\bibfnamefont
  {G.}~\bibnamefont {Vignale}}, \bibinfo {author} {\bibfnamefont {T.~A.}\
  \bibnamefont {Wesolowski}}, \bibinfo {author} {\bibfnamefont
  {X.}~\bibnamefont {Xu}},\ and\ \bibinfo {author} {\bibfnamefont
  {W.}~\bibnamefont {Yang}},\ }\bibfield  {title} {\enquote {\bibinfo {title}
  {{DFT} exchange: {S}haring perspectives on the workhorse of quantum chemistry
  and materials science},}\ }\href {https://doi.org/10.1039/d2cp02827a}
  {\bibfield  {journal} {\bibinfo  {journal} {Phys. Chem. Chem. Phys.}\
  }\textbf {\bibinfo {volume} {24}},\ \bibinfo {pages} {28700} (\bibinfo {year}
  {2022})}\BibitemShut {NoStop}%
\bibitem [{\citenamefont {Huang}\ \emph {et~al.}(2018)\citenamefont {Huang},
  \citenamefont {Zheng}, \citenamefont {Song}, \citenamefont {Chi},
  \citenamefont {Wee},\ and\ \citenamefont {Quek}}]{HZS18}%
  \BibitemOpen
  \bibfield  {author} {\bibinfo {author} {\bibfnamefont {Y.~L.}\ \bibnamefont
  {Huang}}, \bibinfo {author} {\bibfnamefont {Y.~J.}\ \bibnamefont {Zheng}},
  \bibinfo {author} {\bibfnamefont {Z.}~\bibnamefont {Song}}, \bibinfo {author}
  {\bibfnamefont {D.}~\bibnamefont {Chi}}, \bibinfo {author} {\bibfnamefont
  {A.~T.~S.}\ \bibnamefont {Wee}},\ and\ \bibinfo {author} {\bibfnamefont
  {S.~Y.}\ \bibnamefont {Quek}},\ }\bibfield  {title} {\enquote {\bibinfo
  {title} {The organic-2{D} transition metal dichalcogenide heterointerface},}\
  }\href@noop {} {\bibfield  {journal} {\bibinfo  {journal} {Chem. Soc.}\
  }\textbf {\bibinfo {volume} {47}},\ \bibinfo {pages} {3241} (\bibinfo {year}
  {2018})}\BibitemShut {NoStop}%
\bibitem [{\citenamefont {Xie}\ \emph {et~al.}(2020)\citenamefont {Xie},
  \citenamefont {Niu}, \citenamefont {Kim}, \citenamefont {Li},\ and\
  \citenamefont {Yang}}]{XNK20}%
  \BibitemOpen
  \bibfield  {author} {\bibinfo {author} {\bibfnamefont {C.}~\bibnamefont
  {Xie}}, \bibinfo {author} {\bibfnamefont {Z.}~\bibnamefont {Niu}}, \bibinfo
  {author} {\bibfnamefont {D.}~\bibnamefont {Kim}}, \bibinfo {author}
  {\bibfnamefont {M.}~\bibnamefont {Li}},\ and\ \bibinfo {author}
  {\bibfnamefont {P.}~\bibnamefont {Yang}},\ }\bibfield  {title} {\enquote
  {\bibinfo {title} {Surface and interface control in nanoparticle
  catalysis},}\ }\href {https://doi.org/10.1021/acs.chemrev.9b00220} {\bibfield
   {journal} {\bibinfo  {journal} {Chem. Rev.}\ }\textbf {\bibinfo {volume}
  {120}},\ \bibinfo {pages} {1184} (\bibinfo {year} {2020})}\BibitemShut
  {NoStop}%
\bibitem [{\citenamefont {Chen}\ \emph {et~al.}(2020)\citenamefont {Chen},
  \citenamefont {Zhang}, \citenamefont {Li}, \citenamefont {He},\ and\
  \citenamefont {Guo}}]{CZL20}%
  \BibitemOpen
  \bibfield  {author} {\bibinfo {author} {\bibfnamefont {H.}~\bibnamefont
  {Chen}}, \bibinfo {author} {\bibfnamefont {W.}~\bibnamefont {Zhang}},
  \bibinfo {author} {\bibfnamefont {M.}~\bibnamefont {Li}}, \bibinfo {author}
  {\bibfnamefont {G.}~\bibnamefont {He}},\ and\ \bibinfo {author}
  {\bibfnamefont {X.}~\bibnamefont {Guo}},\ }\bibfield  {title} {\enquote
  {\bibinfo {title} {Interface engineering in organic field-effect transistors:
  Principles, applications, and perspectives},}\ }\href
  {https://doi.org/10.1021/acs.chemrev.9b00532} {\bibfield  {journal} {\bibinfo
   {journal} {Chem. Rev.}\ }\textbf {\bibinfo {volume} {120}},\ \bibinfo
  {pages} {2879} (\bibinfo {year} {2020})}\BibitemShut {NoStop}%
\bibitem [{\citenamefont {Gonella}\ \emph {et~al.}(2021)\citenamefont
  {Gonella}, \citenamefont {Backus}, \citenamefont {Nagata}, \citenamefont
  {Bonthuis}, \citenamefont {Loche}, \citenamefont {Schlaich}, \citenamefont
  {Netz}, \citenamefont {Kuhnle}, \citenamefont {McCrum}, \citenamefont
  {Koper}, \citenamefont {Wolf}, \citenamefont {Winter}, \citenamefont
  {Meijer}, \citenamefont {Campen},\ and\ \citenamefont {Bonn}}]{GBN21}%
  \BibitemOpen
  \bibfield  {author} {\bibinfo {author} {\bibfnamefont {G.}~\bibnamefont
  {Gonella}}, \bibinfo {author} {\bibfnamefont {E.~H.~G.}\ \bibnamefont
  {Backus}}, \bibinfo {author} {\bibfnamefont {Y.}~\bibnamefont {Nagata}},
  \bibinfo {author} {\bibfnamefont {D.~J.}\ \bibnamefont {Bonthuis}}, \bibinfo
  {author} {\bibfnamefont {P.}~\bibnamefont {Loche}}, \bibinfo {author}
  {\bibfnamefont {A.}~\bibnamefont {Schlaich}}, \bibinfo {author}
  {\bibfnamefont {R.~R.}\ \bibnamefont {Netz}}, \bibinfo {author}
  {\bibfnamefont {A.}~\bibnamefont {Kuhnle}}, \bibinfo {author} {\bibfnamefont
  {I.~T.}\ \bibnamefont {McCrum}}, \bibinfo {author} {\bibfnamefont {M.~T.~M.}\
  \bibnamefont {Koper}}, \bibinfo {author} {\bibfnamefont {M.}~\bibnamefont
  {Wolf}}, \bibinfo {author} {\bibfnamefont {B.}~\bibnamefont {Winter}},
  \bibinfo {author} {\bibfnamefont {G.}~\bibnamefont {Meijer}}, \bibinfo
  {author} {\bibfnamefont {R.~K.}\ \bibnamefont {Campen}},\ and\ \bibinfo
  {author} {\bibfnamefont {M.}~\bibnamefont {Bonn}},\ }\bibfield  {title}
  {\enquote {\bibinfo {title} {Water at charged interfaces},}\ }\href
  {https://doi.org/10.1038/s41570-021-00293-2} {\bibfield  {journal} {\bibinfo
  {journal} {Nat. Rev. Chem.}\ }\textbf {\bibinfo {volume} {5}},\ \bibinfo
  {pages} {466} (\bibinfo {year} {2021})}\BibitemShut {NoStop}%
\bibitem [{\citenamefont {Lang}\ and\ \citenamefont {Kohn}(1970)}]{LK70}%
  \BibitemOpen
  \bibfield  {author} {\bibinfo {author} {\bibfnamefont {N.~D.}\ \bibnamefont
  {Lang}}\ and\ \bibinfo {author} {\bibfnamefont {W.}~\bibnamefont {Kohn}},\
  }\bibfield  {title} {\enquote {\bibinfo {title} {Theory of metal surfaces:
  Charge density and surface energy},}\ }\href@noop {} {\bibfield  {journal}
  {\bibinfo  {journal} {Phys. Rev. B}\ }\textbf {\bibinfo {volume} {1}},\
  \bibinfo {pages} {4555} (\bibinfo {year} {1970})}\BibitemShut {NoStop}%
\bibitem [{\citenamefont {Lang}\ and\ \citenamefont {Kohn}(1971)}]{LK71}%
  \BibitemOpen
  \bibfield  {author} {\bibinfo {author} {\bibfnamefont {N.~D.}\ \bibnamefont
  {Lang}}\ and\ \bibinfo {author} {\bibfnamefont {W.}~\bibnamefont {Kohn}},\
  }\bibfield  {title} {\enquote {\bibinfo {title} {Theory of metal surfaces:
  Work function},}\ }\href@noop {} {\bibfield  {journal} {\bibinfo  {journal}
  {Phys. Rev. B}\ }\textbf {\bibinfo {volume} {3}},\ \bibinfo {pages} {1215}
  (\bibinfo {year} {1971})}\BibitemShut {NoStop}%
\bibitem [{\citenamefont {Lang}\ and\ \citenamefont {Kohn}(1973)}]{LK73}%
  \BibitemOpen
  \bibfield  {author} {\bibinfo {author} {\bibfnamefont {N.~D.}\ \bibnamefont
  {Lang}}\ and\ \bibinfo {author} {\bibfnamefont {W.}~\bibnamefont {Kohn}},\
  }\bibfield  {title} {\enquote {\bibinfo {title} {Theory of metal surfaces:
  Induced surface charge and image potential},}\ }\href@noop {} {\bibfield
  {journal} {\bibinfo  {journal} {Phys. Rev. B}\ }\textbf {\bibinfo {volume}
  {7}},\ \bibinfo {pages} {3541} (\bibinfo {year} {1973})}\BibitemShut
  {NoStop}%
\bibitem [{\citenamefont {Appelbaum}\ and\ \citenamefont
  {Hamann}(1972)}]{AH72}%
  \BibitemOpen
  \bibfield  {author} {\bibinfo {author} {\bibfnamefont {J.~A.}\ \bibnamefont
  {Appelbaum}}\ and\ \bibinfo {author} {\bibfnamefont {D.~R.}\ \bibnamefont
  {Hamann}},\ }\bibfield  {title} {\enquote {\bibinfo {title} {Variational
  calculation of the image potential near a metal surface},}\ }\href@noop {}
  {\bibfield  {journal} {\bibinfo  {journal} {Phys. Rev. B}\ }\textbf {\bibinfo
  {volume} {6}},\ \bibinfo {pages} {1122} (\bibinfo {year} {1972})}\BibitemShut
  {NoStop}%
\bibitem [{\citenamefont {Inkson}(1971{\natexlab{a}})}]{I71a}%
  \BibitemOpen
  \bibfield  {author} {\bibinfo {author} {\bibfnamefont {J.~C.}\ \bibnamefont
  {Inkson}},\ }\bibfield  {title} {\enquote {\bibinfo {title} {The
  electrostatic image potential in metal semiconductor junctions},}\
  }\href@noop {} {\bibfield  {journal} {\bibinfo  {journal} {J. Phys. C: Solid
  State Phys.}\ }\textbf {\bibinfo {volume} {4}},\ \bibinfo {pages} {591}
  (\bibinfo {year} {1971}{\natexlab{a}})}\BibitemShut {NoStop}%
\bibitem [{\citenamefont {Inkson}(1971{\natexlab{b}})}]{I71b}%
  \BibitemOpen
  \bibfield  {author} {\bibinfo {author} {\bibfnamefont {J.~C.}\ \bibnamefont
  {Inkson}},\ }\bibfield  {title} {\enquote {\bibinfo {title} {The
  electron-electron interaction near an interface},}\ }\href@noop {} {\bibfield
   {journal} {\bibinfo  {journal} {Surf. Sci.}\ }\textbf {\bibinfo {volume}
  {28}},\ \bibinfo {pages} {69} (\bibinfo {year}
  {1971}{\natexlab{b}})}\BibitemShut {NoStop}%
\bibitem [{\citenamefont {Inkson}(1972)}]{I72}%
  \BibitemOpen
  \bibfield  {author} {\bibinfo {author} {\bibfnamefont {J.~C.}\ \bibnamefont
  {Inkson}},\ }\bibfield  {title} {\enquote {\bibinfo {title} {Many-body
  effects at metal-semiconductor junctions: I. {S}urface plasmons and the
  electron-electron screened interaction},}\ }\href@noop {} {\bibfield
  {journal} {\bibinfo  {journal} {J. Phys. C: Solid State Phys.}\ }\textbf
  {\bibinfo {volume} {5}},\ \bibinfo {pages} {2599} (\bibinfo {year}
  {1972})}\BibitemShut {NoStop}%
\bibitem [{\citenamefont {Inkson}(1973)}]{I73}%
  \BibitemOpen
  \bibfield  {author} {\bibinfo {author} {\bibfnamefont {J.~C.}\ \bibnamefont
  {Inkson}},\ }\bibfield  {title} {\enquote {\bibinfo {title} {Many-body
  effects at metal-semiconductor junctions: {II}. {T}he self energy and band
  structure distortion},}\ }\href@noop {} {\bibfield  {journal} {\bibinfo
  {journal} {J. Phys. C: Solid State Phys.}\ }\textbf {\bibinfo {volume} {6}},\
  \bibinfo {pages} {1350} (\bibinfo {year} {1973})}\BibitemShut {NoStop}%
\bibitem [{\citenamefont {Newns}(1969)}]{N69}%
  \BibitemOpen
  \bibfield  {author} {\bibinfo {author} {\bibfnamefont {D.~M.}\ \bibnamefont
  {Newns}},\ }\bibfield  {title} {\enquote {\bibinfo {title} {Fermi-{T}homas
  response of a metal surface to an external point charge},}\ }\href@noop {}
  {\bibfield  {journal} {\bibinfo  {journal} {J. Chem. Phys.}\ }\textbf
  {\bibinfo {volume} {50}},\ \bibinfo {pages} {4572} (\bibinfo {year}
  {1969})}\BibitemShut {NoStop}%
\bibitem [{\citenamefont {Gies}\ and\ \citenamefont {Gerhardts}(1985)}]{GG85}%
  \BibitemOpen
  \bibfield  {author} {\bibinfo {author} {\bibfnamefont {P.}~\bibnamefont
  {Gies}}\ and\ \bibinfo {author} {\bibfnamefont {R.~R.}\ \bibnamefont
  {Gerhardts}},\ }\bibfield  {title} {\enquote {\bibinfo {title}
  {Self-consistent calculation of the electron distribution at a jellium
  surface in a strong static electric field},}\ }\href@noop {} {\bibfield
  {journal} {\bibinfo  {journal} {Phys. Rev. B}\ }\textbf {\bibinfo {volume}
  {31}},\ \bibinfo {pages} {6843(R)} (\bibinfo {year} {1985})}\BibitemShut
  {NoStop}%
\bibitem [{\citenamefont {Gies}\ and\ \citenamefont {Gerhardts}(1986)}]{GG86}%
  \BibitemOpen
  \bibfield  {author} {\bibinfo {author} {\bibfnamefont {P.}~\bibnamefont
  {Gies}}\ and\ \bibinfo {author} {\bibfnamefont {R.~R.}\ \bibnamefont
  {Gerhardts}},\ }\bibfield  {title} {\enquote {\bibinfo {title}
  {Self-consistent calculation of electron-density profiles at strongly charged
  jellium surfaces},}\ }\href@noop {} {\bibfield  {journal} {\bibinfo
  {journal} {Phys. Rev. B}\ }\textbf {\bibinfo {volume} {33}},\ \bibinfo
  {pages} {982} (\bibinfo {year} {1986})}\BibitemShut {NoStop}%
\bibitem [{\citenamefont {Schreier}\ and\ \citenamefont
  {Rebentrost}(1987)}]{SR87}%
  \BibitemOpen
  \bibfield  {author} {\bibinfo {author} {\bibfnamefont {F.}~\bibnamefont
  {Schreier}}\ and\ \bibinfo {author} {\bibfnamefont {F.}~\bibnamefont
  {Rebentrost}},\ }\bibfield  {title} {\enquote {\bibinfo {title}
  {Self-consistent electron densities of a semi-infinite jellium metal surface
  in strong static electrical fields},}\ }\href@noop {} {\bibfield  {journal}
  {\bibinfo  {journal} {J. Phys. C: Solid State Phys.}\ }\textbf {\bibinfo
  {volume} {20}},\ \bibinfo {pages} {2609} (\bibinfo {year}
  {1987})}\BibitemShut {NoStop}%
\bibitem [{\citenamefont {Weber}\ and\ \citenamefont
  {Liebsch}(1987{\natexlab{a}})}]{WL87a}%
  \BibitemOpen
  \bibfield  {author} {\bibinfo {author} {\bibfnamefont {M.}~\bibnamefont
  {Weber}}\ and\ \bibinfo {author} {\bibfnamefont {A.}~\bibnamefont
  {Liebsch}},\ }\bibfield  {title} {\enquote {\bibinfo {title}
  {Density-functional approach to second-harmonic generation at metal
  surfaces},}\ }\href@noop {} {\bibfield  {journal} {\bibinfo  {journal} {Phys.
  Rev. B}\ }\textbf {\bibinfo {volume} {35}},\ \bibinfo {pages} {7411}
  (\bibinfo {year} {1987}{\natexlab{a}})}\BibitemShut {NoStop}%
\bibitem [{\citenamefont {Weber}\ and\ \citenamefont
  {Liebsch}(1987{\natexlab{b}})}]{WL87b}%
  \BibitemOpen
  \bibfield  {author} {\bibinfo {author} {\bibfnamefont {M.~G.}\ \bibnamefont
  {Weber}}\ and\ \bibinfo {author} {\bibfnamefont {A.}~\bibnamefont
  {Liebsch}},\ }\bibfield  {title} {\enquote {\bibinfo {title} {Theory of
  second-harmonic generation by metal overlayers},}\ }\href@noop {} {\bibfield
  {journal} {\bibinfo  {journal} {Phys. Rev. B}\ }\textbf {\bibinfo {volume}
  {36}},\ \bibinfo {pages} {6411} (\bibinfo {year}
  {1987}{\natexlab{b}})}\BibitemShut {NoStop}%
\bibitem [{\citenamefont {Serena}, \citenamefont {Soler},\ and\ \citenamefont
  {Garc\'{i}a}(1988)}]{SSG88}%
  \BibitemOpen
  \bibfield  {author} {\bibinfo {author} {\bibfnamefont {P.~A.}\ \bibnamefont
  {Serena}}, \bibinfo {author} {\bibfnamefont {J.~M.}\ \bibnamefont {Soler}},\
  and\ \bibinfo {author} {\bibfnamefont {N.}~\bibnamefont {Garc\'{i}a}},\
  }\bibfield  {title} {\enquote {\bibinfo {title} {Work function and
  image-plane position of metal surfaces},}\ }\href@noop {} {\bibfield
  {journal} {\bibinfo  {journal} {Phys. Rev. B}\ }\textbf {\bibinfo {volume}
  {37}},\ \bibinfo {pages} {8701} (\bibinfo {year} {1988})}\BibitemShut
  {NoStop}%
\bibitem [{\citenamefont {Ho}, \citenamefont {Harmon},\ and\ \citenamefont
  {Liu}(1980)}]{HHL80}%
  \BibitemOpen
  \bibfield  {author} {\bibinfo {author} {\bibfnamefont {K.-M.}\ \bibnamefont
  {Ho}}, \bibinfo {author} {\bibfnamefont {B.~N.}\ \bibnamefont {Harmon}},\
  and\ \bibinfo {author} {\bibfnamefont {S.~H.}\ \bibnamefont {Liu}},\
  }\bibfield  {title} {\enquote {\bibinfo {title} {Surface-state contribution
  to the electroreflectance of noble metals},}\ }\href@noop {} {\bibfield
  {journal} {\bibinfo  {journal} {Phys. Rev. Lett.}\ }\textbf {\bibinfo
  {volume} {44}},\ \bibinfo {pages} {1531} (\bibinfo {year}
  {1980})}\BibitemShut {NoStop}%
\bibitem [{\citenamefont {Kolb}\ \emph {et~al.}(1981)\citenamefont {Kolb},
  \citenamefont {Boeck}, \citenamefont {Ho},\ and\ \citenamefont
  {Liu}}]{KBHL81}%
  \BibitemOpen
  \bibfield  {author} {\bibinfo {author} {\bibfnamefont {D.~M.}\ \bibnamefont
  {Kolb}}, \bibinfo {author} {\bibfnamefont {W.}~\bibnamefont {Boeck}},
  \bibinfo {author} {\bibfnamefont {K.-M.}\ \bibnamefont {Ho}},\ and\ \bibinfo
  {author} {\bibfnamefont {S.~H.}\ \bibnamefont {Liu}},\ }\bibfield  {title}
  {\enquote {\bibinfo {title} {Observation of surface states on {A}g(100) by
  infrared and visible electroreflectance spectroscopy},}\ }\href@noop {}
  {\bibfield  {journal} {\bibinfo  {journal} {Phys. Rev. Lett.}\ }\textbf
  {\bibinfo {volume} {47}},\ \bibinfo {pages} {1921} (\bibinfo {year}
  {1981})}\BibitemShut {NoStop}%
\bibitem [{\citenamefont {Fu}\ and\ \citenamefont {Ho}(1989)}]{FH89}%
  \BibitemOpen
  \bibfield  {author} {\bibinfo {author} {\bibfnamefont {C.~L.}\ \bibnamefont
  {Fu}}\ and\ \bibinfo {author} {\bibfnamefont {K.~M.}\ \bibnamefont {Ho}},\
  }\bibfield  {title} {\enquote {\bibinfo {title} {External-charge-induced
  surface reconstruction on {A}g(110)},}\ }\href@noop {} {\bibfield  {journal}
  {\bibinfo  {journal} {Phys. Rev. Lett.}\ }\textbf {\bibinfo {volume} {63}},\
  \bibinfo {pages} {1617} (\bibinfo {year} {1989})}\BibitemShut {NoStop}%
\bibitem [{\citenamefont {Aers}\ and\ \citenamefont
  {Inglesfield}(1989)}]{AI89}%
  \BibitemOpen
  \bibfield  {author} {\bibinfo {author} {\bibfnamefont {G.~C.}\ \bibnamefont
  {Aers}}\ and\ \bibinfo {author} {\bibfnamefont {J.~E.}\ \bibnamefont
  {Inglesfield}},\ }\bibfield  {title} {\enquote {\bibinfo {title} {Electric
  field and {A}g(001) surface electronic structure},}\ }\href@noop {}
  {\bibfield  {journal} {\bibinfo  {journal} {Surf. Sci.}\ }\textbf {\bibinfo
  {volume} {217}},\ \bibinfo {pages} {367} (\bibinfo {year}
  {1989})}\BibitemShut {NoStop}%
\bibitem [{\citenamefont {Inglesfield}(1989)}]{I89}%
  \BibitemOpen
  \bibfield  {author} {\bibinfo {author} {\bibfnamefont {J.~E.}\ \bibnamefont
  {Inglesfield}},\ }\bibfield  {title} {\enquote {\bibinfo {title} {The
  screening of an electric field at an {A}l(001) surface},}\ }\href@noop {}
  {\bibfield  {journal} {\bibinfo  {journal} {Surf. Sci.}\ }\textbf {\bibinfo
  {volume} {188}},\ \bibinfo {pages} {L701} (\bibinfo {year}
  {1989})}\BibitemShut {NoStop}%
\bibitem [{\citenamefont {Finnis}(1991)}]{F91}%
  \BibitemOpen
  \bibfield  {author} {\bibinfo {author} {\bibfnamefont {M.~W.}\ \bibnamefont
  {Finnis}},\ }\bibfield  {title} {\enquote {\bibinfo {title} {The interaction
  of a point charge with an aluminium (111) surface},}\ }\href@noop {}
  {\bibfield  {journal} {\bibinfo  {journal} {Surf. Sci.}\ }\textbf {\bibinfo
  {volume} {241}},\ \bibinfo {pages} {61} (\bibinfo {year} {1991})}\BibitemShut
  {NoStop}%
\bibitem [{\citenamefont {Kreuzer}, \citenamefont {Wang},\ and\ \citenamefont
  {Lang}(1992)}]{KWL92}%
  \BibitemOpen
  \bibfield  {author} {\bibinfo {author} {\bibfnamefont {H.~J.}\ \bibnamefont
  {Kreuzer}}, \bibinfo {author} {\bibfnamefont {L.~C.}\ \bibnamefont {Wang}},\
  and\ \bibinfo {author} {\bibfnamefont {N.~D.}\ \bibnamefont {Lang}},\
  }\bibfield  {title} {\enquote {\bibinfo {title} {Self-consistent calculation
  of atomic adsorption on metals in high electric fields},}\ }\href@noop {}
  {\bibfield  {journal} {\bibinfo  {journal} {Phys. Rev. B}\ }\textbf {\bibinfo
  {volume} {45}},\ \bibinfo {pages} {12050} (\bibinfo {year}
  {1992})}\BibitemShut {NoStop}%
\bibitem [{\citenamefont {Lam}\ and\ \citenamefont {Needs}(1992)}]{LN92}%
  \BibitemOpen
  \bibfield  {author} {\bibinfo {author} {\bibfnamefont {S.~C.}\ \bibnamefont
  {Lam}}\ and\ \bibinfo {author} {\bibfnamefont {R.~J.}\ \bibnamefont
  {Needs}},\ }\bibfield  {title} {\enquote {\bibinfo {title} {Field-ion
  microscope tunnelling calculations for the aluminium (111) and (110)
  surfaces},}\ }\href@noop {} {\bibfield  {journal} {\bibinfo  {journal} {Surf.
  Sci.}\ }\textbf {\bibinfo {volume} {277}},\ \bibinfo {pages} {173} (\bibinfo
  {year} {1992})}\BibitemShut {NoStop}%
\bibitem [{\citenamefont {Lam}\ and\ \citenamefont {Needs}(1993)}]{LN93}%
  \BibitemOpen
  \bibfield  {author} {\bibinfo {author} {\bibfnamefont {S.~C.}\ \bibnamefont
  {Lam}}\ and\ \bibinfo {author} {\bibfnamefont {R.~J.}\ \bibnamefont
  {Needs}},\ }\bibfield  {title} {\enquote {\bibinfo {title} {First-principles
  calculations of the screening of electric fields at the aluminium(111) and
  (110) surfaces},}\ }\href@noop {} {\bibfield  {journal} {\bibinfo  {journal}
  {J. Phys.: Condens. Matter}\ }\textbf {\bibinfo {volume} {5}},\ \bibinfo
  {pages} {2101} (\bibinfo {year} {1993})}\BibitemShut {NoStop}%
\bibitem [{\citenamefont {Penn}(1962)}]{P62}%
  \BibitemOpen
  \bibfield  {author} {\bibinfo {author} {\bibfnamefont {D.~R.}\ \bibnamefont
  {Penn}},\ }\bibfield  {title} {\enquote {\bibinfo {title}
  {Wave-number-dependent dielectric function of semiconductors},}\ }\href@noop
  {} {\bibfield  {journal} {\bibinfo  {journal} {Phys. Rev.}\ }\textbf
  {\bibinfo {volume} {128}},\ \bibinfo {pages} {2093} (\bibinfo {year}
  {1962})}\BibitemShut {NoStop}%
\bibitem [{\citenamefont {Neaton}, \citenamefont {Hybertsen},\ and\
  \citenamefont {Louie}(2006)}]{NHL06}%
  \BibitemOpen
  \bibfield  {author} {\bibinfo {author} {\bibfnamefont {J.~B.}\ \bibnamefont
  {Neaton}}, \bibinfo {author} {\bibfnamefont {M.~S.}\ \bibnamefont
  {Hybertsen}},\ and\ \bibinfo {author} {\bibfnamefont {S.~G.}\ \bibnamefont
  {Louie}},\ }\bibfield  {title} {\enquote {\bibinfo {title} {Renormalization
  of molecular electronic levels at metal-molecule interfaces},}\ }\href
  {https://doi.org/10.1103/PhysRevLett.97.216405} {\bibfield  {journal}
  {\bibinfo  {journal} {Phys. Rev. Lett.}\ }\textbf {\bibinfo {volume} {97}},\
  \bibinfo {pages} {216405} (\bibinfo {year} {2006})}\BibitemShut {NoStop}%
\bibitem [{\citenamefont {Quek}\ \emph {et~al.}(2007)\citenamefont {Quek},
  \citenamefont {Venkataraman}, \citenamefont {Choi}, \citenamefont {Louie},
  \citenamefont {Hybertsen},\ and\ \citenamefont {Neaton}}]{QVCL07}%
  \BibitemOpen
  \bibfield  {author} {\bibinfo {author} {\bibfnamefont {S.~Y.}\ \bibnamefont
  {Quek}}, \bibinfo {author} {\bibfnamefont {L.}~\bibnamefont {Venkataraman}},
  \bibinfo {author} {\bibfnamefont {H.~J.}\ \bibnamefont {Choi}}, \bibinfo
  {author} {\bibfnamefont {S.~G.}\ \bibnamefont {Louie}}, \bibinfo {author}
  {\bibfnamefont {M.~S.}\ \bibnamefont {Hybertsen}},\ and\ \bibinfo {author}
  {\bibfnamefont {J.~B.}\ \bibnamefont {Neaton}},\ }\bibfield  {title}
  {\enquote {\bibinfo {title} {Amine-gold linked single-molecule circuits:
  {E}xperiment and theory},}\ }\href@noop {} {\bibfield  {journal} {\bibinfo
  {journal} {Nano Lett.}\ }\textbf {\bibinfo {volume} {7}},\ \bibinfo {pages}
  {3477} (\bibinfo {year} {2007})}\BibitemShut {NoStop}%
\bibitem [{\citenamefont {Liu}\ \emph {et~al.}(2014)\citenamefont {Liu},
  \citenamefont {Wei}, \citenamefont {Yoon}, \citenamefont {Adak},
  \citenamefont {Ponce}, \citenamefont {Jiang}, \citenamefont {Jang},
  \citenamefont {Campos}, \citenamefont {Venkataraman},\ and\ \citenamefont
  {Neaton}}]{LWYA14}%
  \BibitemOpen
  \bibfield  {author} {\bibinfo {author} {\bibfnamefont {Z.-F.}\ \bibnamefont
  {Liu}}, \bibinfo {author} {\bibfnamefont {S.}~\bibnamefont {Wei}}, \bibinfo
  {author} {\bibfnamefont {H.}~\bibnamefont {Yoon}}, \bibinfo {author}
  {\bibfnamefont {O.}~\bibnamefont {Adak}}, \bibinfo {author} {\bibfnamefont
  {I.}~\bibnamefont {Ponce}}, \bibinfo {author} {\bibfnamefont
  {Y.}~\bibnamefont {Jiang}}, \bibinfo {author} {\bibfnamefont {W.-D.}\
  \bibnamefont {Jang}}, \bibinfo {author} {\bibfnamefont {L.~M.}\ \bibnamefont
  {Campos}}, \bibinfo {author} {\bibfnamefont {L.}~\bibnamefont
  {Venkataraman}},\ and\ \bibinfo {author} {\bibfnamefont {J.~B.}\ \bibnamefont
  {Neaton}},\ }\bibfield  {title} {\enquote {\bibinfo {title} {Control of
  single-molecule junction conductance of porphyrins via a transition-metal
  center},}\ }\href@noop {} {\bibfield  {journal} {\bibinfo  {journal} {Nano
  Lett.}\ }\textbf {\bibinfo {volume} {14}},\ \bibinfo {pages} {5365} (\bibinfo
  {year} {2014})}\BibitemShut {NoStop}%
\bibitem [{\citenamefont {Li}, \citenamefont {Lu},\ and\ \citenamefont
  {Galli}(2009)}]{LLG09}%
  \BibitemOpen
  \bibfield  {author} {\bibinfo {author} {\bibfnamefont {Y.}~\bibnamefont
  {Li}}, \bibinfo {author} {\bibfnamefont {D.}~\bibnamefont {Lu}},\ and\
  \bibinfo {author} {\bibfnamefont {G.}~\bibnamefont {Galli}},\ }\bibfield
  {title} {\enquote {\bibinfo {title} {Calculation of quasi-particle energies
  of aromatic self-assembled monolayers on {A}u(111)},}\ }\href@noop {}
  {\bibfield  {journal} {\bibinfo  {journal} {J. Chem. Theory Comput.}\
  }\textbf {\bibinfo {volume} {5}},\ \bibinfo {pages} {881} (\bibinfo {year}
  {2009})}\BibitemShut {NoStop}%
\bibitem [{\citenamefont {Egger}\ \emph {et~al.}(2015)\citenamefont {Egger},
  \citenamefont {Liu}, \citenamefont {Neaton},\ and\ \citenamefont
  {Kronik}}]{ELNK15}%
  \BibitemOpen
  \bibfield  {author} {\bibinfo {author} {\bibfnamefont {D.~A.}\ \bibnamefont
  {Egger}}, \bibinfo {author} {\bibfnamefont {Z.-F.}\ \bibnamefont {Liu}},
  \bibinfo {author} {\bibfnamefont {J.~B.}\ \bibnamefont {Neaton}},\ and\
  \bibinfo {author} {\bibfnamefont {L.}~\bibnamefont {Kronik}},\ }\bibfield
  {title} {\enquote {\bibinfo {title} {Reliable energy level alignment at
  physisorbed molecule-metal interfaces from density functional theory},}\
  }\href@noop {} {\bibfield  {journal} {\bibinfo  {journal} {Nano Lett.}\
  }\textbf {\bibinfo {volume} {15}},\ \bibinfo {pages} {2448} (\bibinfo {year}
  {2015})}\BibitemShut {NoStop}%
\bibitem [{\citenamefont {Thygesen}\ and\ \citenamefont {Rubio}(2009)}]{TR09}%
  \BibitemOpen
  \bibfield  {author} {\bibinfo {author} {\bibfnamefont {K.~S.}\ \bibnamefont
  {Thygesen}}\ and\ \bibinfo {author} {\bibfnamefont {A.}~\bibnamefont
  {Rubio}},\ }\bibfield  {title} {\enquote {\bibinfo {title} {Renormalization
  of molecular quasiparticle levels at metal-molecule interfaces: Trends across
  binding regimes},}\ }\href@noop {} {\bibfield  {journal} {\bibinfo  {journal}
  {Phys. Rev. Lett.}\ }\textbf {\bibinfo {volume} {102}},\ \bibinfo {pages}
  {046802} (\bibinfo {year} {2009})}\BibitemShut {NoStop}%
\bibitem [{\citenamefont {Anderson}(1961)}]{A61}%
  \BibitemOpen
  \bibfield  {author} {\bibinfo {author} {\bibfnamefont {P.~W.}\ \bibnamefont
  {Anderson}},\ }\bibfield  {title} {\enquote {\bibinfo {title} {Localized
  magnetic states in metals},}\ }\href@noop {} {\bibfield  {journal} {\bibinfo
  {journal} {Phys. Rev.}\ }\textbf {\bibinfo {volume} {124}},\ \bibinfo {pages}
  {41} (\bibinfo {year} {1961})}\BibitemShut {NoStop}%
\bibitem [{\citenamefont {Garcia-Lastra}\ \emph {et~al.}(2009)\citenamefont
  {Garcia-Lastra}, \citenamefont {Rostgaard}, \citenamefont {Rubio},\ and\
  \citenamefont {Thygesen}}]{GRRT09}%
  \BibitemOpen
  \bibfield  {author} {\bibinfo {author} {\bibfnamefont {J.~M.}\ \bibnamefont
  {Garcia-Lastra}}, \bibinfo {author} {\bibfnamefont {C.}~\bibnamefont
  {Rostgaard}}, \bibinfo {author} {\bibfnamefont {A.}~\bibnamefont {Rubio}},\
  and\ \bibinfo {author} {\bibfnamefont {K.~S.}\ \bibnamefont {Thygesen}},\
  }\bibfield  {title} {\enquote {\bibinfo {title} {Polarization-induced
  renormalization of molecular levels at metallic and semiconducting
  surfaces},}\ }\href@noop {} {\bibfield  {journal} {\bibinfo  {journal} {Phys.
  Rev. B}\ }\textbf {\bibinfo {volume} {80}},\ \bibinfo {pages} {245427}
  (\bibinfo {year} {2009})}\BibitemShut {NoStop}%
\bibitem [{\citenamefont {Strange}\ and\ \citenamefont
  {Thygesen}(2012)}]{ST12}%
  \BibitemOpen
  \bibfield  {author} {\bibinfo {author} {\bibfnamefont {M.}~\bibnamefont
  {Strange}}\ and\ \bibinfo {author} {\bibfnamefont {K.~S.}\ \bibnamefont
  {Thygesen}},\ }\bibfield  {title} {\enquote {\bibinfo {title}
  {Image-charge-induced localization of molecular orbitals at metal-molecule
  interfaces: Self-consistent ${GW}$ calculations},}\ }\href
  {https://doi.org/10.1103/physrevb.86.195121} {\bibfield  {journal} {\bibinfo
  {journal} {Phys. Rev. B}\ }\textbf {\bibinfo {volume} {86}},\ \bibinfo
  {pages} {195121} (\bibinfo {year} {2012})}\BibitemShut {NoStop}%
\bibitem [{\citenamefont {Bardeen}(1936)}]{B36}%
  \BibitemOpen
  \bibfield  {author} {\bibinfo {author} {\bibfnamefont {J.}~\bibnamefont
  {Bardeen}},\ }\bibfield  {title} {\enquote {\bibinfo {title} {Theory of the
  work function. {II}. {T}he surface double layer},}\ }\href@noop {} {\bibfield
   {journal} {\bibinfo  {journal} {Phys. Rev.}\ }\textbf {\bibinfo {volume}
  {49}},\ \bibinfo {pages} {653} (\bibinfo {year} {1936})}\BibitemShut
  {NoStop}%
\bibitem [{\citenamefont {Juretschke}(1953)}]{J53}%
  \BibitemOpen
  \bibfield  {author} {\bibinfo {author} {\bibfnamefont {H.~J.}\ \bibnamefont
  {Juretschke}},\ }\bibfield  {title} {\enquote {\bibinfo {title} {Exchange
  potential in the surface region of a free-electron metal},}\ }\href@noop {}
  {\bibfield  {journal} {\bibinfo  {journal} {Phys. Rev.}\ }\textbf {\bibinfo
  {volume} {92}},\ \bibinfo {pages} {1140} (\bibinfo {year}
  {1953})}\BibitemShut {NoStop}%
\bibitem [{\citenamefont {Inglesfield}\ and\ \citenamefont
  {Moore}(1978)}]{IM78}%
  \BibitemOpen
  \bibfield  {author} {\bibinfo {author} {\bibfnamefont {J.~E.}\ \bibnamefont
  {Inglesfield}}\ and\ \bibinfo {author} {\bibfnamefont {I.~D.}\ \bibnamefont
  {Moore}},\ }\bibfield  {title} {\enquote {\bibinfo {title} {The
  exchange-correlation hole at a surface},}\ }\href@noop {} {\bibfield
  {journal} {\bibinfo  {journal} {Solid State Commun.}\ }\textbf {\bibinfo
  {volume} {26}},\ \bibinfo {pages} {867} (\bibinfo {year} {1978})}\BibitemShut
  {NoStop}%
\bibitem [{\citenamefont {Sahni}\ and\ \citenamefont {Bohnen}(1984)}]{SB84}%
  \BibitemOpen
  \bibfield  {author} {\bibinfo {author} {\bibfnamefont {V.}~\bibnamefont
  {Sahni}}\ and\ \bibinfo {author} {\bibfnamefont {K.-P.}\ \bibnamefont
  {Bohnen}},\ }\bibfield  {title} {\enquote {\bibinfo {title} {Exchange charge
  density at metallic surfaces},}\ }\href@noop {} {\bibfield  {journal}
  {\bibinfo  {journal} {Phys. Rev. B}\ }\textbf {\bibinfo {volume} {29}},\
  \bibinfo {pages} {1045} (\bibinfo {year} {1984})}\BibitemShut {NoStop}%
\bibitem [{\citenamefont {Sahni}\ and\ \citenamefont {Bohnen}(1985)}]{SB85}%
  \BibitemOpen
  \bibfield  {author} {\bibinfo {author} {\bibfnamefont {V.}~\bibnamefont
  {Sahni}}\ and\ \bibinfo {author} {\bibfnamefont {K.-P.}\ \bibnamefont
  {Bohnen}},\ }\bibfield  {title} {\enquote {\bibinfo {title} {Image charge at
  a metal surface},}\ }\href@noop {} {\bibfield  {journal} {\bibinfo  {journal}
  {Phys. Rev. B}\ }\textbf {\bibinfo {volume} {31}},\ \bibinfo {pages} {7651}
  (\bibinfo {year} {1985})}\BibitemShut {NoStop}%
\bibitem [{\citenamefont {Harbola}\ and\ \citenamefont {Sahni}(1988)}]{HS88}%
  \BibitemOpen
  \bibfield  {author} {\bibinfo {author} {\bibfnamefont {M.~K.}\ \bibnamefont
  {Harbola}}\ and\ \bibinfo {author} {\bibfnamefont {V.}~\bibnamefont
  {Sahni}},\ }\bibfield  {title} {\enquote {\bibinfo {title} {Structure of the
  {F}ermi hole at surfaces},}\ }\href@noop {} {\bibfield  {journal} {\bibinfo
  {journal} {Phys. Rev. B}\ }\textbf {\bibinfo {volume} {37}},\ \bibinfo
  {pages} {745} (\bibinfo {year} {1988})}\BibitemShut {NoStop}%
\bibitem [{\citenamefont {Serena}, \citenamefont {Soler},\ and\ \citenamefont
  {Garc\'{i}a}(1986)}]{SSG86}%
  \BibitemOpen
  \bibfield  {author} {\bibinfo {author} {\bibfnamefont {P.~A.}\ \bibnamefont
  {Serena}}, \bibinfo {author} {\bibfnamefont {J.~M.}\ \bibnamefont {Soler}},\
  and\ \bibinfo {author} {\bibfnamefont {N.}~\bibnamefont {Garc\'{i}a}},\
  }\bibfield  {title} {\enquote {\bibinfo {title} {Self-consistent image
  potential in a metal surface},}\ }\href@noop {} {\bibfield  {journal}
  {\bibinfo  {journal} {Phys. Rev. B}\ }\textbf {\bibinfo {volume} {34}},\
  \bibinfo {pages} {6767} (\bibinfo {year} {1986})}\BibitemShut {NoStop}%
\bibitem [{\citenamefont {Acioli}\ and\ \citenamefont {Ceperley}(1996)}]{AC96}%
  \BibitemOpen
  \bibfield  {author} {\bibinfo {author} {\bibfnamefont {P.~H.}\ \bibnamefont
  {Acioli}}\ and\ \bibinfo {author} {\bibfnamefont {D.~M.}\ \bibnamefont
  {Ceperley}},\ }\bibfield  {title} {\enquote {\bibinfo {title} {Diffusion
  {M}onte {C}arlo study of jellium surfaces: Electronic densities and pair
  correlation functions},}\ }\href@noop {} {\bibfield  {journal} {\bibinfo
  {journal} {Phys. Rev. B}\ }\textbf {\bibinfo {volume} {54}},\ \bibinfo
  {pages} {17199} (\bibinfo {year} {1996})}\BibitemShut {NoStop}%
\bibitem [{\citenamefont {Hsing}, \citenamefont {Chou},\ and\ \citenamefont
  {Lee}(2006)}]{HCL06}%
  \BibitemOpen
  \bibfield  {author} {\bibinfo {author} {\bibfnamefont {C.~R.}\ \bibnamefont
  {Hsing}}, \bibinfo {author} {\bibfnamefont {M.~Y.}\ \bibnamefont {Chou}},\
  and\ \bibinfo {author} {\bibfnamefont {T.~K.}\ \bibnamefont {Lee}},\
  }\bibfield  {title} {\enquote {\bibinfo {title} {Exchange-correlation energy
  in molecules: A variational quantum {M}onte {C}arlo study},}\ }\href@noop {}
  {\bibfield  {journal} {\bibinfo  {journal} {Phys. Rev. A}\ }\textbf {\bibinfo
  {volume} {74}},\ \bibinfo {pages} {032507} (\bibinfo {year}
  {2006})}\BibitemShut {NoStop}%
\bibitem [{\citenamefont {Hood}\ \emph {et~al.}(1998)\citenamefont {Hood},
  \citenamefont {Chou}, \citenamefont {Williamson}, \citenamefont {Rajagopal},\
  and\ \citenamefont {Needs}}]{HCWR98}%
  \BibitemOpen
  \bibfield  {author} {\bibinfo {author} {\bibfnamefont {R.~Q.}\ \bibnamefont
  {Hood}}, \bibinfo {author} {\bibfnamefont {M.~Y.}\ \bibnamefont {Chou}},
  \bibinfo {author} {\bibfnamefont {A.~J.}\ \bibnamefont {Williamson}},
  \bibinfo {author} {\bibfnamefont {G.}~\bibnamefont {Rajagopal}},\ and\
  \bibinfo {author} {\bibfnamefont {R.~J.}\ \bibnamefont {Needs}},\ }\bibfield
  {title} {\enquote {\bibinfo {title} {Exchange and correlation in silicon},}\
  }\href@noop {} {\bibfield  {journal} {\bibinfo  {journal} {Phys. Rev. B}\
  }\textbf {\bibinfo {volume} {57}},\ \bibinfo {pages} {8972} (\bibinfo {year}
  {1998})}\BibitemShut {NoStop}%
\bibitem [{\citenamefont {Rushton}, \citenamefont {Tozer},\ and\ \citenamefont
  {Clark}(2002)}]{RTC02}%
  \BibitemOpen
  \bibfield  {author} {\bibinfo {author} {\bibfnamefont {P.~P.}\ \bibnamefont
  {Rushton}}, \bibinfo {author} {\bibfnamefont {D.~J.}\ \bibnamefont {Tozer}},\
  and\ \bibinfo {author} {\bibfnamefont {S.~J.}\ \bibnamefont {Clark}},\
  }\bibfield  {title} {\enquote {\bibinfo {title} {Nonlocal density-functional
  description of exchange and correlation in silicon},}\ }\href@noop {}
  {\bibfield  {journal} {\bibinfo  {journal} {Phys. Rev. B}\ }\textbf {\bibinfo
  {volume} {65}},\ \bibinfo {pages} {235203} (\bibinfo {year}
  {2002})}\BibitemShut {NoStop}%
\bibitem [{\citenamefont {Garc\'{i}a-Gonz\'{a}lez}\ \emph
  {et~al.}(2000)\citenamefont {Garc\'{i}a-Gonz\'{a}lez}, \citenamefont
  {Alvarellos}, \citenamefont {Chac\'{o}n},\ and\ \citenamefont
  {Tarazona}}]{GACT00}%
  \BibitemOpen
  \bibfield  {author} {\bibinfo {author} {\bibfnamefont {P.}~\bibnamefont
  {Garc\'{i}a-Gonz\'{a}lez}}, \bibinfo {author} {\bibfnamefont {J.~E.}\
  \bibnamefont {Alvarellos}}, \bibinfo {author} {\bibfnamefont
  {E.}~\bibnamefont {Chac\'{o}n}},\ and\ \bibinfo {author} {\bibfnamefont
  {P.}~\bibnamefont {Tarazona}},\ }\bibfield  {title} {\enquote {\bibinfo
  {title} {Image potential and the exchange-correlation weighted density
  approximation functional},}\ }\href@noop {} {\bibfield  {journal} {\bibinfo
  {journal} {Phys. Rev. B}\ }\textbf {\bibinfo {volume} {62}},\ \bibinfo
  {pages} {16063} (\bibinfo {year} {2000})}\BibitemShut {NoStop}%
\bibitem [{\citenamefont {Gunnarsson}, \citenamefont {Jonson},\ and\
  \citenamefont {Lundqvist}(1979)}]{GJL79}%
  \BibitemOpen
  \bibfield  {author} {\bibinfo {author} {\bibfnamefont {O.}~\bibnamefont
  {Gunnarsson}}, \bibinfo {author} {\bibfnamefont {M.}~\bibnamefont {Jonson}},\
  and\ \bibinfo {author} {\bibfnamefont {B.~I.}\ \bibnamefont {Lundqvist}},\
  }\bibfield  {title} {\enquote {\bibinfo {title} {Descriptions of exchange and
  correlation effects in inhomogeneous electron systems},}\ }\href@noop {}
  {\bibfield  {journal} {\bibinfo  {journal} {Phys. Rev. B}\ }\textbf {\bibinfo
  {volume} {20}},\ \bibinfo {pages} {3136} (\bibinfo {year}
  {1979})}\BibitemShut {NoStop}%
\bibitem [{\citenamefont {Jochym}\ and\ \citenamefont {Clark}(2007)}]{JC07}%
  \BibitemOpen
  \bibfield  {author} {\bibinfo {author} {\bibfnamefont {D.~B.}\ \bibnamefont
  {Jochym}}\ and\ \bibinfo {author} {\bibfnamefont {S.~J.}\ \bibnamefont
  {Clark}},\ }\bibfield  {title} {\enquote {\bibinfo {title}
  {Exchange-correlation holes in metal surfaces using nonlocal
  density-functional theory},}\ }\href@noop {} {\bibfield  {journal} {\bibinfo
  {journal} {Phys. Rev. B}\ }\textbf {\bibinfo {volume} {76}},\ \bibinfo
  {pages} {075411} (\bibinfo {year} {2007})}\BibitemShut {NoStop}%
\bibitem [{\citenamefont {Nekovee}\ and\ \citenamefont {Pitarke}(2001)}]{NP01}%
  \BibitemOpen
  \bibfield  {author} {\bibinfo {author} {\bibfnamefont {M.}~\bibnamefont
  {Nekovee}}\ and\ \bibinfo {author} {\bibfnamefont {J.}~\bibnamefont
  {Pitarke}},\ }\bibfield  {title} {\enquote {\bibinfo {title} {Recent progress
  in the computational many-body theory of metal surfaces},}\ }\href@noop {}
  {\bibfield  {journal} {\bibinfo  {journal} {Comput. Phys. Commun.}\ }\textbf
  {\bibinfo {volume} {137}},\ \bibinfo {pages} {123} (\bibinfo {year}
  {2001})}\BibitemShut {NoStop}%
\bibitem [{\citenamefont {Constantin}\ and\ \citenamefont
  {Pitarke}(2009)}]{CP09}%
  \BibitemOpen
  \bibfield  {author} {\bibinfo {author} {\bibfnamefont {L.~A.}\ \bibnamefont
  {Constantin}}\ and\ \bibinfo {author} {\bibfnamefont {J.~M.}\ \bibnamefont
  {Pitarke}},\ }\bibfield  {title} {\enquote {\bibinfo {title} {The many-body
  exchange-correlation hole at metal surfaces},}\ }\href@noop {} {\bibfield
  {journal} {\bibinfo  {journal} {J. Chem. Theory Comput.}\ }\textbf {\bibinfo
  {volume} {5}},\ \bibinfo {pages} {895} (\bibinfo {year} {2009})}\BibitemShut
  {NoStop}%
\bibitem [{\citenamefont {Pitarke}\ and\ \citenamefont {Eguiluz}(1998)}]{PE98}%
  \BibitemOpen
  \bibfield  {author} {\bibinfo {author} {\bibfnamefont {J.~M.}\ \bibnamefont
  {Pitarke}}\ and\ \bibinfo {author} {\bibfnamefont {A.~G.}\ \bibnamefont
  {Eguiluz}},\ }\bibfield  {title} {\enquote {\bibinfo {title} {Surface energy
  of a bounded electron gas: Analysis of the accuracy of the local-density
  approximation via ab initio self-consistent-field calculations},}\
  }\href@noop {} {\bibfield  {journal} {\bibinfo  {journal} {Phys. Rev. B}\
  }\textbf {\bibinfo {volume} {57}},\ \bibinfo {pages} {6329} (\bibinfo {year}
  {1998})}\BibitemShut {NoStop}%
\bibitem [{\citenamefont {Pitarke}\ and\ \citenamefont {Eguiluz}(2001)}]{PE01}%
  \BibitemOpen
  \bibfield  {author} {\bibinfo {author} {\bibfnamefont {J.~M.}\ \bibnamefont
  {Pitarke}}\ and\ \bibinfo {author} {\bibfnamefont {A.~G.}\ \bibnamefont
  {Eguiluz}},\ }\bibfield  {title} {\enquote {\bibinfo {title} {Jellium surface
  energy beyond the local-density approximation: Self-consistent-field
  calculations},}\ }\href@noop {} {\bibfield  {journal} {\bibinfo  {journal}
  {Phys. Rev. B}\ }\textbf {\bibinfo {volume} {63}},\ \bibinfo {pages} {045116}
  (\bibinfo {year} {2001})}\BibitemShut {NoStop}%
\bibitem [{\citenamefont {Sham}\ and\ \citenamefont
  {Schl\"{u}ter}(1983)}]{SS83}%
  \BibitemOpen
  \bibfield  {author} {\bibinfo {author} {\bibfnamefont {L.~J.}\ \bibnamefont
  {Sham}}\ and\ \bibinfo {author} {\bibfnamefont {M.}~\bibnamefont
  {Schl\"{u}ter}},\ }\bibfield  {title} {\enquote {\bibinfo {title}
  {Density-functional theory of the energy gap},}\ }\href@noop {} {\bibfield
  {journal} {\bibinfo  {journal} {Phys. Rev. Lett.}\ }\textbf {\bibinfo
  {volume} {51}},\ \bibinfo {pages} {1888} (\bibinfo {year}
  {1983})}\BibitemShut {NoStop}%
\bibitem [{\citenamefont {Sham}(1985)}]{S85}%
  \BibitemOpen
  \bibfield  {author} {\bibinfo {author} {\bibfnamefont {L.~J.}\ \bibnamefont
  {Sham}},\ }\bibfield  {title} {\enquote {\bibinfo {title} {Exchange and
  correlation in density-functional theory},}\ }\href@noop {} {\bibfield
  {journal} {\bibinfo  {journal} {Phys. Rev. B}\ }\textbf {\bibinfo {volume}
  {32}},\ \bibinfo {pages} {3876} (\bibinfo {year} {1985})}\BibitemShut
  {NoStop}%
\bibitem [{\citenamefont {Rudnick}(1970)}]{R70}%
  \BibitemOpen
  \bibfield  {author} {\bibinfo {author} {\bibfnamefont {J.~A.}\ \bibnamefont
  {Rudnick}},\ }\href@noop {} {Ph.D. thesis},\ \bibinfo  {school} {University
  of California, San Diego} (\bibinfo {year} {1970})\BibitemShut {NoStop}%
\bibitem [{\citenamefont {Eguiluz}\ \emph {et~al.}(1992)\citenamefont
  {Eguiluz}, \citenamefont {Heinrichsmeier}, \citenamefont {Fleszar},\ and\
  \citenamefont {Hanke}}]{EHFH92}%
  \BibitemOpen
  \bibfield  {author} {\bibinfo {author} {\bibfnamefont {A.~G.}\ \bibnamefont
  {Eguiluz}}, \bibinfo {author} {\bibfnamefont {M.}~\bibnamefont
  {Heinrichsmeier}}, \bibinfo {author} {\bibfnamefont {A.}~\bibnamefont
  {Fleszar}},\ and\ \bibinfo {author} {\bibfnamefont {W.}~\bibnamefont
  {Hanke}},\ }\bibfield  {title} {\enquote {\bibinfo {title} {First-principles
  evaluation of the surface barrier for a {K}ohn-{S}ham electron at a metal
  surface},}\ }\href@noop {} {\bibfield  {journal} {\bibinfo  {journal} {Phys.
  Rev. Lett.}\ }\textbf {\bibinfo {volume} {68}},\ \bibinfo {pages} {1359}
  (\bibinfo {year} {1992})}\BibitemShut {NoStop}%
\bibitem [{\citenamefont {Ceperley}\ and\ \citenamefont {Alder}(1980)}]{CA80}%
  \BibitemOpen
  \bibfield  {author} {\bibinfo {author} {\bibfnamefont {D.~M.}\ \bibnamefont
  {Ceperley}}\ and\ \bibinfo {author} {\bibfnamefont {B.~J.}\ \bibnamefont
  {Alder}},\ }\bibfield  {title} {\enquote {\bibinfo {title} {Ground state of
  the electron gas by a stochastic method},}\ }\href@noop {} {\bibfield
  {journal} {\bibinfo  {journal} {Phys. Rev. Lett.}\ }\textbf {\bibinfo
  {volume} {45}},\ \bibinfo {pages} {566} (\bibinfo {year} {1980})}\BibitemShut
  {NoStop}%
\bibitem [{\citenamefont {White}\ \emph {et~al.}(1998)\citenamefont {White},
  \citenamefont {Godby}, \citenamefont {Rieger},\ and\ \citenamefont
  {Needs}}]{WGRN98}%
  \BibitemOpen
  \bibfield  {author} {\bibinfo {author} {\bibfnamefont {I.~D.}\ \bibnamefont
  {White}}, \bibinfo {author} {\bibfnamefont {R.~W.}\ \bibnamefont {Godby}},
  \bibinfo {author} {\bibfnamefont {M.~M.}\ \bibnamefont {Rieger}},\ and\
  \bibinfo {author} {\bibfnamefont {R.~J.}\ \bibnamefont {Needs}},\ }\bibfield
  {title} {\enquote {\bibinfo {title} {Dynamic image potential at an {A}l(111)
  surface},}\ }\href@noop {} {\bibfield  {journal} {\bibinfo  {journal} {Phys.
  Rev. Lett.}\ }\textbf {\bibinfo {volume} {80}},\ \bibinfo {pages} {4265}
  (\bibinfo {year} {1998})}\BibitemShut {NoStop}%
\bibitem [{\citenamefont {Eguiluz}\ and\ \citenamefont {Hanke}(1989)}]{EH89}%
  \BibitemOpen
  \bibfield  {author} {\bibinfo {author} {\bibfnamefont {A.~G.}\ \bibnamefont
  {Eguiluz}}\ and\ \bibinfo {author} {\bibfnamefont {W.}~\bibnamefont
  {Hanke}},\ }\bibfield  {title} {\enquote {\bibinfo {title} {Evaluation of the
  exchange-correlation potential at a metal surface from many-body perturbation
  theory},}\ }\href@noop {} {\bibfield  {journal} {\bibinfo  {journal} {Phys.
  Rev. B}\ }\textbf {\bibinfo {volume} {39}},\ \bibinfo {pages} {10433}
  (\bibinfo {year} {1989})}\BibitemShut {NoStop}%
\bibitem [{\citenamefont {Almbladh}\ and\ \citenamefont {von
  Barth}(1985)}]{AB85}%
  \BibitemOpen
  \bibfield  {author} {\bibinfo {author} {\bibfnamefont {C.-O.}\ \bibnamefont
  {Almbladh}}\ and\ \bibinfo {author} {\bibfnamefont {U.}~\bibnamefont {von
  Barth}},\ }\bibfield  {title} {\enquote {\bibinfo {title} {Exact results for
  the charge and spin densities, exchange-correlation potentials, and
  density-functional eigenvalues},}\ }\href@noop {} {\bibfield  {journal}
  {\bibinfo  {journal} {Phys. Rev. B}\ }\textbf {\bibinfo {volume} {31}},\
  \bibinfo {pages} {3231} (\bibinfo {year} {1985})}\BibitemShut {NoStop}%
\bibitem [{\citenamefont {Harbola}\ and\ \citenamefont
  {Sahni}(1989{\natexlab{a}})}]{HS89a}%
  \BibitemOpen
  \bibfield  {author} {\bibinfo {author} {\bibfnamefont {M.~K.}\ \bibnamefont
  {Harbola}}\ and\ \bibinfo {author} {\bibfnamefont {V.}~\bibnamefont
  {Sahni}},\ }\bibfield  {title} {\enquote {\bibinfo {title}
  {Quantum-mechanical interpretation of the exchange-correlation potential of
  {K}ohn-{S}ham density-functional theory},}\ }\href@noop {} {\bibfield
  {journal} {\bibinfo  {journal} {Phys. Rev. Lett.}\ }\textbf {\bibinfo
  {volume} {62}},\ \bibinfo {pages} {489} (\bibinfo {year}
  {1989}{\natexlab{a}})}\BibitemShut {NoStop}%
\bibitem [{\citenamefont {Harbola}\ and\ \citenamefont
  {Sahni}(1989{\natexlab{b}})}]{HS89b}%
  \BibitemOpen
  \bibfield  {author} {\bibinfo {author} {\bibfnamefont {M.~K.}\ \bibnamefont
  {Harbola}}\ and\ \bibinfo {author} {\bibfnamefont {V.}~\bibnamefont
  {Sahni}},\ }\bibfield  {title} {\enquote {\bibinfo {title}
  {Quantum-mechanical origin of the asymptotic effective potential at metal
  surfaces},}\ }\href@noop {} {\bibfield  {journal} {\bibinfo  {journal} {Phys.
  Rev. B}\ }\textbf {\bibinfo {volume} {39}},\ \bibinfo {pages} {10437}
  (\bibinfo {year} {1989}{\natexlab{b}})}\BibitemShut {NoStop}%
\bibitem [{\citenamefont {Solomatin}\ and\ \citenamefont {Sahni}(1996)}]{SS96}%
  \BibitemOpen
  \bibfield  {author} {\bibinfo {author} {\bibfnamefont {A.}~\bibnamefont
  {Solomatin}}\ and\ \bibinfo {author} {\bibfnamefont {V.}~\bibnamefont
  {Sahni}},\ }\bibfield  {title} {\enquote {\bibinfo {title} {Analytical
  asymptotic structure of the exchange and correlation potentials at a metal
  surface},}\ }\href@noop {} {\bibfield  {journal} {\bibinfo  {journal} {Phys.
  Lett. A}\ }\textbf {\bibinfo {volume} {212}},\ \bibinfo {pages} {263}
  (\bibinfo {year} {1996})}\BibitemShut {NoStop}%
\bibitem [{\citenamefont {Slater}(1951)}]{S51}%
  \BibitemOpen
  \bibfield  {author} {\bibinfo {author} {\bibfnamefont {J.~C.}\ \bibnamefont
  {Slater}},\ }\bibfield  {title} {\enquote {\bibinfo {title} {A simplification
  of the {H}artree-{F}ock method},}\ }\href@noop {} {\bibfield  {journal}
  {\bibinfo  {journal} {Phys. Rev.}\ }\textbf {\bibinfo {volume} {81}},\
  \bibinfo {pages} {385} (\bibinfo {year} {1951})}\BibitemShut {NoStop}%
\bibitem [{\citenamefont {Seidl}\ \emph {et~al.}(1996)\citenamefont {Seidl},
  \citenamefont {G\"orling}, \citenamefont {Vogl}, \citenamefont {Majewski},\
  and\ \citenamefont {Levy}}]{SGVM96}%
  \BibitemOpen
  \bibfield  {author} {\bibinfo {author} {\bibfnamefont {A.}~\bibnamefont
  {Seidl}}, \bibinfo {author} {\bibfnamefont {A.}~\bibnamefont {G\"orling}},
  \bibinfo {author} {\bibfnamefont {P.}~\bibnamefont {Vogl}}, \bibinfo {author}
  {\bibfnamefont {J.~A.}\ \bibnamefont {Majewski}},\ and\ \bibinfo {author}
  {\bibfnamefont {M.}~\bibnamefont {Levy}},\ }\bibfield  {title} {\enquote
  {\bibinfo {title} {Generalized {K}ohn-{S}ham schemes and the band-gap
  problem},}\ }\href {https://doi.org/10.1103/PhysRevB.53.3764} {\bibfield
  {journal} {\bibinfo  {journal} {Phys. Rev. B}\ }\textbf {\bibinfo {volume}
  {53}},\ \bibinfo {pages} {3764} (\bibinfo {year} {1996})}\BibitemShut
  {NoStop}%
\bibitem [{\citenamefont {Yang}, \citenamefont {Cohen},\ and\ \citenamefont
  {Mori-S\'{a}nchez}(2012)}]{YCM12}%
  \BibitemOpen
  \bibfield  {author} {\bibinfo {author} {\bibfnamefont {W.}~\bibnamefont
  {Yang}}, \bibinfo {author} {\bibfnamefont {A.~J.}\ \bibnamefont {Cohen}},\
  and\ \bibinfo {author} {\bibfnamefont {P.}~\bibnamefont {Mori-S\'{a}nchez}},\
  }\bibfield  {title} {\enquote {\bibinfo {title} {Derivative discontinuity,
  bandgap and lowest unoccupied molecular orbital in density functional
  theory},}\ }\href@noop {} {\bibfield  {journal} {\bibinfo  {journal} {J.
  Chem. Phys.}\ }\textbf {\bibinfo {volume} {36}},\ \bibinfo {pages} {204111}
  (\bibinfo {year} {2012})}\BibitemShut {NoStop}%
\bibitem [{\citenamefont {Perdew}\ \emph {et~al.}(2017)\citenamefont {Perdew},
  \citenamefont {Yang}, \citenamefont {Burke}, \citenamefont {Yang},
  \citenamefont {Gross}, \citenamefont {Scheffler}, \citenamefont {Scuseria},
  \citenamefont {Henderson}, \citenamefont {Zhang}, \citenamefont {Ruzsinszky},
  \citenamefont {Peng}, \citenamefont {Sun}, \citenamefont {Trushin},\ and\
  \citenamefont {Goerling}}]{PYBY17}%
  \BibitemOpen
  \bibfield  {author} {\bibinfo {author} {\bibfnamefont {J.~P.}\ \bibnamefont
  {Perdew}}, \bibinfo {author} {\bibfnamefont {W.}~\bibnamefont {Yang}},
  \bibinfo {author} {\bibfnamefont {K.}~\bibnamefont {Burke}}, \bibinfo
  {author} {\bibfnamefont {Z.}~\bibnamefont {Yang}}, \bibinfo {author}
  {\bibfnamefont {E.~K.~U.}\ \bibnamefont {Gross}}, \bibinfo {author}
  {\bibfnamefont {M.}~\bibnamefont {Scheffler}}, \bibinfo {author}
  {\bibfnamefont {G.~E.}\ \bibnamefont {Scuseria}}, \bibinfo {author}
  {\bibfnamefont {T.~M.}\ \bibnamefont {Henderson}}, \bibinfo {author}
  {\bibfnamefont {I.~Y.}\ \bibnamefont {Zhang}}, \bibinfo {author}
  {\bibfnamefont {A.}~\bibnamefont {Ruzsinszky}}, \bibinfo {author}
  {\bibfnamefont {H.}~\bibnamefont {Peng}}, \bibinfo {author} {\bibfnamefont
  {J.}~\bibnamefont {Sun}}, \bibinfo {author} {\bibfnamefont {E.}~\bibnamefont
  {Trushin}},\ and\ \bibinfo {author} {\bibfnamefont {A.}~\bibnamefont
  {Goerling}},\ }\bibfield  {title} {\enquote {\bibinfo {title} {Understanding
  band gaps of solids in generalized {K}ohn-{S}ham theory},}\ }\href
  {https://doi.org/10.1073/pnas.1621352114} {\bibfield  {journal} {\bibinfo
  {journal} {Proc. Natl. Acad. Sci. U.S.A.}\ }\textbf {\bibinfo {volume}
  {114}},\ \bibinfo {pages} {2801} (\bibinfo {year} {2017})}\BibitemShut
  {NoStop}%
\bibitem [{\citenamefont {Perdew}\ and\ \citenamefont
  {Ruzsinszky}(2018)}]{PR18}%
  \BibitemOpen
  \bibfield  {author} {\bibinfo {author} {\bibfnamefont {J.~P.}\ \bibnamefont
  {Perdew}}\ and\ \bibinfo {author} {\bibfnamefont {A.}~\bibnamefont
  {Ruzsinszky}},\ }\bibfield  {title} {\enquote {\bibinfo {title}
  {Density-functional energy gaps of solids demystified},}\ }\href@noop {}
  {\bibfield  {journal} {\bibinfo  {journal} {Eur. Phys. J. B}\ }\textbf
  {\bibinfo {volume} {91}},\ \bibinfo {pages} {108} (\bibinfo {year}
  {2018})}\BibitemShut {NoStop}%
\bibitem [{\citenamefont {Becke}(1993)}]{B93}%
  \BibitemOpen
  \bibfield  {author} {\bibinfo {author} {\bibfnamefont {A.~D.}\ \bibnamefont
  {Becke}},\ }\bibfield  {title} {\enquote {\bibinfo {title}
  {Density-functional thermochemistry. {III}. {T}he role of exact exchange},}\
  }\href {https://doi.org/10.1063/1.464913} {\bibfield  {journal} {\bibinfo
  {journal} {J. Chem. Phys.}\ }\textbf {\bibinfo {volume} {98}},\ \bibinfo
  {pages} {5648} (\bibinfo {year} {1993})}\BibitemShut {NoStop}%
\bibitem [{\citenamefont {Baer}, \citenamefont {Livshits},\ and\ \citenamefont
  {Salzner}(2010)}]{BLS10}%
  \BibitemOpen
  \bibfield  {author} {\bibinfo {author} {\bibfnamefont {R.}~\bibnamefont
  {Baer}}, \bibinfo {author} {\bibfnamefont {E.}~\bibnamefont {Livshits}},\
  and\ \bibinfo {author} {\bibfnamefont {U.}~\bibnamefont {Salzner}},\
  }\bibfield  {title} {\enquote {\bibinfo {title} {Tuned range-separated
  hybrids in density functional theory},}\ }\href
  {https://doi.org/10.1146/annurev.physchem.012809.103321} {\bibfield
  {journal} {\bibinfo  {journal} {Annu. Rev. Phys. Chem.}\ }\textbf {\bibinfo
  {volume} {61}},\ \bibinfo {pages} {85} (\bibinfo {year} {2010})}\BibitemShut
  {NoStop}%
\bibitem [{\citenamefont {Kronik}\ \emph {et~al.}(2012)\citenamefont {Kronik},
  \citenamefont {Stein}, \citenamefont {Refaely-Abramson},\ and\ \citenamefont
  {Baer}}]{KSRB12}%
  \BibitemOpen
  \bibfield  {author} {\bibinfo {author} {\bibfnamefont {L.}~\bibnamefont
  {Kronik}}, \bibinfo {author} {\bibfnamefont {T.}~\bibnamefont {Stein}},
  \bibinfo {author} {\bibfnamefont {S.}~\bibnamefont {Refaely-Abramson}},\ and\
  \bibinfo {author} {\bibfnamefont {R.}~\bibnamefont {Baer}},\ }\bibfield
  {title} {\enquote {\bibinfo {title} {Excitation gaps of finite-sized systems
  from optimally tuned range-separated hybrid functionals},}\ }\href@noop {}
  {\bibfield  {journal} {\bibinfo  {journal} {J. Chem. Theory Comput.}\
  }\textbf {\bibinfo {volume} {8}},\ \bibinfo {pages} {1515} (\bibinfo {year}
  {2012})}\BibitemShut {NoStop}%
\bibitem [{\citenamefont {Koerzdoerfer}\ and\ \citenamefont
  {Bredas}(2014)}]{KB14}%
  \BibitemOpen
  \bibfield  {author} {\bibinfo {author} {\bibfnamefont {T.}~\bibnamefont
  {Koerzdoerfer}}\ and\ \bibinfo {author} {\bibfnamefont {J.-L.}\ \bibnamefont
  {Bredas}},\ }\bibfield  {title} {\enquote {\bibinfo {title} {Organic
  electronic materials: Recent advances in the {DFT} description of the ground
  and excited states using tuned range-separated hybrid functionals},}\
  }\href@noop {} {\bibfield  {journal} {\bibinfo  {journal} {Acc. Chem. Res.}\
  }\textbf {\bibinfo {volume} {47}},\ \bibinfo {pages} {3284} (\bibinfo {year}
  {2014})}\BibitemShut {NoStop}%
\bibitem [{\citenamefont {Heyd}, \citenamefont {Scuseria},\ and\ \citenamefont
  {Ernzerhof}(2003)}]{HSE03}%
  \BibitemOpen
  \bibfield  {author} {\bibinfo {author} {\bibfnamefont {J.}~\bibnamefont
  {Heyd}}, \bibinfo {author} {\bibfnamefont {G.~E.}\ \bibnamefont {Scuseria}},\
  and\ \bibinfo {author} {\bibfnamefont {M.}~\bibnamefont {Ernzerhof}},\
  }\bibfield  {title} {\enquote {\bibinfo {title} {Hybrid functionals based on
  a screened coulomb potential},}\ }\href {https://doi.org/10.1063/1.1564060}
  {\bibfield  {journal} {\bibinfo  {journal} {J. Chem. Phys.}\ }\textbf
  {\bibinfo {volume} {118}},\ \bibinfo {pages} {8207} (\bibinfo {year}
  {2003})}\BibitemShut {NoStop}%
\bibitem [{\citenamefont {Toulouse}, \citenamefont {Colonna},\ and\
  \citenamefont {Savin}(2004)}]{TCS04}%
  \BibitemOpen
  \bibfield  {author} {\bibinfo {author} {\bibfnamefont {J.}~\bibnamefont
  {Toulouse}}, \bibinfo {author} {\bibfnamefont {F.}~\bibnamefont {Colonna}},\
  and\ \bibinfo {author} {\bibfnamefont {A.}~\bibnamefont {Savin}},\ }\bibfield
   {title} {\enquote {\bibinfo {title} {Long-range-short-range separation of
  the electron-electron interaction in density-functional theory},}\
  }\href@noop {} {\bibfield  {journal} {\bibinfo  {journal} {Phys. Rev. A}\
  }\textbf {\bibinfo {volume} {70}},\ \bibinfo {pages} {062505} (\bibinfo
  {year} {2004})}\BibitemShut {NoStop}%
\bibitem [{\citenamefont {Yanai}, \citenamefont {Tew},\ and\ \citenamefont
  {Handy}(2004)}]{YTH04}%
  \BibitemOpen
  \bibfield  {author} {\bibinfo {author} {\bibfnamefont {T.}~\bibnamefont
  {Yanai}}, \bibinfo {author} {\bibfnamefont {D.~P.}\ \bibnamefont {Tew}},\
  and\ \bibinfo {author} {\bibfnamefont {N.~C.}\ \bibnamefont {Handy}},\
  }\bibfield  {title} {\enquote {\bibinfo {title} {A new hybrid
  exchange-correlation functional using the {Coulomb}-attenuating method
  ({CAM-B3LYP})},}\ }\href@noop {} {\bibfield  {journal} {\bibinfo  {journal}
  {Chem. Phys. Lett.}\ }\textbf {\bibinfo {volume} {393}},\ \bibinfo {pages}
  {51} (\bibinfo {year} {2004})}\BibitemShut {NoStop}%
\bibitem [{\citenamefont {Shimazaki}\ and\ \citenamefont {Asai}(2008)}]{SA08}%
  \BibitemOpen
  \bibfield  {author} {\bibinfo {author} {\bibfnamefont {T.}~\bibnamefont
  {Shimazaki}}\ and\ \bibinfo {author} {\bibfnamefont {Y.}~\bibnamefont
  {Asai}},\ }\bibfield  {title} {\enquote {\bibinfo {title} {Band structure
  calculations based on screened {F}ock exchange method},}\ }\href@noop {}
  {\bibfield  {journal} {\bibinfo  {journal} {Chem. Phys. Lett.}\ }\textbf
  {\bibinfo {volume} {466}},\ \bibinfo {pages} {91} (\bibinfo {year}
  {2008})}\BibitemShut {NoStop}%
\bibitem [{\citenamefont {Bechstedt}\ \emph {et~al.}(1992)\citenamefont
  {Bechstedt}, \citenamefont {Del~Sole}, \citenamefont {Cappellini},\ and\
  \citenamefont {Reining}}]{BSCR92}%
  \BibitemOpen
  \bibfield  {author} {\bibinfo {author} {\bibfnamefont {F.}~\bibnamefont
  {Bechstedt}}, \bibinfo {author} {\bibfnamefont {R.}~\bibnamefont {Del~Sole}},
  \bibinfo {author} {\bibfnamefont {G.}~\bibnamefont {Cappellini}},\ and\
  \bibinfo {author} {\bibfnamefont {L.}~\bibnamefont {Reining}},\ }\bibfield
  {title} {\enquote {\bibinfo {title} {An efficient method for calculating
  quasiparticle energies in semiconductors},}\ }\href@noop {} {\bibfield
  {journal} {\bibinfo  {journal} {Solid State Commun.}\ }\textbf {\bibinfo
  {volume} {84}},\ \bibinfo {pages} {765} (\bibinfo {year} {1992})}\BibitemShut
  {NoStop}%
\bibitem [{\citenamefont {Cappellini}\ \emph {et~al.}(1993)\citenamefont
  {Cappellini}, \citenamefont {Del~Sole}, \citenamefont {Reining},\ and\
  \citenamefont {Bechstedt}}]{CDLB93}%
  \BibitemOpen
  \bibfield  {author} {\bibinfo {author} {\bibfnamefont {G.}~\bibnamefont
  {Cappellini}}, \bibinfo {author} {\bibfnamefont {R.}~\bibnamefont
  {Del~Sole}}, \bibinfo {author} {\bibfnamefont {L.}~\bibnamefont {Reining}},\
  and\ \bibinfo {author} {\bibfnamefont {F.}~\bibnamefont {Bechstedt}},\
  }\bibfield  {title} {\enquote {\bibinfo {title} {Model dielectric function
  for semiconductors},}\ }\href@noop {} {\bibfield  {journal} {\bibinfo
  {journal} {Phys. Rev. B}\ }\textbf {\bibinfo {volume} {47}},\ \bibinfo
  {pages} {9892} (\bibinfo {year} {1993})}\BibitemShut {NoStop}%
\bibitem [{\citenamefont {Shimazaki}\ and\ \citenamefont {Asai}(2009)}]{SA09}%
  \BibitemOpen
  \bibfield  {author} {\bibinfo {author} {\bibfnamefont {T.}~\bibnamefont
  {Shimazaki}}\ and\ \bibinfo {author} {\bibfnamefont {Y.}~\bibnamefont
  {Asai}},\ }\bibfield  {title} {\enquote {\bibinfo {title} {First principles
  band structure calculations based on self-consistent screened
  {H}artree-{F}ock exchange potential},}\ }\href@noop {} {\bibfield  {journal}
  {\bibinfo  {journal} {J. Chem. Phys.}\ }\textbf {\bibinfo {volume} {130}},\
  \bibinfo {pages} {164702} (\bibinfo {year} {2009})}\BibitemShut {NoStop}%
\bibitem [{\citenamefont {Shimazaki}\ and\ \citenamefont {Asai}(2010)}]{SA10}%
  \BibitemOpen
  \bibfield  {author} {\bibinfo {author} {\bibfnamefont {T.}~\bibnamefont
  {Shimazaki}}\ and\ \bibinfo {author} {\bibfnamefont {Y.}~\bibnamefont
  {Asai}},\ }\bibfield  {title} {\enquote {\bibinfo {title} {Energy band
  structure calculations based on screened {H}artree-{F}ock exchange method:
  {Si, AlP, AlAs, GaP, and GaAs}},}\ }\href@noop {} {\bibfield  {journal}
  {\bibinfo  {journal} {J. Chem. Phys.}\ }\textbf {\bibinfo {volume} {132}},\
  \bibinfo {pages} {224105} (\bibinfo {year} {2010})}\BibitemShut {NoStop}%
\bibitem [{\citenamefont {Refaely-Abramson}\ \emph {et~al.}(2012)\citenamefont
  {Refaely-Abramson}, \citenamefont {Sharifzadeh}, \citenamefont {Govind},
  \citenamefont {Autschbach}, \citenamefont {Neaton}, \citenamefont {Baer},\
  and\ \citenamefont {Kronik}}]{RSGA12}%
  \BibitemOpen
  \bibfield  {author} {\bibinfo {author} {\bibfnamefont {S.}~\bibnamefont
  {Refaely-Abramson}}, \bibinfo {author} {\bibfnamefont {S.}~\bibnamefont
  {Sharifzadeh}}, \bibinfo {author} {\bibfnamefont {N.}~\bibnamefont {Govind}},
  \bibinfo {author} {\bibfnamefont {J.}~\bibnamefont {Autschbach}}, \bibinfo
  {author} {\bibfnamefont {J.~B.}\ \bibnamefont {Neaton}}, \bibinfo {author}
  {\bibfnamefont {R.}~\bibnamefont {Baer}},\ and\ \bibinfo {author}
  {\bibfnamefont {L.}~\bibnamefont {Kronik}},\ }\bibfield  {title} {\enquote
  {\bibinfo {title} {Quasiparticle spectra from a nonempirical optimally tuned
  range-separated hybrid density functional},}\ }\href@noop {} {\bibfield
  {journal} {\bibinfo  {journal} {Phys. Rev. Lett.}\ }\textbf {\bibinfo
  {volume} {109}},\ \bibinfo {pages} {226405} (\bibinfo {year}
  {2012})}\BibitemShut {NoStop}%
\bibitem [{\citenamefont {Rohrdanz}\ and\ \citenamefont
  {Herbert}(2008)}]{RH08}%
  \BibitemOpen
  \bibfield  {author} {\bibinfo {author} {\bibfnamefont {M.~A.}\ \bibnamefont
  {Rohrdanz}}\ and\ \bibinfo {author} {\bibfnamefont {J.~M.}\ \bibnamefont
  {Herbert}},\ }\bibfield  {title} {\enquote {\bibinfo {title} {Simultaneous
  benchmarking of ground- and excited-state properties with
  long-range-corrected density functional theory},}\ }\href@noop {} {\bibfield
  {journal} {\bibinfo  {journal} {J. Chem. Phys.}\ }\textbf {\bibinfo {volume}
  {129}},\ \bibinfo {pages} {034107} (\bibinfo {year} {2008})}\BibitemShut
  {NoStop}%
\bibitem [{\citenamefont {Refaely-Abramson}\ \emph {et~al.}(2013)\citenamefont
  {Refaely-Abramson}, \citenamefont {Sharifzadeh}, \citenamefont {Jain},
  \citenamefont {Baer}, \citenamefont {Neaton},\ and\ \citenamefont
  {Kronik}}]{RSJB13}%
  \BibitemOpen
  \bibfield  {author} {\bibinfo {author} {\bibfnamefont {S.}~\bibnamefont
  {Refaely-Abramson}}, \bibinfo {author} {\bibfnamefont {S.}~\bibnamefont
  {Sharifzadeh}}, \bibinfo {author} {\bibfnamefont {M.}~\bibnamefont {Jain}},
  \bibinfo {author} {\bibfnamefont {R.}~\bibnamefont {Baer}}, \bibinfo {author}
  {\bibfnamefont {J.~B.}\ \bibnamefont {Neaton}},\ and\ \bibinfo {author}
  {\bibfnamefont {L.}~\bibnamefont {Kronik}},\ }\bibfield  {title} {\enquote
  {\bibinfo {title} {Gap renormalization of molecular crystals from
  density-functional theory},}\ }\href@noop {} {\bibfield  {journal} {\bibinfo
  {journal} {Phys. Rev. B}\ }\textbf {\bibinfo {volume} {88}},\ \bibinfo
  {pages} {081204(R)} (\bibinfo {year} {2013})}\BibitemShut {NoStop}%
\bibitem [{\citenamefont {Liu}\ \emph {et~al.}(2017)\citenamefont {Liu},
  \citenamefont {Egger}, \citenamefont {Refaely-Abramson}, \citenamefont
  {Kronik},\ and\ \citenamefont {Neaton}}]{LERK17}%
  \BibitemOpen
  \bibfield  {author} {\bibinfo {author} {\bibfnamefont {Z.-F.}\ \bibnamefont
  {Liu}}, \bibinfo {author} {\bibfnamefont {D.~A.}\ \bibnamefont {Egger}},
  \bibinfo {author} {\bibfnamefont {S.}~\bibnamefont {Refaely-Abramson}},
  \bibinfo {author} {\bibfnamefont {L.}~\bibnamefont {Kronik}},\ and\ \bibinfo
  {author} {\bibfnamefont {J.~B.}\ \bibnamefont {Neaton}},\ }\bibfield  {title}
  {\enquote {\bibinfo {title} {Energy level alignment at molecule-metal
  interfaces from an optimally tuned range-separated hybrid functional},}\
  }\href@noop {} {\bibfield  {journal} {\bibinfo  {journal} {J. Chem. Phys.}\
  }\textbf {\bibinfo {volume} {146}},\ \bibinfo {pages} {092326} (\bibinfo
  {year} {2017})}\BibitemShut {NoStop}%
\bibitem [{\citenamefont {Biller}\ \emph {et~al.}(2011)\citenamefont {Biller},
  \citenamefont {Tamblyn}, \citenamefont {Neaton},\ and\ \citenamefont
  {Kronik}}]{BTNK11}%
  \BibitemOpen
  \bibfield  {author} {\bibinfo {author} {\bibfnamefont {A.}~\bibnamefont
  {Biller}}, \bibinfo {author} {\bibfnamefont {I.}~\bibnamefont {Tamblyn}},
  \bibinfo {author} {\bibfnamefont {J.~B.}\ \bibnamefont {Neaton}},\ and\
  \bibinfo {author} {\bibfnamefont {L.}~\bibnamefont {Kronik}},\ }\bibfield
  {title} {\enquote {\bibinfo {title} {Electronic level alignment at a
  metal-molecule interface from a short-range hybrid functional},}\ }\href@noop
  {} {\bibfield  {journal} {\bibinfo  {journal} {J. Chem. Phys.}\ }\textbf
  {\bibinfo {volume} {135}},\ \bibinfo {pages} {164706} (\bibinfo {year}
  {2011})}\BibitemShut {NoStop}%
\bibitem [{\citenamefont {Zhou}\ \emph {et~al.}(2021)\citenamefont {Zhou},
  \citenamefont {Liu}, \citenamefont {Marks},\ and\ \citenamefont
  {Darancet}}]{ZLMD21}%
  \BibitemOpen
  \bibfield  {author} {\bibinfo {author} {\bibfnamefont {Q.}~\bibnamefont
  {Zhou}}, \bibinfo {author} {\bibfnamefont {Z.-F.}\ \bibnamefont {Liu}},
  \bibinfo {author} {\bibfnamefont {T.~J.}\ \bibnamefont {Marks}},\ and\
  \bibinfo {author} {\bibfnamefont {P.}~\bibnamefont {Darancet}},\ }\bibfield
  {title} {\enquote {\bibinfo {title} {Range-separated hybrid functionals for
  mixed dimensional heterojunctions: Application to
  phthalocyanines/{MoS}$_2$},}\ }\href@noop {} {\bibfield  {journal} {\bibinfo
  {journal} {APL Mater.}\ }\textbf {\bibinfo {volume} {9}},\ \bibinfo {pages}
  {121112} (\bibinfo {year} {2021})}\BibitemShut {NoStop}%
\bibitem [{\citenamefont {Noori}\ \emph {et~al.}(2019)\citenamefont {Noori},
  \citenamefont {Cheng}, \citenamefont {Xuan},\ and\ \citenamefont
  {Quek}}]{NCXQ19}%
  \BibitemOpen
  \bibfield  {author} {\bibinfo {author} {\bibfnamefont {K.}~\bibnamefont
  {Noori}}, \bibinfo {author} {\bibfnamefont {N.~L.~Q.}\ \bibnamefont {Cheng}},
  \bibinfo {author} {\bibfnamefont {F.}~\bibnamefont {Xuan}},\ and\ \bibinfo
  {author} {\bibfnamefont {S.~Y.}\ \bibnamefont {Quek}},\ }\bibfield  {title}
  {\enquote {\bibinfo {title} {Dielectric screening by {2D} substrates},}\
  }\href@noop {} {\bibfield  {journal} {\bibinfo  {journal} {2D Mater.}\
  }\textbf {\bibinfo {volume} {6}},\ \bibinfo {pages} {035036} (\bibinfo {year}
  {2019})}\BibitemShut {NoStop}%
\bibitem [{\citenamefont {Cho}\ and\ \citenamefont {Berkelbach}(2018)}]{CB18}%
  \BibitemOpen
  \bibfield  {author} {\bibinfo {author} {\bibfnamefont {Y.}~\bibnamefont
  {Cho}}\ and\ \bibinfo {author} {\bibfnamefont {T.~C.}\ \bibnamefont
  {Berkelbach}},\ }\bibfield  {title} {\enquote {\bibinfo {title}
  {Environmentally sensitive theory of electronic and optical transitions in
  atomically thin semiconductors},}\ }\href@noop {} {\bibfield  {journal}
  {\bibinfo  {journal} {Phys. Rev. B}\ }\textbf {\bibinfo {volume} {97}},\
  \bibinfo {pages} {041409} (\bibinfo {year} {2018})}\BibitemShut {NoStop}%
\bibitem [{\citenamefont {Marques}\ \emph {et~al.}(2011)\citenamefont
  {Marques}, \citenamefont {Vidal}, \citenamefont {Oliveira}, \citenamefont
  {Reining},\ and\ \citenamefont {Botti}}]{MVOR11}%
  \BibitemOpen
  \bibfield  {author} {\bibinfo {author} {\bibfnamefont {M.~A.~L.}\
  \bibnamefont {Marques}}, \bibinfo {author} {\bibfnamefont {J.}~\bibnamefont
  {Vidal}}, \bibinfo {author} {\bibfnamefont {M.~J.~T.}\ \bibnamefont
  {Oliveira}}, \bibinfo {author} {\bibfnamefont {L.}~\bibnamefont {Reining}},\
  and\ \bibinfo {author} {\bibfnamefont {S.}~\bibnamefont {Botti}},\ }\bibfield
   {title} {\enquote {\bibinfo {title} {Density-based mixing parameter for
  hybrid functionals},}\ }\href {https://doi.org/10.1103/physrevb.83.035119}
  {\bibfield  {journal} {\bibinfo  {journal} {Phys. Rev. B}\ }\textbf {\bibinfo
  {volume} {83}},\ \bibinfo {pages} {035119} (\bibinfo {year}
  {2011})}\BibitemShut {NoStop}%
\bibitem [{\citenamefont {Koller}, \citenamefont {Blaha},\ and\ \citenamefont
  {Tran}(2013)}]{KBT13}%
  \BibitemOpen
  \bibfield  {author} {\bibinfo {author} {\bibfnamefont {D.}~\bibnamefont
  {Koller}}, \bibinfo {author} {\bibfnamefont {P.}~\bibnamefont {Blaha}},\ and\
  \bibinfo {author} {\bibfnamefont {F.}~\bibnamefont {Tran}},\ }\bibfield
  {title} {\enquote {\bibinfo {title} {Hybrid functionals for solids with an
  optimized {H}artree-{F}ock mixing parameter},}\ }\href@noop {} {\bibfield
  {journal} {\bibinfo  {journal} {J. Phys.: Condens. Matter}\ }\textbf
  {\bibinfo {volume} {25}},\ \bibinfo {pages} {435503} (\bibinfo {year}
  {2013})}\BibitemShut {NoStop}%
\bibitem [{\citenamefont {Shimazaki}\ and\ \citenamefont
  {Nakajima}(2014)}]{SN14}%
  \BibitemOpen
  \bibfield  {author} {\bibinfo {author} {\bibfnamefont {T.}~\bibnamefont
  {Shimazaki}}\ and\ \bibinfo {author} {\bibfnamefont {T.}~\bibnamefont
  {Nakajima}},\ }\bibfield  {title} {\enquote {\bibinfo {title}
  {Dielectric-dependent screened {H}artree-{F}ock exchange potential and
  {S}later-formula with {C}oulomb-hole interaction for energy band structure
  calculations},}\ }\href@noop {} {\bibfield  {journal} {\bibinfo  {journal}
  {J. Chem. Phys.}\ }\textbf {\bibinfo {volume} {141}},\ \bibinfo {pages}
  {114109} (\bibinfo {year} {2014})}\BibitemShut {NoStop}%
\bibitem [{\citenamefont {Skone}, \citenamefont {Govoni},\ and\ \citenamefont
  {Galli}(2014)}]{SGG14}%
  \BibitemOpen
  \bibfield  {author} {\bibinfo {author} {\bibfnamefont {J.~H.}\ \bibnamefont
  {Skone}}, \bibinfo {author} {\bibfnamefont {M.}~\bibnamefont {Govoni}},\ and\
  \bibinfo {author} {\bibfnamefont {G.}~\bibnamefont {Galli}},\ }\bibfield
  {title} {\enquote {\bibinfo {title} {Self-consistent hybrid functional for
  condensed systems},}\ }\href {https://doi.org/10.1103/physrevb.89.195112}
  {\bibfield  {journal} {\bibinfo  {journal} {Phys. Rev. B}\ }\textbf {\bibinfo
  {volume} {89}},\ \bibinfo {pages} {195112} (\bibinfo {year}
  {2014})}\BibitemShut {NoStop}%
\bibitem [{\citenamefont {Ferrero}\ \emph {et~al.}(2008)\citenamefont
  {Ferrero}, \citenamefont {R\'{e}rat}, \citenamefont {Orlando}, \citenamefont
  {Dovesi},\ and\ \citenamefont {Bush}}]{FROD08}%
  \BibitemOpen
  \bibfield  {author} {\bibinfo {author} {\bibfnamefont {M.}~\bibnamefont
  {Ferrero}}, \bibinfo {author} {\bibfnamefont {M.}~\bibnamefont {R\'{e}rat}},
  \bibinfo {author} {\bibfnamefont {R.}~\bibnamefont {Orlando}}, \bibinfo
  {author} {\bibfnamefont {R.}~\bibnamefont {Dovesi}},\ and\ \bibinfo {author}
  {\bibfnamefont {I.~J.}\ \bibnamefont {Bush}},\ }\bibfield  {title} {\enquote
  {\bibinfo {title} {Coupled perturbed {K}ohn-{S}ham calculation of static
  polarizabilities of periodic compounds},}\ }\href@noop {} {\bibfield
  {journal} {\bibinfo  {journal} {J. Phys.: Conf. Ser.}\ }\textbf {\bibinfo
  {volume} {117}},\ \bibinfo {pages} {012016} (\bibinfo {year}
  {2008})}\BibitemShut {NoStop}%
\bibitem [{\citenamefont {Skone}, \citenamefont {Govoni},\ and\ \citenamefont
  {Galli}(2016)}]{SGG16}%
  \BibitemOpen
  \bibfield  {author} {\bibinfo {author} {\bibfnamefont {J.~H.}\ \bibnamefont
  {Skone}}, \bibinfo {author} {\bibfnamefont {M.}~\bibnamefont {Govoni}},\ and\
  \bibinfo {author} {\bibfnamefont {G.}~\bibnamefont {Galli}},\ }\bibfield
  {title} {\enquote {\bibinfo {title} {Nonempirical range-separated hybrid
  functionals for solids and molecules},}\ }\href
  {https://doi.org/10.1103/physrevb.93.235106} {\bibfield  {journal} {\bibinfo
  {journal} {Phys. Rev. B}\ }\textbf {\bibinfo {volume} {93}},\ \bibinfo
  {pages} {235106} (\bibinfo {year} {2016})}\BibitemShut {NoStop}%
\bibitem [{\citenamefont {Wilson}, \citenamefont {Gygi},\ and\ \citenamefont
  {Galli}(2008)}]{WGG08}%
  \BibitemOpen
  \bibfield  {author} {\bibinfo {author} {\bibfnamefont {H.~F.}\ \bibnamefont
  {Wilson}}, \bibinfo {author} {\bibfnamefont {F.}~\bibnamefont {Gygi}},\ and\
  \bibinfo {author} {\bibfnamefont {G.}~\bibnamefont {Galli}},\ }\bibfield
  {title} {\enquote {\bibinfo {title} {Efficient iterative method for
  calculations of dielectric matrices},}\ }\href@noop {} {\bibfield  {journal}
  {\bibinfo  {journal} {Phys. Rev. B}\ }\textbf {\bibinfo {volume} {78}},\
  \bibinfo {pages} {113303} (\bibinfo {year} {2008})}\BibitemShut {NoStop}%
\bibitem [{\citenamefont {Brawand}\ \emph {et~al.}(2016)\citenamefont
  {Brawand}, \citenamefont {V\"{o}r\"{o}s}, \citenamefont {Govoni},\ and\
  \citenamefont {Galli}}]{BVGG16}%
  \BibitemOpen
  \bibfield  {author} {\bibinfo {author} {\bibfnamefont {N.~P.}\ \bibnamefont
  {Brawand}}, \bibinfo {author} {\bibfnamefont {M.}~\bibnamefont
  {V\"{o}r\"{o}s}}, \bibinfo {author} {\bibfnamefont {M.}~\bibnamefont
  {Govoni}},\ and\ \bibinfo {author} {\bibfnamefont {G.}~\bibnamefont
  {Galli}},\ }\bibfield  {title} {\enquote {\bibinfo {title} {Generalization of
  dielectric-dependent hybrid functionals to finite systems},}\ }\href@noop {}
  {\bibfield  {journal} {\bibinfo  {journal} {Phys. Rev. X}\ }\textbf {\bibinfo
  {volume} {6}},\ \bibinfo {pages} {041002} (\bibinfo {year}
  {2016})}\BibitemShut {NoStop}%
\bibitem [{\citenamefont {Zheng}, \citenamefont {Govoni},\ and\ \citenamefont
  {Galli}(2019)}]{ZGG19}%
  \BibitemOpen
  \bibfield  {author} {\bibinfo {author} {\bibfnamefont {H.}~\bibnamefont
  {Zheng}}, \bibinfo {author} {\bibfnamefont {M.}~\bibnamefont {Govoni}},\ and\
  \bibinfo {author} {\bibfnamefont {G.}~\bibnamefont {Galli}},\ }\bibfield
  {title} {\enquote {\bibinfo {title} {Dielectric-dependent hybrid functionals
  for heterogeneous materials},}\ }\href
  {https://doi.org/10.1103/physrevmaterials.3.073803} {\bibfield  {journal}
  {\bibinfo  {journal} {Phys. Rev. Mater.}\ }\textbf {\bibinfo {volume} {3}},\
  \bibinfo {pages} {073803} (\bibinfo {year} {2019})}\BibitemShut {NoStop}%
\bibitem [{\citenamefont {Gerosa}\ \emph
  {et~al.}(2015{\natexlab{a}})\citenamefont {Gerosa}, \citenamefont {Bottani},
  \citenamefont {Caramella}, \citenamefont {Onida}, \citenamefont
  {Di~Valentin},\ and\ \citenamefont {Pacchioni}}]{GBCO15a}%
  \BibitemOpen
  \bibfield  {author} {\bibinfo {author} {\bibfnamefont {M.}~\bibnamefont
  {Gerosa}}, \bibinfo {author} {\bibfnamefont {C.~E.}\ \bibnamefont {Bottani}},
  \bibinfo {author} {\bibfnamefont {L.}~\bibnamefont {Caramella}}, \bibinfo
  {author} {\bibfnamefont {G.}~\bibnamefont {Onida}}, \bibinfo {author}
  {\bibfnamefont {C.}~\bibnamefont {Di~Valentin}},\ and\ \bibinfo {author}
  {\bibfnamefont {G.}~\bibnamefont {Pacchioni}},\ }\bibfield  {title} {\enquote
  {\bibinfo {title} {Electronic structure and phase stability of oxide
  semiconductors: Performance of dielectric-dependent hybrid functional {DFT},
  benchmarked against ${GW}$ band structure calculations and experiments},}\
  }\href {https://doi.org/10.1103/physrevb.91.155201} {\bibfield  {journal}
  {\bibinfo  {journal} {Phys. Rev. B}\ }\textbf {\bibinfo {volume} {91}},\
  \bibinfo {pages} {155201} (\bibinfo {year} {2015}{\natexlab{a}})}\BibitemShut
  {NoStop}%
\bibitem [{\citenamefont {Gerosa}\ \emph {et~al.}(2018)\citenamefont {Gerosa},
  \citenamefont {Bottani}, \citenamefont {Di~Valentin}, \citenamefont {Onida},\
  and\ \citenamefont {Pacchioni}}]{GBDO18}%
  \BibitemOpen
  \bibfield  {author} {\bibinfo {author} {\bibfnamefont {M.}~\bibnamefont
  {Gerosa}}, \bibinfo {author} {\bibfnamefont {C.~E.}\ \bibnamefont {Bottani}},
  \bibinfo {author} {\bibfnamefont {C.}~\bibnamefont {Di~Valentin}}, \bibinfo
  {author} {\bibfnamefont {G.}~\bibnamefont {Onida}},\ and\ \bibinfo {author}
  {\bibfnamefont {G.}~\bibnamefont {Pacchioni}},\ }\bibfield  {title} {\enquote
  {\bibinfo {title} {Accuracy of dielectric-dependent hybrid functionals in the
  prediction of optoelectronic properties of metal oxide semiconductors: {A}
  comprehensive comparison with many-body ${GW}$ and experiments},}\
  }\href@noop {} {\bibfield  {journal} {\bibinfo  {journal} {J. Phys.: Condens.
  Matter}\ }\textbf {\bibinfo {volume} {30}},\ \bibinfo {pages} {044003}
  (\bibinfo {year} {2018})}\BibitemShut {NoStop}%
\bibitem [{\citenamefont {Chen}\ \emph {et~al.}(2018)\citenamefont {Chen},
  \citenamefont {Miceli}, \citenamefont {Rignanese},\ and\ \citenamefont
  {Pasquarello}}]{CMRP18}%
  \BibitemOpen
  \bibfield  {author} {\bibinfo {author} {\bibfnamefont {W.}~\bibnamefont
  {Chen}}, \bibinfo {author} {\bibfnamefont {G.}~\bibnamefont {Miceli}},
  \bibinfo {author} {\bibfnamefont {G.-M.}\ \bibnamefont {Rignanese}},\ and\
  \bibinfo {author} {\bibfnamefont {A.}~\bibnamefont {Pasquarello}},\
  }\bibfield  {title} {\enquote {\bibinfo {title} {Nonempirical
  dielectric-dependent hybrid functional with range separation for
  semiconductors and insulators},}\ }\href
  {https://doi.org/10.1103/physrevmaterials.2.073803} {\bibfield  {journal}
  {\bibinfo  {journal} {Phys. Rev. Mater.}\ }\textbf {\bibinfo {volume} {2}},\
  \bibinfo {pages} {073803} (\bibinfo {year} {2018})}\BibitemShut {NoStop}%
\bibitem [{\citenamefont {Bischoff}\ \emph {et~al.}(2019)\citenamefont
  {Bischoff}, \citenamefont {Wiktor}, \citenamefont {Chen},\ and\ \citenamefont
  {Pasquarello}}]{BWCP19}%
  \BibitemOpen
  \bibfield  {author} {\bibinfo {author} {\bibfnamefont {T.}~\bibnamefont
  {Bischoff}}, \bibinfo {author} {\bibfnamefont {J.}~\bibnamefont {Wiktor}},
  \bibinfo {author} {\bibfnamefont {W.}~\bibnamefont {Chen}},\ and\ \bibinfo
  {author} {\bibfnamefont {A.}~\bibnamefont {Pasquarello}},\ }\bibfield
  {title} {\enquote {\bibinfo {title} {Nonempirical hybrid functionals for band
  gaps of inorganic metal-halide perovskites},}\ }\href@noop {} {\bibfield
  {journal} {\bibinfo  {journal} {Phys. Rev. Mater.}\ }\textbf {\bibinfo
  {volume} {3}},\ \bibinfo {pages} {123802} (\bibinfo {year}
  {2019})}\BibitemShut {NoStop}%
\bibitem [{\citenamefont {Jana}\ \emph {et~al.}(2020)\citenamefont {Jana},
  \citenamefont {Patra}, \citenamefont {{\'{S}}miga}, \citenamefont
  {Constantin},\ and\ \citenamefont {Samal}}]{JPSC20}%
  \BibitemOpen
  \bibfield  {author} {\bibinfo {author} {\bibfnamefont {S.}~\bibnamefont
  {Jana}}, \bibinfo {author} {\bibfnamefont {B.}~\bibnamefont {Patra}},
  \bibinfo {author} {\bibfnamefont {S.}~\bibnamefont {{\'{S}}miga}}, \bibinfo
  {author} {\bibfnamefont {L.~A.}\ \bibnamefont {Constantin}},\ and\ \bibinfo
  {author} {\bibfnamefont {P.}~\bibnamefont {Samal}},\ }\bibfield  {title}
  {\enquote {\bibinfo {title} {Improved solid stability from a screened
  range-separated hybrid functional by satisfying semiclassical atom theory and
  local density linear response},}\ }\href
  {https://doi.org/10.1103/physrevb.102.155107} {\bibfield  {journal} {\bibinfo
   {journal} {Phys. Rev. B}\ }\textbf {\bibinfo {volume} {102}},\ \bibinfo
  {pages} {155107} (\bibinfo {year} {2020})}\BibitemShut {NoStop}%
\bibitem [{\citenamefont {Liu}\ \emph {et~al.}(2020)\citenamefont {Liu},
  \citenamefont {Franchini}, \citenamefont {Marsman},\ and\ \citenamefont
  {Kresse}}]{LFMK20}%
  \BibitemOpen
  \bibfield  {author} {\bibinfo {author} {\bibfnamefont {P.}~\bibnamefont
  {Liu}}, \bibinfo {author} {\bibfnamefont {C.}~\bibnamefont {Franchini}},
  \bibinfo {author} {\bibfnamefont {M.}~\bibnamefont {Marsman}},\ and\ \bibinfo
  {author} {\bibfnamefont {G.}~\bibnamefont {Kresse}},\ }\bibfield  {title}
  {\enquote {\bibinfo {title} {Assessing model-dielectric-dependent hybrid
  functionals on the antiferromagnetic transition-metal monoxides {MnO, FeO,
  CoO, and NiO}},}\ }\href@noop {} {\bibfield  {journal} {\bibinfo  {journal}
  {J. Phys.: Condens. Matter}\ }\textbf {\bibinfo {volume} {32}},\ \bibinfo
  {pages} {015502} (\bibinfo {year} {2020})}\BibitemShut {NoStop}%
\bibitem [{\citenamefont {Skelton}\ \emph {et~al.}(2020)\citenamefont
  {Skelton}, \citenamefont {Gunn}, \citenamefont {Metz},\ and\ \citenamefont
  {Parker}}]{SGMP20}%
  \BibitemOpen
  \bibfield  {author} {\bibinfo {author} {\bibfnamefont {J.~M.}\ \bibnamefont
  {Skelton}}, \bibinfo {author} {\bibfnamefont {D.~S.~D.}\ \bibnamefont
  {Gunn}}, \bibinfo {author} {\bibfnamefont {S.}~\bibnamefont {Metz}},\ and\
  \bibinfo {author} {\bibfnamefont {S.~C.}\ \bibnamefont {Parker}},\ }\bibfield
   {title} {\enquote {\bibinfo {title} {Accuracy of hybrid functionals with
  non-self-consistent {K}ohn-{S}ham orbitals for predicting the properties of
  semiconductors},}\ }\href@noop {} {\bibfield  {journal} {\bibinfo  {journal}
  {J. Chem. Theory Comput.}\ }\textbf {\bibinfo {volume} {16}},\ \bibinfo
  {pages} {3543} (\bibinfo {year} {2020})}\BibitemShut {NoStop}%
\bibitem [{\citenamefont {Gerosa}\ \emph
  {et~al.}(2015{\natexlab{b}})\citenamefont {Gerosa}, \citenamefont {Bottani},
  \citenamefont {Caramella}, \citenamefont {Onida}, \citenamefont
  {Di~Valentin},\ and\ \citenamefont {Pacchioni}}]{GBCO15b}%
  \BibitemOpen
  \bibfield  {author} {\bibinfo {author} {\bibfnamefont {M.}~\bibnamefont
  {Gerosa}}, \bibinfo {author} {\bibfnamefont {C.~E.}\ \bibnamefont {Bottani}},
  \bibinfo {author} {\bibfnamefont {L.}~\bibnamefont {Caramella}}, \bibinfo
  {author} {\bibfnamefont {G.}~\bibnamefont {Onida}}, \bibinfo {author}
  {\bibfnamefont {C.}~\bibnamefont {Di~Valentin}},\ and\ \bibinfo {author}
  {\bibfnamefont {G.}~\bibnamefont {Pacchioni}},\ }\bibfield  {title} {\enquote
  {\bibinfo {title} {Defect calculations in semiconductors through a
  dielectric-dependent hybrid {DFT} functional: The case of oxygen vacancies in
  metal oxides},}\ }\href@noop {} {\bibfield  {journal} {\bibinfo  {journal}
  {J. Chem. Phys.}\ }\textbf {\bibinfo {volume} {143}},\ \bibinfo {pages}
  {134702} (\bibinfo {year} {2015}{\natexlab{b}})}\BibitemShut {NoStop}%
\bibitem [{\citenamefont {Das}\ \emph {et~al.}(2019)\citenamefont {Das},
  \citenamefont {Di~Liberto}, \citenamefont {Tosoni},\ and\ \citenamefont
  {Pacchioni}}]{DDTP19}%
  \BibitemOpen
  \bibfield  {author} {\bibinfo {author} {\bibfnamefont {T.}~\bibnamefont
  {Das}}, \bibinfo {author} {\bibfnamefont {G.}~\bibnamefont {Di~Liberto}},
  \bibinfo {author} {\bibfnamefont {S.}~\bibnamefont {Tosoni}},\ and\ \bibinfo
  {author} {\bibfnamefont {G.}~\bibnamefont {Pacchioni}},\ }\bibfield  {title}
  {\enquote {\bibinfo {title} {Band gap of 3{D} metal oxides and quasi-2{D}
  materials from hybrid density functional theory: Are dielectric-dependent
  functionals superior?}}\ }\href {https://doi.org/10.1021/acs.jctc.9b00545}
  {\bibfield  {journal} {\bibinfo  {journal} {J. Chem. Theory Comput.}\
  }\textbf {\bibinfo {volume} {15}},\ \bibinfo {pages} {6294} (\bibinfo {year}
  {2019})}\BibitemShut {NoStop}%
\bibitem [{\citenamefont {H\"{a}fner}\ and\ \citenamefont
  {Bredow}(2020)}]{HB20}%
  \BibitemOpen
  \bibfield  {author} {\bibinfo {author} {\bibfnamefont {M.}~\bibnamefont
  {H\"{a}fner}}\ and\ \bibinfo {author} {\bibfnamefont {T.}~\bibnamefont
  {Bredow}},\ }\bibfield  {title} {\enquote {\bibinfo {title} {${F}$ and ${M}$
  centers in alkali halides: A theoretical study applying self-consistent
  dielectric-dependent hybrid density functional theory},}\ }\href@noop {}
  {\bibfield  {journal} {\bibinfo  {journal} {Phys. Rev. B}\ }\textbf {\bibinfo
  {volume} {102}},\ \bibinfo {pages} {184108} (\bibinfo {year}
  {2020})}\BibitemShut {NoStop}%
\bibitem [{\citenamefont {Hochheim}\ and\ \citenamefont {Bredow}(2018)}]{HB18}%
  \BibitemOpen
  \bibfield  {author} {\bibinfo {author} {\bibfnamefont {M.}~\bibnamefont
  {Hochheim}}\ and\ \bibinfo {author} {\bibfnamefont {T.}~\bibnamefont
  {Bredow}},\ }\bibfield  {title} {\enquote {\bibinfo {title} {Band-edge levels
  of the {NaCl}(100) surface: Self-consistent hybrid density functional theory
  compared to many-body perturbation theory},}\ }\href@noop {} {\bibfield
  {journal} {\bibinfo  {journal} {Phys. Rev. B}\ }\textbf {\bibinfo {volume}
  {97}},\ \bibinfo {pages} {235447} (\bibinfo {year} {2018})}\BibitemShut
  {NoStop}%
\bibitem [{\citenamefont {Hinuma}\ \emph {et~al.}(2017)\citenamefont {Hinuma},
  \citenamefont {Kumagai}, \citenamefont {Tanaka},\ and\ \citenamefont
  {Oba}}]{HKTO17}%
  \BibitemOpen
  \bibfield  {author} {\bibinfo {author} {\bibfnamefont {Y.}~\bibnamefont
  {Hinuma}}, \bibinfo {author} {\bibfnamefont {Y.}~\bibnamefont {Kumagai}},
  \bibinfo {author} {\bibfnamefont {I.}~\bibnamefont {Tanaka}},\ and\ \bibinfo
  {author} {\bibfnamefont {F.}~\bibnamefont {Oba}},\ }\bibfield  {title}
  {\enquote {\bibinfo {title} {Band alignment of semiconductors and insulators
  using dielectric-dependent hybrid functionals: Toward high-throughput
  evaluation},}\ }\href {https://doi.org/10.1103/physrevb.95.075302} {\bibfield
   {journal} {\bibinfo  {journal} {Phys. Rev. B}\ }\textbf {\bibinfo {volume}
  {95}},\ \bibinfo {pages} {075302} (\bibinfo {year} {2017})}\BibitemShut
  {NoStop}%
\bibitem [{\citenamefont {Hinuma}, \citenamefont {Gake},\ and\ \citenamefont
  {Oba}(2019)}]{HGO19}%
  \BibitemOpen
  \bibfield  {author} {\bibinfo {author} {\bibfnamefont {Y.}~\bibnamefont
  {Hinuma}}, \bibinfo {author} {\bibfnamefont {T.}~\bibnamefont {Gake}},\ and\
  \bibinfo {author} {\bibfnamefont {F.}~\bibnamefont {Oba}},\ }\bibfield
  {title} {\enquote {\bibinfo {title} {Band alignment at surfaces and
  heterointerfaces of {Al$_2$O$_3$, Ga$_2$O$_3$, In$_2$O$_3$}, and related
  group-{III} oxide polymorphs: A first-principles study},}\ }\href
  {https://doi.org/10.1103/physrevmaterials.3.084605} {\bibfield  {journal}
  {\bibinfo  {journal} {Phys. Rev. Mater.}\ }\textbf {\bibinfo {volume} {3}},\
  \bibinfo {pages} {084605} (\bibinfo {year} {2019})}\BibitemShut {NoStop}%
\bibitem [{\citenamefont {Ni}\ \emph {et~al.}(2022)\citenamefont {Ni},
  \citenamefont {Hu}, \citenamefont {Li}, \citenamefont {Jia}, \citenamefont
  {Qin}, \citenamefont {Li}, \citenamefont {Zhou}, \citenamefont {Zha},
  \citenamefont {Ren},\ and\ \citenamefont {Wang}}]{NHLJ22}%
  \BibitemOpen
  \bibfield  {author} {\bibinfo {author} {\bibfnamefont {J.}~\bibnamefont
  {Ni}}, \bibinfo {author} {\bibfnamefont {S.}~\bibnamefont {Hu}}, \bibinfo
  {author} {\bibfnamefont {M.}~\bibnamefont {Li}}, \bibinfo {author}
  {\bibfnamefont {F.}~\bibnamefont {Jia}}, \bibinfo {author} {\bibfnamefont
  {G.}~\bibnamefont {Qin}}, \bibinfo {author} {\bibfnamefont {Q.}~\bibnamefont
  {Li}}, \bibinfo {author} {\bibfnamefont {Z.}~\bibnamefont {Zhou}}, \bibinfo
  {author} {\bibfnamefont {F.}~\bibnamefont {Zha}}, \bibinfo {author}
  {\bibfnamefont {W.}~\bibnamefont {Ren}},\ and\ \bibinfo {author}
  {\bibfnamefont {Y.}~\bibnamefont {Wang}},\ }\bibfield  {title} {\enquote
  {\bibinfo {title} {Accurate band offset prediction of {Sc$_2$O$_3$/GaN} and
  $\theta$‐{Al$_2$O$_3$/GaN} heterojunctions using a dielectric-dependent
  hybrid functional},}\ }\href@noop {} {\bibfield  {journal} {\bibinfo
  {journal} {ACS Appl. Electron. Mater.}\ }\textbf {\bibinfo {volume} {4}},\
  \bibinfo {pages} {2747} (\bibinfo {year} {2022})}\BibitemShut {NoStop}%
\bibitem [{\citenamefont {Cruz}, \citenamefont {Lam},\ and\ \citenamefont
  {Burke}(1998)}]{CLB98}%
  \BibitemOpen
  \bibfield  {author} {\bibinfo {author} {\bibfnamefont {F.~G.}\ \bibnamefont
  {Cruz}}, \bibinfo {author} {\bibfnamefont {K.-C.}\ \bibnamefont {Lam}},\ and\
  \bibinfo {author} {\bibfnamefont {K.}~\bibnamefont {Burke}},\ }\bibfield
  {title} {\enquote {\bibinfo {title} {Exchange-correlation energy density from
  virial theorem},}\ }\href@noop {} {\bibfield  {journal} {\bibinfo  {journal}
  {J. Phys. Chem. A}\ }\textbf {\bibinfo {volume} {102}},\ \bibinfo {pages}
  {4911} (\bibinfo {year} {1998})}\BibitemShut {NoStop}%
\bibitem [{\citenamefont {Jaramillo}, \citenamefont {Scuseria},\ and\
  \citenamefont {Ernzerhof}(2003)}]{JES03}%
  \BibitemOpen
  \bibfield  {author} {\bibinfo {author} {\bibfnamefont {J.}~\bibnamefont
  {Jaramillo}}, \bibinfo {author} {\bibfnamefont {G.~E.}\ \bibnamefont
  {Scuseria}},\ and\ \bibinfo {author} {\bibfnamefont {M.}~\bibnamefont
  {Ernzerhof}},\ }\bibfield  {title} {\enquote {\bibinfo {title} {Local hybrid
  functionals},}\ }\href {https://doi.org/10.1063/1.1528936} {\bibfield
  {journal} {\bibinfo  {journal} {J. Chem. Phys.}\ }\textbf {\bibinfo {volume}
  {118}},\ \bibinfo {pages} {1068} (\bibinfo {year} {2003})}\BibitemShut
  {NoStop}%
\bibitem [{\citenamefont {Burke}, \citenamefont {Cruz},\ and\ \citenamefont
  {Lam}(1998)}]{BCL98}%
  \BibitemOpen
  \bibfield  {author} {\bibinfo {author} {\bibfnamefont {K.}~\bibnamefont
  {Burke}}, \bibinfo {author} {\bibfnamefont {F.~G.}\ \bibnamefont {Cruz}},\
  and\ \bibinfo {author} {\bibfnamefont {K.-C.}\ \bibnamefont {Lam}},\
  }\bibfield  {title} {\enquote {\bibinfo {title} {Unambiguous
  exchange-correlation energy density},}\ }\href
  {https://doi.org/10.1063/1.477479} {\bibfield  {journal} {\bibinfo  {journal}
  {J. Chem. Phys.}\ }\textbf {\bibinfo {volume} {109}},\ \bibinfo {pages}
  {8161} (\bibinfo {year} {1998})}\BibitemShut {NoStop}%
\bibitem [{\citenamefont {Bahmann}\ \emph {et~al.}(2007)\citenamefont
  {Bahmann}, \citenamefont {Rodenberg}, \citenamefont {Arbuznikov},\ and\
  \citenamefont {Kaupp}}]{BRAK07}%
  \BibitemOpen
  \bibfield  {author} {\bibinfo {author} {\bibfnamefont {H.}~\bibnamefont
  {Bahmann}}, \bibinfo {author} {\bibfnamefont {A.}~\bibnamefont {Rodenberg}},
  \bibinfo {author} {\bibfnamefont {A.~V.}\ \bibnamefont {Arbuznikov}},\ and\
  \bibinfo {author} {\bibfnamefont {M.}~\bibnamefont {Kaupp}},\ }\bibfield
  {title} {\enquote {\bibinfo {title} {A thermochemically competitive local
  hybrid functional without gradient corrections},}\ }\href
  {https://doi.org/10.1063/1.2429058} {\bibfield  {journal} {\bibinfo
  {journal} {J. Chem. Phys.}\ }\textbf {\bibinfo {volume} {126}},\ \bibinfo
  {pages} {011103} (\bibinfo {year} {2007})}\BibitemShut {NoStop}%
\bibitem [{\citenamefont {Perdew}\ \emph {et~al.}(2008)\citenamefont {Perdew},
  \citenamefont {Staroverov}, \citenamefont {Tao},\ and\ \citenamefont
  {Scuseria}}]{PSTS08}%
  \BibitemOpen
  \bibfield  {author} {\bibinfo {author} {\bibfnamefont {J.~P.}\ \bibnamefont
  {Perdew}}, \bibinfo {author} {\bibfnamefont {V.~N.}\ \bibnamefont
  {Staroverov}}, \bibinfo {author} {\bibfnamefont {J.}~\bibnamefont {Tao}},\
  and\ \bibinfo {author} {\bibfnamefont {G.~E.}\ \bibnamefont {Scuseria}},\
  }\bibfield  {title} {\enquote {\bibinfo {title} {Density functional with full
  exact exchange, balanced nonlocality of correlation, and constraint
  satisfaction},}\ }\href {https://doi.org/10.1103/physreva.78.052513}
  {\bibfield  {journal} {\bibinfo  {journal} {Phys. Rev. A}\ }\textbf {\bibinfo
  {volume} {78}},\ \bibinfo {pages} {052513} (\bibinfo {year}
  {2008})}\BibitemShut {NoStop}%
\bibitem [{\citenamefont {Schmidt}\ \emph {et~al.}(2014)\citenamefont
  {Schmidt}, \citenamefont {Kraisler}, \citenamefont {Makmal}, \citenamefont
  {Kronik},\ and\ \citenamefont {Kuemmel}}]{SKMK14}%
  \BibitemOpen
  \bibfield  {author} {\bibinfo {author} {\bibfnamefont {T.}~\bibnamefont
  {Schmidt}}, \bibinfo {author} {\bibfnamefont {E.}~\bibnamefont {Kraisler}},
  \bibinfo {author} {\bibfnamefont {A.}~\bibnamefont {Makmal}}, \bibinfo
  {author} {\bibfnamefont {L.}~\bibnamefont {Kronik}},\ and\ \bibinfo {author}
  {\bibfnamefont {S.}~\bibnamefont {Kuemmel}},\ }\bibfield  {title} {\enquote
  {\bibinfo {title} {A self-interaction-free local hybrid functional: Accurate
  binding energies vis-a-vis accurate ionization potentials from {K}ohn-{S}ham
  eigenvalues},}\ }\href {https://doi.org/10.1063/1.4865942} {\bibfield
  {journal} {\bibinfo  {journal} {J. Chem. Phys.}\ }\textbf {\bibinfo {volume}
  {140}},\ \bibinfo {pages} {18A510} (\bibinfo {year} {2014})}\BibitemShut
  {NoStop}%
\bibitem [{\citenamefont {Johnson}(2014)}]{J14}%
  \BibitemOpen
  \bibfield  {author} {\bibinfo {author} {\bibfnamefont {E.~R.}\ \bibnamefont
  {Johnson}},\ }\bibfield  {title} {\enquote {\bibinfo {title} {Local-hybrid
  functional based on the correlation length},}\ }\href
  {https://doi.org/10.1063/1.4896302} {\bibfield  {journal} {\bibinfo
  {journal} {J. Chem. Phys.}\ }\textbf {\bibinfo {volume} {141}},\ \bibinfo
  {pages} {124120} (\bibinfo {year} {2014})}\BibitemShut {NoStop}%
\bibitem [{\citenamefont {Holzer}\ and\ \citenamefont {Franzke}(2022)}]{HF22}%
  \BibitemOpen
  \bibfield  {author} {\bibinfo {author} {\bibfnamefont {C.}~\bibnamefont
  {Holzer}}\ and\ \bibinfo {author} {\bibfnamefont {Y.~J.}\ \bibnamefont
  {Franzke}},\ }\bibfield  {title} {\enquote {\bibinfo {title} {A local hybrid
  exchange functional approximation from first principles},}\ }\href
  {https://doi.org/10.1063/5.0100439} {\bibfield  {journal} {\bibinfo
  {journal} {J. Chem. Phys}\ }\textbf {\bibinfo {volume} {157}},\ \bibinfo
  {pages} {034108} (\bibinfo {year} {2022})}\BibitemShut {NoStop}%
\bibitem [{\citenamefont {Janesko}, \citenamefont {Krukau},\ and\ \citenamefont
  {Scuseria}(2008)}]{JKS08}%
  \BibitemOpen
  \bibfield  {author} {\bibinfo {author} {\bibfnamefont {B.~G.}\ \bibnamefont
  {Janesko}}, \bibinfo {author} {\bibfnamefont {A.~V.}\ \bibnamefont
  {Krukau}},\ and\ \bibinfo {author} {\bibfnamefont {G.~E.}\ \bibnamefont
  {Scuseria}},\ }\bibfield  {title} {\enquote {\bibinfo {title}
  {Self-consistent generalized {K}ohn-{S}ham local hybrid functionals of
  screened exchange: Combining local and range-separated hybridization},}\
  }\href@noop {} {\bibfield  {journal} {\bibinfo  {journal} {J. Chem. Phys.}\
  }\textbf {\bibinfo {volume} {129}},\ \bibinfo {pages} {124110} (\bibinfo
  {year} {2008})}\BibitemShut {NoStop}%
\bibitem [{\citenamefont {Haunschild}\ and\ \citenamefont
  {Scuseria}(2010)}]{HS10}%
  \BibitemOpen
  \bibfield  {author} {\bibinfo {author} {\bibfnamefont {R.}~\bibnamefont
  {Haunschild}}\ and\ \bibinfo {author} {\bibfnamefont {G.~E.}\ \bibnamefont
  {Scuseria}},\ }\bibfield  {title} {\enquote {\bibinfo {title}
  {Range-separated local hybrids},}\ }\href {https://doi.org/10.1063/1.3451078}
  {\bibfield  {journal} {\bibinfo  {journal} {J. Chem. Phys.}\ }\textbf
  {\bibinfo {volume} {132}},\ \bibinfo {pages} {224106} (\bibinfo {year}
  {2010})}\BibitemShut {NoStop}%
\bibitem [{\citenamefont {Krukau}\ \emph {et~al.}(2008)\citenamefont {Krukau},
  \citenamefont {Scuseria}, \citenamefont {Perdew},\ and\ \citenamefont
  {Savin}}]{KSPS08}%
  \BibitemOpen
  \bibfield  {author} {\bibinfo {author} {\bibfnamefont {A.~V.}\ \bibnamefont
  {Krukau}}, \bibinfo {author} {\bibfnamefont {G.~E.}\ \bibnamefont
  {Scuseria}}, \bibinfo {author} {\bibfnamefont {J.~P.}\ \bibnamefont
  {Perdew}},\ and\ \bibinfo {author} {\bibfnamefont {A.}~\bibnamefont
  {Savin}},\ }\bibfield  {title} {\enquote {\bibinfo {title} {Hybrid
  functionals with local range separation},}\ }\href
  {https://doi.org/10.1063/1.2978377} {\bibfield  {journal} {\bibinfo
  {journal} {J. Chem. Phys.}\ }\textbf {\bibinfo {volume} {129}},\ \bibinfo
  {pages} {124103} (\bibinfo {year} {2008})}\BibitemShut {NoStop}%
\bibitem [{\citenamefont {Klawohn}\ and\ \citenamefont {Bahmann}(2020)}]{KB20}%
  \BibitemOpen
  \bibfield  {author} {\bibinfo {author} {\bibfnamefont {S.}~\bibnamefont
  {Klawohn}}\ and\ \bibinfo {author} {\bibfnamefont {H.}~\bibnamefont
  {Bahmann}},\ }\bibfield  {title} {\enquote {\bibinfo {title} {Self-consistent
  implementation of hybrid functionals with local range separation},}\
  }\href@noop {} {\bibfield  {journal} {\bibinfo  {journal} {J. Chem. Theory
  Comput.}\ }\textbf {\bibinfo {volume} {16}},\ \bibinfo {pages} {953}
  (\bibinfo {year} {2020})}\BibitemShut {NoStop}%
\bibitem [{\citenamefont {Maier}, \citenamefont {Arbuznikov},\ and\
  \citenamefont {Kaupp}(2018)}]{MAK18}%
  \BibitemOpen
  \bibfield  {author} {\bibinfo {author} {\bibfnamefont {T.~M.}\ \bibnamefont
  {Maier}}, \bibinfo {author} {\bibfnamefont {A.~V.}\ \bibnamefont
  {Arbuznikov}},\ and\ \bibinfo {author} {\bibfnamefont {M.}~\bibnamefont
  {Kaupp}},\ }\bibfield  {title} {\enquote {\bibinfo {title} {Local hybrid
  functionals: Theory, implementation, and performance of an emerging new tool
  in quantum chemistry and beyond},}\ }\href
  {https://doi.org/10.1002/wcms.1378} {\bibfield  {journal} {\bibinfo
  {journal} {WIREs Comput. Mol. Sci.}\ }\textbf {\bibinfo {volume} {9}},\
  \bibinfo {pages} {e1378} (\bibinfo {year} {2018})}\BibitemShut {NoStop}%
\bibitem [{\citenamefont {Borlido}, \citenamefont {Marques},\ and\
  \citenamefont {Botti}(2018)}]{BMB18}%
  \BibitemOpen
  \bibfield  {author} {\bibinfo {author} {\bibfnamefont {P.}~\bibnamefont
  {Borlido}}, \bibinfo {author} {\bibfnamefont {M.~A.~L.}\ \bibnamefont
  {Marques}},\ and\ \bibinfo {author} {\bibfnamefont {S.}~\bibnamefont
  {Botti}},\ }\bibfield  {title} {\enquote {\bibinfo {title} {Local hybrid
  density functional for interfaces},}\ }\href
  {https://doi.org/10.1021/acs.jctc.7b00853} {\bibfield  {journal} {\bibinfo
  {journal} {J. Chem. Theory Comput.}\ }\textbf {\bibinfo {volume} {14}},\
  \bibinfo {pages} {939} (\bibinfo {year} {2018})}\BibitemShut {NoStop}%
\bibitem [{\citenamefont {Shimazaki}\ and\ \citenamefont
  {Nakajima}(2015)}]{SN15a}%
  \BibitemOpen
  \bibfield  {author} {\bibinfo {author} {\bibfnamefont {T.}~\bibnamefont
  {Shimazaki}}\ and\ \bibinfo {author} {\bibfnamefont {T.}~\bibnamefont
  {Nakajima}},\ }\bibfield  {title} {\enquote {\bibinfo {title} {Theoretical
  study of a screened {H}artree-{F}ock exchange potential using
  position-dependent atomic dielectric constants},}\ }\href@noop {} {\bibfield
  {journal} {\bibinfo  {journal} {J. Chem. Phys.}\ }\textbf {\bibinfo {volume}
  {142}},\ \bibinfo {pages} {074109} (\bibinfo {year} {2015})}\BibitemShut
  {NoStop}%
\bibitem [{\citenamefont {Rohlfing}(2010)}]{R10}%
  \BibitemOpen
  \bibfield  {author} {\bibinfo {author} {\bibfnamefont {M.}~\bibnamefont
  {Rohlfing}},\ }\bibfield  {title} {\enquote {\bibinfo {title} {Electronic
  excitations from a perturbative {LDA+$GdW$} approach},}\ }\href@noop {}
  {\bibfield  {journal} {\bibinfo  {journal} {Phys. Rev. B}\ }\textbf {\bibinfo
  {volume} {82}},\ \bibinfo {pages} {205127} (\bibinfo {year}
  {2010})}\BibitemShut {NoStop}%
\bibitem [{\citenamefont {Tran}\ and\ \citenamefont {Blaha}(2009)}]{TB09}%
  \BibitemOpen
  \bibfield  {author} {\bibinfo {author} {\bibfnamefont {F.}~\bibnamefont
  {Tran}}\ and\ \bibinfo {author} {\bibfnamefont {P.}~\bibnamefont {Blaha}},\
  }\bibfield  {title} {\enquote {\bibinfo {title} {Accurate band gaps of
  semiconductors and insulators with a semilocal exchange-correlation
  potential},}\ }\href@noop {} {\bibfield  {journal} {\bibinfo  {journal}
  {Phys. Rev. Lett.}\ }\textbf {\bibinfo {volume} {102}},\ \bibinfo {pages}
  {226401} (\bibinfo {year} {2009})}\BibitemShut {NoStop}%
\bibitem [{\citenamefont {Sorouri}, \citenamefont {Foulkes},\ and\
  \citenamefont {Hine}(2006)}]{SFH06}%
  \BibitemOpen
  \bibfield  {author} {\bibinfo {author} {\bibfnamefont {A.}~\bibnamefont
  {Sorouri}}, \bibinfo {author} {\bibfnamefont {W.~M.~C.}\ \bibnamefont
  {Foulkes}},\ and\ \bibinfo {author} {\bibfnamefont {N.~D.~M.}\ \bibnamefont
  {Hine}},\ }\bibfield  {title} {\enquote {\bibinfo {title} {Accurate and
  efficient method for the treatment of exchange in a plane-wave basis},}\
  }\href@noop {} {\bibfield  {journal} {\bibinfo  {journal} {J. Chem. Phys.}\
  }\textbf {\bibinfo {volume} {124}},\ \bibinfo {pages} {064105} (\bibinfo
  {year} {2006})}\BibitemShut {NoStop}%
\bibitem [{\citenamefont {King-Smith}\ and\ \citenamefont
  {Vanderbilt}(1993)}]{KV93}%
  \BibitemOpen
  \bibfield  {author} {\bibinfo {author} {\bibfnamefont {R.~D.}\ \bibnamefont
  {King-Smith}}\ and\ \bibinfo {author} {\bibfnamefont {D.}~\bibnamefont
  {Vanderbilt}},\ }\bibfield  {title} {\enquote {\bibinfo {title} {Theory of
  polarization of crystalline solids},}\ }\href@noop {} {\bibfield  {journal}
  {\bibinfo  {journal} {Phys. Rev. B}\ }\textbf {\bibinfo {volume} {47}},\
  \bibinfo {pages} {1651} (\bibinfo {year} {1993})}\BibitemShut {NoStop}%
\bibitem [{\citenamefont {Rauch}, \citenamefont {Marques},\ and\ \citenamefont
  {Botti}(2020)}]{RMB20}%
  \BibitemOpen
  \bibfield  {author} {\bibinfo {author} {\bibfnamefont {T.}~\bibnamefont
  {Rauch}}, \bibinfo {author} {\bibfnamefont {M.~A.~L.}\ \bibnamefont
  {Marques}},\ and\ \bibinfo {author} {\bibfnamefont {S.}~\bibnamefont
  {Botti}},\ }\bibfield  {title} {\enquote {\bibinfo {title} {Local modified
  {B}ecke-{J}ohnson exchange-correlation potential for interfaces, surfaces,
  and two-dimensional materials},}\ }\href@noop {} {\bibfield  {journal}
  {\bibinfo  {journal} {J. Chem. Theory Comput.}\ }\textbf {\bibinfo {volume}
  {16}},\ \bibinfo {pages} {2654} (\bibinfo {year} {2020})}\BibitemShut
  {NoStop}%
\bibitem [{\citenamefont {Becke}\ and\ \citenamefont {Roussel}(1989)}]{BR89}%
  \BibitemOpen
  \bibfield  {author} {\bibinfo {author} {\bibfnamefont {A.~D.}\ \bibnamefont
  {Becke}}\ and\ \bibinfo {author} {\bibfnamefont {M.~R.}\ \bibnamefont
  {Roussel}},\ }\bibfield  {title} {\enquote {\bibinfo {title} {Exchange holes
  in inhomogeneous systems: A coordinate-space model},}\ }\href@noop {}
  {\bibfield  {journal} {\bibinfo  {journal} {Phys. Rev. A}\ }\textbf {\bibinfo
  {volume} {39}},\ \bibinfo {pages} {3761} (\bibinfo {year}
  {1989})}\BibitemShut {NoStop}%
\bibitem [{\citenamefont {Becke}\ and\ \citenamefont {Johnson}(2006)}]{BJ06}%
  \BibitemOpen
  \bibfield  {author} {\bibinfo {author} {\bibfnamefont {A.~D.}\ \bibnamefont
  {Becke}}\ and\ \bibinfo {author} {\bibfnamefont {E.~R.}\ \bibnamefont
  {Johnson}},\ }\bibfield  {title} {\enquote {\bibinfo {title} {A simple
  effective potential for exchange},}\ }\href@noop {} {\bibfield  {journal}
  {\bibinfo  {journal} {J. Chem. Phys.}\ }\textbf {\bibinfo {volume} {124}},\
  \bibinfo {pages} {221101} (\bibinfo {year} {2006})}\BibitemShut {NoStop}%
\end{thebibliography}%
\end{document}